\documentclass[12pt]{article} 
\pdfoutput=1

\usepackage{amsmath,amssymb,amsfonts}
\usepackage{color}
\definecolor{darkblue}{rgb}{0.1,0.1,.7}
\usepackage[colorlinks, linkcolor=darkblue, citecolor=darkblue, urlcolor=darkblue, linktocpage]{hyperref} 
\usepackage[square, comma, compress,numbers]{natbib}
\usepackage[]{amsmath}
\usepackage[]{graphicx}
\usepackage[]{latexsym}
\usepackage{geometry}
\usepackage{amscd}
\usepackage[all,cmtip]{xy}
\usepackage{mathrsfs}
\usepackage{bbold}
\usepackage{subfigure}
\usepackage[margin=10pt,font=small,labelfont=bf]{caption}
\usepackage{dsdshorthand}
\usepackage{simplewick}
\usepackage{changepage}
\usepackage[]{algorithm2e}
\usepackage{booktabs,multirow}
\newcommand{\es}[2] {\begin{equation} \label{#1} \begin{split} #2 \end{split} \end{equation}}
\newcommand\ext{\mathrm{ext}}
\newcommand\RP{\mathbb{RP}}

\newcommand\zMinus{{\bf 0}^-}

\begin{document}

\vspace*{-.6in} \thispagestyle{empty}
\begin{flushright}
CALT-TH-2019-051
\end{flushright}
\vspace{.2in} {\Large
\begin{center}
{\bf 
Carving out OPE space \\
and precise $O(2)$ model critical exponents
}
\end{center}
}
\vspace{.2in}
\begin{center}
{\bf 
Shai M. Chester$^{a}$,
Walter Landry$^{b,c}$,
Junyu Liu$^{c,d}$,
David Poland$^{c,e}$,\\
David Simmons-Duffin$^{c}$,
Ning Su$^{f}$,
Alessandro Vichi$^{f,g}$} 
\\
\vspace{.2in} 
$^a$ {\it Department of Particle Physics and Astrophysics, Weizmann Institute of Science, Rehovot, Israel}\\
$^b$ {\it Simons Collaboration on the Nonperturbative Bootstrap}\\
$^c$ {\it Walter Burke Institute for Theoretical Physics, Caltech, Pasadena, CA 91125, USA}\\
$^d$ {\it Institute for Quantum Information and Matter, Caltech, Pasadena, CA 91125, USA}\\
$^e$ {\it Department of Physics, Yale University, New Haven, CT 06520, USA}\\
$^f$ {\it Institute of Physics,
\'Ecole Polytechnique F\'ed\'erale de Lausanne (EPFL),\\
 CH-1015 Lausanne, Switzerland}\\
 $^g$ {\it Department of Physics, University of Pisa, I-56127 Pisa, Italy}

\end{center}

\vspace{.2in}

\begin{abstract}
We develop new tools for isolating CFTs using the numerical bootstrap. A ``cutting surface" algorithm for scanning OPE coefficients makes it possible to find islands in high-dimensional spaces. Together with recent progress in large-scale semidefinite programming, this enables bootstrap studies of much larger systems of correlation functions than was previously practical. We apply these methods to correlation functions of charge-0, 1, and 2 scalars in the 3d $O(2)$ model, computing new precise values for scaling dimensions and OPE coefficients in this theory. Our new determinations of scaling dimensions are consistent with and improve upon existing Monte Carlo simulations, sharpening the existing decades-old $8\sigma$ discrepancy between theory and experiment.
\end{abstract}

\newpage

\tableofcontents

\newpage

\section{Introduction}
\label{sec:introduction}

\subsection{Large-scale bootstrap problems}
\label{sec:largeBootstrap}

Numerical bootstrap methods  \cite{Rattazzi:2008pe,Rychkov:2009ij} (see \cite{Poland:2018epd,Chester:2019wfx} for recent reviews) can help achieve two important goals: (1) make general statements about the space of all CFTs, and (2) isolate specific theories and compute their observables to high precision. In this work, we introduce new tools for isolating theories, and apply them to the 3d critical $O(2)$ model.

To isolate a theory with the numerical bootstrap, one must choose a set of crossing symmetry equations and make reasonable assumptions about the spectrum of the theory. By analyzing the crossing equations using convex optimization, one obtains exclusion plots in the space of CFT data. In favorable circumstances, such exclusion plots contain small islands around the theory of interest --- we then say that we have ``isolated" the theory \cite{Kos:2014bka,Kos:2015mba,Kos:2016ysd,Rong:2018okz,Agmon:2019imm}. It is unknown in general which crossing equations and assumptions are needed to isolate a given theory. However, it is clearly important to incorporate as much information about the target theory as possible. In practice, this means we would like to study large systems of correlation functions involving multiple scalars \cite{Li:2016wdp,Nakayama:2016jhq,Li:2017ddj,Behan:2018hfx,Kousvos:2018rhl,Kousvos:2019hgc}, fermions \cite{Iliesiu:2015qra,Iliesiu:2017nrv,Karateev:2019pvw}, currents \cite{Dymarsky:2017xzb,Reehorst:2019pzi}, stress tensors \cite{Dymarsky:2017yzx}, various global symmetry representations \cite{Rattazzi:2010yc,Vichi:2011ux,Poland:2011ey,Kos:2013tga,Berkooz:2014yda,Nakayama:2014lva,Caracciolo:2014cxa,Nakayama:2014sba,Chester:2014gqa,Nakayama:2014yia,Chester:2015qca,Chester:2015lej,Chester:2016wrc,Nakayama:2016knq,Iha:2016ppj,Nakayama:2017vdd,Rong:2017cow,Chester:2017vdh,Stergiou:2018gjj,Li:2018lyb,Rong:2019qer}, etc.. There are many indications that such large-scale bootstrap problems could help isolate myriad interesting theories.\footnote{See for instance \cite{Rychkov:2011et,ElShowk:2012ht,Gaiotto:2013nva,El-Showk:2014dwa,Chang:2017cdx,Li:2017kck,Hasegawa:2018yqg,Gowdigere:2018lxz,Stergiou:2019dcv} or \cite{Poland:2010wg,Beem:2013qxa,Alday:2013opa,Alday:2014qfa,Chester:2014fya,Beem:2014zpa,Bobev:2015jxa,Beem:2015aoa,Poland:2015mta,Lemos:2015awa,Lin:2015wcg,Lin:2016gcl,Bae:2016jpi,Lemos:2016xke,Beem:2016wfs,Cornagliotto:2017dup,Chang:2017xmr,Cornagliotto:2017snu,Agmon:2017xes,Baggio:2017mas,Liendo:2018ukf,Atanasov:2018kqw,Chang:2019dzt} for supersymmetric studies. Other analysis can also be found in  \cite{Rattazzi:2010gj,Caracciolo:2009bx,Liendo:2012hy,ElShowk:2012hu,Nakayama:2016cim,Echeverri:2016ztu,Cappelli:2018vir}.}

Until recently, our ability to study large systems of correlation functions has been limited. One tool that will facilitate going beyond previous studies is a new version of the semidefinite program solver {\tt SDPB} \cite{Simmons-Duffin:2015qma}, which can now run on hundreds of cores across multiple machines \cite{Landry:2019qug}.

Besides solving big semidefinite programs, another issue that arises in large-scale bootstrap studies is the difficulty of searching high-dimensional spaces. More crossing equations are parametrized by more input data, including scaling dimensions and OPE coefficients. If some input data is unknown, then we must scan over it to make an exclusion plot. For example, to study correlation functions of the scalars $\s$ and $\e$ in the 3d Ising model, we must scan over their scaling dimensions $\De_\s$ and $\De_\e$.  It was shown in \cite{Kos:2016ysd} that it is also beneficial to scan over the OPE coefficient ratio $\l_{\s\s \e}/\l_{\e\e\e}$. Specifically, the island in the space of scaling dimensions and OPE coefficient ratios is smaller than the island in the space of scaling dimensions alone. To study an even larger system of correlation functions, one must scan over an even larger set of scaling dimensions and OPE coefficients.

One of the main contributions in this work is an efficient ``cutting surface" algorithm for scanning over OPE coefficients. Because OPE coefficients enter quadratically in the crossing equations, our algorithm can scan a region of volume $V$ in OPE coefficient space in time $\log V$. We also explain how to use our algorithm in conjunction with hot-starting \cite{Go:2019lke}, and introduce efficient methods for scanning over scaling dimensions.

 \begin{table*}[ht]
\centering
\begin{tabular}{@{}cc|cc@{}}
	\toprule
CFT data & method & value & ref \\
	\midrule
$\Delta_s$ & EXP & 1.50946(22) & \cite{Lipa:2003zz}\\
& MC & 1.51122(15) &\cite{Hasenbusch:2019jkj} \\ 
 & CB & 1.51136({\bf22}) & \\
	 \midrule
 $\Delta_\phi$ & MC & 0.519050(40) & \cite{Hasenbusch:2019jkj} \\
 & CB & 0.519088({\bf22}) & \\
	 \midrule
 $\Delta_t$ & MC &1.2361(11) & \cite{PhysRevB.84.125136} \\
 & CB &1.23629({\bf11}) & \\
 \midrule
$\lambda_{\f\f s}$ & CB & $0.687126(27^*)$ \\
$\lambda_{sss}$ & CB & $0.830914(32^*)$ \\
$\lambda_{tts}$ & CB & $1.25213(14^*)$ \\
$\lambda_{\f\f t}$ & CB & $1.213408(65^*)$\\
$C_J/C_J^\mathrm{free}$ &CB & $0.904395(28^*)$\\
$C_T/C_T^\mathrm{free}$ &CB & $0.944056(15^*)$\\
\bottomrule 
\end{tabular}
	\caption{Comparison of conformal bootstrap (CB) results with previous determinations from Monte Carlo (MC) or experiment (EXP). We denote the leading charge 0, 1, and 2 scalars by $s,\f,t$, respectively. Bold uncertainties correspond to  rigorous intervals from bootstrap bounds. Uncertainties marked with a $^*$ indicate that the value is estimated non-rigorously by sampling points, see sections~\ref{sec:opescanresults} and \ref{sec:centralcharges}. 
	\label{tab:results}}
\end{table*}

We apply our methods to study correlation functions of the lowest-dimension charge-0, charge-1, and charge-2 scalars in the three-dimensional critical $O(2)$ model. The 3d $O(2)$ model is one of the most studied renormalization group (RG) fixed points, both theoretically and experimentally. It describes phase transitions in numerous physical systems, including ferromagnets and antiferromagnets with easy-plane anisotropy, from which it also inherits the name of the $XY$ universality class. Unfortunately, experimental results and Monte Carlo results for the critical exponents of the $O(2)$ model have been in $8\sigma$ tension for two decades. We have computed the critical exponents to high precision (with rigorous error bars). We find excellent agreement with Monte Carlo results, and a clear discrepancy with experiment. In addition, we compute numerous other scaling dimensions and OPE coefficients in the $O(2)$ model. Our results, together with comparisons to other methods, are summarized in table~\ref{tab:results}.

\subsection{Experimental and theoretical approaches to the 3d $O(2)$ model}

In the remainder of this introduction, we provide an account of past approaches to the 3d $O(2)$ model, including a history of the discrepancy between experiment and Monte Carlo. We also describe past bootstrap studies of the $O(2)$ model and motivate the calculation in this work.

The simplest continuum field theory in the $O(2)$ universality class is the theory of a scalar field $\vec \phi$ transforming in the fundamental representation of $O(2)$, with Lagrangian
\be
\label{eq:o2theory}
\cL &= \frac 1 2 |\ptl \vec \phi|^2 + \frac {1}{2} m^2 |\vec \phi|^2 + \frac{g}{4!} |\vec\phi|^4.
\ee
A large negative mass-squared for the scalar induces spontaneous symmetry breaking, leading to the ordered phase, while a large positive mass-squared leads to the disordered phase. The critical point is achieved by tuning the UV mass so that the IR correlation length diverges. Critical exponents are linked to operator dimensions at the fixed point by the simple relations
\be
\Delta_\phi = \frac{1+\eta}2 \,,\qquad \Delta_s = 3-\frac1{\nu}\,.
\ee 
Here, $s\sim |\vec{\phi}|^2$ denotes the lowest-dimension charge-0 scalar. 

\subsubsection{The $\lambda$-point experiment}

\begin{figure}[ht!]
\begin{center}
\includegraphics[scale=.8]{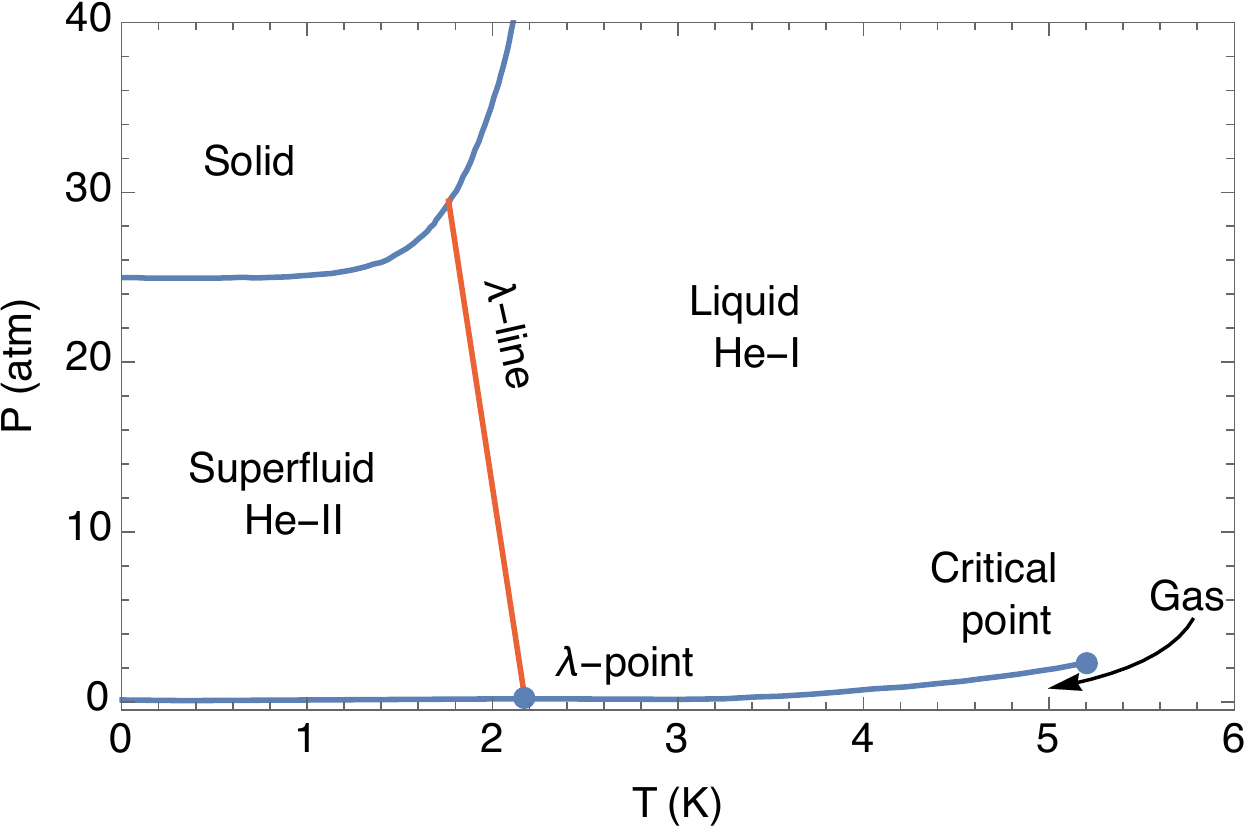}
\caption{Schematic representation of the $^4$He phase diagram. Figure taken from \cite{tilley1990superfluidity}.} 
\label{fig:phase-diagram}
\end{center}
\end{figure}

Perhaps the most intriguing experimental representative of the $O(2)$ universality class is the superfluid transition in ${}^4$He along the so-called $\lambda$-line, see Fig.~\ref{fig:phase-diagram}.  Several features make this system ideal for experimental tests of critical phenomena. Firstly, the transition is second-order along the entire $\lambda$-line. This should be compared, for instance, with the liquid-vapor transition in water\footnote{Which however belongs to the Ising universality class.} where the critical point occurs at a single point on the temperature-pressure plane.\footnote{The reason that the critical regime of liquid ${}^4$He has codimension-1 on the temperature-pressure plane is that $O(2)$ symmetry is present microscopically. It arises from phase rotations of the collective wavefunction of the superfluid condensate, which is an exact symmetry. This symmetry protects against deformations by the $\phi$ operator, and allows only a single relevant deformation: the lowest-dimension charge-0 scalar $s$.} Secondly, the steep slope of the $\lambda$-line makes the critical temperature weakly dependent on the pressure. Thirdly, that compressibility is weakly divergent at the critical point and one side of the phase transition is a superfluid state (thus free of temperature gradients) renders the system less subject to gravitational effects, which still represent the major limitation for Earth-bound experiments. 

The Earth's gravitational field creates a challenge for precise measurements of critical points in fluids. Gravity has two main effects \cite{Moldover:1979zz}: (1) it induces a density gradient in the fluid, making the system inhomogeneous; (2) more dramatically, it prevents fluctuations from growing indefinitely, making the correlation length effectively finite (gravitational rounding). Because of gravitational effects, most Earth-bound critical systems can only be tuned to $|t|\gtrsim 10^{-4}$, where $t$ is the reduced temperature $t=1-T/T_c$. Due to its favorable properties, the superfluid transition of ${}^4$He can instead reach $|t|\simeq 10^{-7}$. To get even closer to the critical regime, the $\lambda$-point experiment was conducted on the Space Shuttle Columbia in 1992 \cite{Lipa:1996zz}. The micro-gravity environment allowed the experiment to reach $|t|\simeq 5\x 10^{-9}$. In \cite{Lipa:2000zz,Lipa:2003zz}, by fitting measurements from the $\l$-point experiment, the following value of the critical exponent $\nu$ was obtained:
\be
\label{eq:experimentalnu}
\nu^\text{EXP}  = 0.6709(1) \,.
\ee

\subsubsection{Monte Carlo results}

Over the past few decades, Monte Carlo (MC) simulations of lattice models in the $O(2)$ universality class have provided the most precise theoretical predictions for critical exponents in the $O(2)$ model. The most recent determination using purely MC techniques \cite{Hasenbusch:2019jkj} gives
\be\label{eq:nuMC}
\nu^\text{MC} =0.67169(7)  \,.
\ee
We refer to \cite{Pelissetto:2000ek} and references therein for older results.\footnote{A determination that post-dates the review \cite{Pelissetto:2000ek} is $\nu^\textrm{MC}=0.6717(3)$ in~\cite{Burovski_2006}. A more recent computation using pseudo-$\epsilon$ expansion methods was performed in~\cite{Sokolov:2014mfa} giving $\nu^{\textrm{p$\epsilon$}} = 0.6706(12)$, which is closer to the experimental result. Another determination was recently obtained in \cite{Xu:2019mvy} using only MC techniques, $\nu^\text{MC}=0.67183(18)$. The latter determination and the value in (\ref{eq:nuMC}) do not entirely overlap.} The above value is fully in agreement with the second most precise theoretical  determination of $\nu$: in \cite{Campostrini:2006ms} MC simulations were combined with (uncontrolled) high temperature (HT) expansion\footnote{In the HT expansion the generating functional  
\be
Z(J) = \displaystyle\sum_{\<ij\>}e^{-\beta H + \vec{J}_i \cdot \vec{S}_i}
\ee
is expanded in powers of the inverse temperature $\beta$. Each term in the expansion can then be interpreted  as a graphical sum. Each graph consists of vertices (lattice sites) connected by bonds, each of which is associated with a factor $\beta$. The graphs enumeration becomes a combinatoric problem and can be automatized (see \cite{Pelissetto:2000ek} for a list of available HT series). Once the series is known to a sufficiently large order, it can be Borel resummed and extended down to the critical temperature. In \cite{Campostrini:2006ms} they used a 22nd order expansion.} computations to obtain
\be\label{eq:nuMC+HT}
\nu^{\text{MC}+\text{HT}}=0.6717(1) \,.
\ee
Unfortunately, comparison of the MC results (\ref{eq:nuMC}) and (\ref{eq:nuMC+HT}) with the experimental determination (\ref{eq:experimentalnu}) reveals a large discrepancy of approximatively $8\sigma$. The obvious questions is: which one is correct?

\subsubsection{The conformal bootstrap}

The numerical conformal bootstrap offers a rigorous and independent method to resolve this controversy. Three dimensional $O(N)$-models were first studied with bootstrap methods in \cite{Kos:2013tga} by considering the correlation function $\<\phi_i\phi_j \phi_k\phi_l\>$, where $\phi_i$ is the lowest-dimension scalar transforming in the vector representation of $O(N)$. That work showed that the $O(N)$-models occupy special places in the space of three dimensional CFTs: they saturate bounds on scalar operator dimensions, and their presence is signaled by a ``kink" in those bounds: a change of slope along an otherwise smooth boundary.
 
 A rigorous determination of the critical exponents $\nu$ and $\eta$ was later obtained in \cite{Kos:2015mba} by studying all nontrivial four-point functions containing $\phi_i$ and the lowest-dimension singlet scalar $s$, furthermore imposing that $\phi_i$ and $s$ are the only relevant scalars with their respective $O(N)$ representations. The resulting bounds on $(\Delta_\phi,\Delta_s)$ carve out an isolated island where the $O(N)$ model lives, together with a detached region where all other $O(N)$-symmetric CFTs satisfying these relevancy assumptions must live. 
 
 The computation of \cite{Kos:2015mba} was further improved for the cases $N=1,2,3$ in \cite{Kos:2016ysd}. The latter work used essentially the same setup of \cite{Kos:2015mba}, but additionally explored the power of scanning over OPE coefficients. Specifically, the authors asked the following question: in the space $(\Delta_\phi,\Delta_s,\theta)$, where $\theta$ parametrizes the ratio between two three point functions coefficients $\tan(\theta)\simeq\lambda_{sss}/\lambda_{\phi\phi s}$, what is the region consistent with crossing symmetry? It turned out that this apparently simple upgrade has a huge effect, but still not enough to make a conclusive statement about the MC/experiments discrepancy. 
 
A complementary approach for the case $N=2$ was initiated in \cite{Reehorst:2019pzi}, which studied the system of correlators involving the field $\phi_i$ and the conserved current associated to the global $O(2)$ symmetry. Although the determination of critical exponents was not competitive with previous bootstrap analysis, this framework gives access to new CFT-data, in particular quantities related to transport properties near the quantum critical point.\footnote{As a future direction it would be very interesting to combine this analysis with the techniques developed in this work to study the mixed system of a conserved current and multiple scalars.}

In this work, we study a larger system of correlation functions using numerical bootstrap techniques: in addition to $\phi_i$ and $s$, we incorporate the lowest-dimension charge-2 scalar $t_{ij} \sim \phi_{(i}\phi_{j)}$. A motivation for this choice is the idea that there exist strong constraints among the low-twist data of a CFT. For example, in \cite{Simmons-Duffin:2016wlq,Albayrak:2019gnz}, it was shown using the lightcone bootstrap that crossing symmetry for the operators $\s,\e$ in the 3d Ising model can be approximately recast as a set of constraints for a small amount of low-twist data, namely $\De_\s,\De_\e, \l_{\s\s\e},\l_{\e\e\e},$ and $c_T$. This immediately points to a deficiency in previous bootstrap studies of the $O(2)$ model. The operator $t_{ij}$ is expected to have lower dimension than $s$ ($\De_t\approx 1.2$, while $\De_s\approx 1.5$). Thus, it makes sense to include it in the set of crossing equations we study.
 
\begin{figure}[ht!]
\begin{center}
\includegraphics[scale=.4]{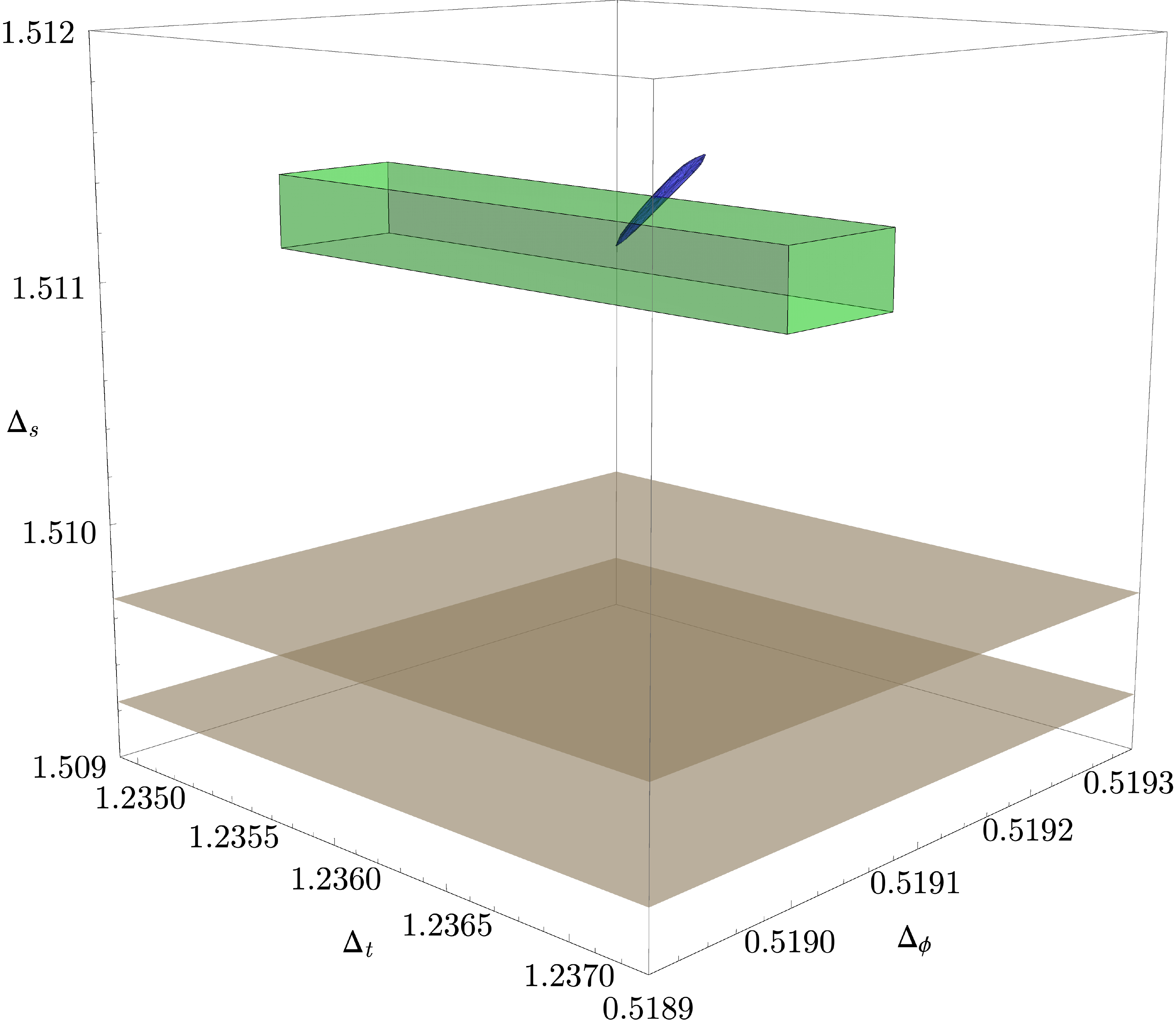}
\caption{3d region corresponding to our new $O(2)$ island using the $\{\phi_i, s, t_{ij}\}$ system and OPE scans at $\Lambda = 43$ (blue). The result is compared with the best fit values of $\Delta_{s}$ to ${}^4$He data~\cite{Lipa:2003zz} (brown planes) and the region for $\{\Delta_{\phi}, \Delta_s, \Delta_t\}$ reported by the Monte Carlo studies~\cite{PhysRevB.84.125136,Hasenbusch:2019jkj} (green box).} 
\label{fig:3dIsland-with-OPE-untransformed}
\end{center}
\end{figure} 
 
As mentioned in section~\ref{sec:largeBootstrap}, studying the larger set of crossing equations involving $\{\phi,s,t\}$ requires searching over more input data: the operator dimensions $\{\De_\f,\De_s,\De_t\}$, and the OPE coefficients $\{\l_{sss}, \l_{\f\f s}, \l_{tts}, \l_{\f\f t}\}$ (more precisely their ratios). Our new search algorithms are crucial for scanning this space efficiently. In figure~\ref{fig:3dIsland-with-OPE-untransformed}, we show the resulting island in the space of scaling dimensions $\De_\f,\De_s,\De_t$, and compare to Monte Carlo and experimental determinations. Our determination is consistent with Monte Carlo simulations and inconsistent with the results of the $\l$-point experiment.

\subsection{Structure of this work}

This work is structured as follows. In section~\ref{sec:setup}, we describe the system of correlation functions in the $O(2)$ model that we study, together with previously known information about its spectrum. In section~\ref{sec:methods}, we introduce new search methods: a ``cutting surface" algorithm for scanning over OPE coefficients, tricks for hot-starting, and Delaunay-triangulation methods for searching in dimension space. In section~\ref{sec:results}, we present results for scaling dimensions and OPE coefficients in the $O(2)$ model. Appendix~\ref{app:code} provides links to the code used in this work and appendix~\ref{app:software} contains technical details of our software and hardware setup. Other appendices provide details about the crossing equations of the $O(2)$ model and specific points that we have tested.

\section{The $O(2)$ Model}
\label{sec:setup}
\subsection{Crossing equations}
\label{cross}

We begin by describing the representation theory of $O(2)\cong U(1)\ltimes \mathbb{Z}_2$. The irreducible representations of $O(2)$ are:
\begin{itemize}
\item The trivial representation ${\bf 0^+}$.
\item The sign representation ${\bf 0^-}$, in which $U(1)$ acts trivially and the nontrivial element of $\Z_2$ acts by $-1$.
\item For each $q\in \Z_{>0}$, a unique two-dimensional irreducible representation ${\bf q}$. The states of $\bq$ have $U(1)$ charges $\pm q$ and are exchanged by $\Z_2$.
\end{itemize}
Tensor products of these irreps are given by
\es{tensor}{
&{\bf q_1}\otimes {\bf q_2}=({\bf q_1+q_2}) \oplus |{\bf q_1-q_2}|\,,\\ 
&{\bf q}\otimes {\bf q}=({\bf2q})_s \oplus {\bf0}^+_s\oplus {\bf0}^-_a\,,\\ 
&{\bf 0}^\pm\otimes {\bf q}={\bf q}\,,\\ 
&{\bf 0}^{\pm}\otimes {\bf 0}^{\pm}={\bf 0}^+_s\,,\\
&{\bf 0}^{\pm}\otimes {\bf 0}^{\mp}={\bf 0}^-\,,\\
}
where $s/a$ denotes the symmetric/antisymmetric part of the tensor product, in the case of identical irreps. For any irrep $\cR$ of $O(2)$, we define $q$ as the highest $U(1)$ charge in the representation.

Operators $\cO_{{\bf q}}(x)$ in irrep ${\bf q}$ can be written in terms of $O(2)$ fundamental indices $i=1,2$ as rank-$q$ symmetric traceless tensors $\cO^{i_1\dots i_q}(x)$. It is convenient to contract these with auxiliary polarization vectors $y^i$ that are defined to be null, $y\cdot y=0$, so that 
\es{polDef}{
\cO(x,y)\equiv\cO^{i_1\dots i_q}(x)y_{i_1}\cdots y_{i_q}\,.
}
The singlet operator $\cO_{{\bf0}^+}(x)$ has no indices or $y$'s. The $\mathbb{Z}_2$-odd operator $\cO_{{\bf0}^-}(x)$ could be written with antisymmetric indices $\cO^{[i_1 i_2]}(x)$. Alternatively, we can take into account the $O(2)$ dependence of correlation functions that include $\cO_{{\bf0}^-}(x)$ in an index-free manner by requiring that all pairs of distinct $y_1,y_2$ must be contracted as 
\es{cont}{
y_1\cdot y_2\equiv y_1^iy_2^j\delta_{ij}\,,\qquad y_1\wedge y_2\equiv \epsilon_{ij} y_1^iy_2^j\,, 
}
where the number of $\wedge$'s must be zero/one if an even/odd number of $\cO_{\bf0^-}(x)$'s appear.

Tensor structures for correlation functions of charged operators can be factorized into ``flavor" tensor structures for the $O(2)$ polarization vectors $y_i$ and ``kinematic" tensor structures that encode spacetime dependence.
For two-point functions, we have
\be
\<\cO^{\mu_1\cdots\mu_J}(x_1,y_1) \cO_{\nu_1\cdots\nu_J}(x_2,y_2)\> &= c_\cO (y_1\.y_2)^q \frac{I^{(\mu_1}_{(\nu_1}(x_{12}) \cdots I^{\mu_J)}_{\nu_J)}(x_{12})}{x_{12}^{2\De}},\\
I^\mu_\nu(x) &= \de^\mu_\nu - \frac{2x^\mu x_\nu}{x^2},\nn
\ee
where $\De$ and $J$ are the dimension and spin of $\cO$, and $q$ is the maximal $U(1)$ charge of the $O(2)$ representation of $\cO$. Here, $c_\cO$ is a constant that we usually set to $1$.

The most general three-point function we need in this work is between two scalars and a spin-$J$ operator. It takes the form
\be
&\<\vf_1(x_1,y_1)\vf_2(x_2,y_2) \cO_3^{\mu_1\cdots\mu_J}(x_3,y_3)\> \nn\\
&= \l_{\vf_1\vf_2\cO_3} T_{\cR_1\cR_2\cR_3}(y_1,y_2,y_3) \frac{Z^{(\mu_1}\cdots Z^{\mu_J)}-\mathrm{traces}}{x_{12}^{\De_1+\De_2-\De_3} x_{23}^{\De_2+\De_3-\De_1} x_{31}^{\De_3+\De_1-\De_2}},
\ee
where
\be
Z^\mu &= \frac{|x_{13}||x_{23}|}{|x_{12}|} \p{\frac{x_{13}^\mu}{x_{13}^2} - \frac{x_{23}^\mu}{x_{23}^2}}.
\ee
Here, $\cR_i$ is the $O(2)$ representation of the operator at position $x_i$.
Our conventions for flavor three-point structures are
\be
\label{eq:flavorthreept}
T_{\bq_1\bq_2\bq_3}(y_1,y_2,y_3) &= (y_1\.y_2)^{\frac{q_1+q_2-q_3}{2}} (y_2\.y_3)^{\frac{q_2+q_3-q_1}{2}} (y_3\.y_1)^{\frac{q_3+q_1-q_2}{2}}\,, && (q_1,q_2,q_3> 0)\nn\\
T_{\bq \bq {\bf 0}^+}(y_1,y_2,y_3) &= (y_1\.y_2)^q, \nn\\
T_{\bq \bq {\bf 0}^-}(y_1,y_2,y_3) &= (y_1\wedge y_2)(y_1\.y_2)^{q-1}, && (q>0) \nn\\
T_{{\bf 0}^+ {\bf 0}^+ {\bf 0}^+}(y_1,y_2,y_3) &= 1,
\ee
where we only list structures that will be needed below. In the first line, we have either $q_3=q_1+q_2$ or $q_3=|q_1-q_2|$, in accordance with the rules for tensor products.

In general, four-point functions of scalars operators $\varphi^i(x_i,y_i)$, where $i$ here labels each operator that transforms in $O(2)$ irrep $\mathcal{R}_i$, can be expanded in the $s$-channel in terms of conformal blocks as\footnote{Our conformal blocks are normalized as in the second line of table 1 in \cite{Poland:2018epd}.}
 \be
 \label{4point}
&\left\langle { \varphi^1_{\mathcal{R}_1}(x_1,y_1) \varphi^2_{\mathcal{R}_2}(x_2,y_2)  \varphi^3_{\mathcal{R}_3}(x_3,y_3)  \varphi^4_{\mathcal{R}_4}(x_4,y_4) } \right\rangle   \nn\\
&=\frac{\left(\frac{x_{24}}{x_{14}}\right)^{{\Delta_{12}}}  \left(\frac{x_{14}}{x_{13}}\right)^{{\Delta_{34}}} }{x_{12}^{\Delta_1+\Delta_2}x_{34}^{\Delta_3+\Delta_4}} \sum_{\cO}(-1)^\ell\lambda_{\varphi_1\varphi_2\cO}\lambda_{\varphi_3\varphi_4\cO}T^\cR_{\mathcal{R}_1\mathcal{R}_2\mathcal{R}_3\mathcal{R}_4}(y_i)g^{\Delta_{12},\Delta_{34}}_{\Delta,\ell}(u,v) \,,
\ee
 where $\Delta_{ij}\equiv\Delta_i-\Delta_j$, the conformal cross ratios $u,v$ are 
 \es{uv}{
 u \equiv \frac{{x_{12}^2x_{34}^2}}{{x_{13}^2x_{24}^2}},\qquad v \equiv \frac{{x_{14}^2x_{23}^2}}{{x_{13}^2x_{24}^2}}\,,
 }
and the operators $\cO$ that appear both OPEs $\varphi^1\times\varphi^2$ and $\varphi^3\times\varphi^4$ have scaling dimension $\Delta$, spin $\ell$, and transform in an irrep $\cR$ that appears in both tensor products $\mathcal{R}_1\otimes\mathcal{R}_2$ and $\mathcal{R}_3\otimes\mathcal{R}_4$. For each $\cR$, the $O(2)$ structure $T^\cR_{\mathcal{R}_1\mathcal{R}_2\mathcal{R}_3\mathcal{R}_4}(y_i)$ is a polynomial in $y_i$ for $i=1,2,3,4$, that can be derived from contracting appropriate 3-point functions as described in appendix \ref{Ts}. If $\varphi^1=\varphi^2$ (or $\varphi^3=\varphi^4$), then Bose symmetry requires that $\cO$ have only even/odd $\ell$ for $R$ in the symmetric/antisymmetric product of $\mathcal{R}_1\otimes\mathcal{R}_2$ (or $\mathcal{R}_3\otimes\mathcal{R}_4$).

We are interested in four-point functions of the lowest dimension scalar operators transforming in the ${\bf0^+}$, ${\bf1}$, and ${\bf2}$ representations, which we will denote following \cite{Kos:2013tga,Kos:2015mba,Kos:2016ysd} as $s$, $\phi$, and $t$, respectively.\footnote{The singlet $S$, traceless symmetric $T$, vector $V$ and antisymmetric $A$ irreps considered in previous $O(N)$ bootstrap papers \cite{Kos:2013tga,Kos:2015mba,Kos:2016ysd} correspond for $O(2)$ to the ${\bf 0^+}$, ${\bf 2}$, ${\bf 1}$, and ${\bf 0^-}$ irreps, respectively.} These operators are normalized via their two point functions as
\es{2points}{
\langle s(x_1) s(x_2) \rangle=\frac{1}{x_{12}^{2\Delta_s}}\,,\quad
\langle \phi(x_1,y_1) \phi(x_2,y_2) \rangle&=\frac{y_1\cdot y_2}{x_{12}^{2\Delta_\phi}}\,,\quad
\langle t(x_1,y_1) t(x_2,y_2) \rangle=\frac{(y_1\cdot y_2)^2}{x_{12}^{2\Delta_t}}\,,\\
}
where $x_{12}\equiv |x_1-x_2|$. In table \ref{table} we list the 4-point functions of $s$, $\phi$, and $t$ that are allowed by $O(2)$ symmetry\footnote{These 4-point functions, and the resulting crossing equations, are identical for a theory with just $SO(2)$ symmetry. The only difference between $O(2)$ and $SO(2)$ is that for the latter ${\bf 0^+}\cong {\bf 0^-}$ and $\epsilon_{ij}$ is now an invariant tensor, so one would need to consider correlators of operators with ${\bf 0^-}$, such as $\langle \cO_{{\bf 0^+}} \cO_{{\bf 0^+}} \cO_{{\bf 0^-}} \cO_{{\bf 0^-}}\rangle$, to distinguish between $O(2)$ and $SO(2)$.} whose $s$ and $t$-channel configuration lead to independent crossing equations, along with the irreps and spins of the operators that appear in the OPE, and the number of crossing equations that they yield.
\begin{table}
\begin{center}
\begin{tabular}{@{}c|c|c|c@{}}
	\toprule
 4-pnt& $s$-channel & $t$-channel&Eqs\\
 \midrule 
$\langle\phi\phi\phi\phi\rangle$&  $(\ell^+,{\bf0^+})$, $(\ell^-,{\bf0^-})$, $(\ell^+,{\bf2})$&same&  3   \\
 \midrule
 $\langle tttt\rangle$&   $(\ell^+,{\bf0^+})$, $(\ell^-,{\bf0^-})$, $(\ell^+,{\bf4})$&same&  3   \\
 \midrule
 $\langle t\phi t\phi\rangle$&   $(\ell^\pm,{\bf1})$, $(\ell^\pm,{\bf3})$ &same&  2   \\
 \midrule
 $\langle tt\phi\phi\rangle$&   $(\ell^+,{\bf0^+})$, $(\ell^-,{\bf0^-})$&$(\ell^\pm,{\bf1})$,$(\ell^\pm,{\bf3})$&  4   \\
 \midrule
 $\langle ssss\rangle$&  $(\ell^+,{\bf0^+})$&same&  1   \\
 \midrule
 $\langle\phi s\phi s\rangle$&  $(\ell^\pm,{\bf1})$&same&  1   \\
 \midrule
 $\langle tsts\rangle$&  $(\ell^\pm,{\bf2})$&same&  1   \\
 \midrule
 $\langle ttss\rangle$&  $(\ell^+,{\bf0^+})$& $(\ell^\pm,{\bf2})$ &  2   \\
 \midrule
 $\langle\phi\phi ss\rangle$&   $(\ell^+,{\bf0^+})$& $(\ell^\pm,{\bf1})$ &  2   \\
 \midrule
 $\langle\phi s\phi t\rangle$&   $(\ell^\pm,{\bf1})$& same &  1   \\
 \midrule
 $\langle s\phi\phi t\rangle$&   $(\ell^\pm,{\bf1})$& $(\ell^+,{\bf2})$ &  2   \\
 \bottomrule
\end{tabular}
\caption{Four-point function configurations that give independent crossing equations under equating their $s$- and $t$-channel, along with the even/odd spins that appear for each irrep in each channel, and the number of crossing equations that each configuration yields.}
\label{table}
\end{center}
\end{table}
 These 4-point functions can be written explicitly as in \eqref{4point}, where the explicit $O(2)$ structures  $T^\cR_{\mathcal{R}_1\mathcal{R}_2\mathcal{R}_3\mathcal{R}_4}(y_i)$ are computed in appendix \ref{Ts}. Equating each of these $s$-channel 4-point functions with their respective $t$-channels yields the crossing equations
 \es{crossing}{
&\sum_{ \cO_{\bf0^+},\ell^+}  \begin{pmatrix} \lambda_{ss\cO_{\bf0^+}} & \lambda_{\phi\phi\cO_{\bf0^+}} & \lambda_{tt\cO_{\bf0^+}} \end{pmatrix} \vec V_{{\bf0^+},\Delta,\ell^+}  \begin{pmatrix}  \lambda_{ss\cO_{\bf0^+}} \\ \lambda_{\phi\phi\cO_{\bf0^+}} \\ \lambda_{tt\cO_{\bf0^+}}  \end{pmatrix} 
+
 \sum_{ \cO_{\bf0^-},\ell^-}  \begin{pmatrix} \lambda_{\phi\phi \cO_{\bf0^-}} & \lambda_{tt\cO_{\bf0^-}} \end{pmatrix} \vec V_{{\bf0^-},\Delta,\ell^-}  \begin{pmatrix} \lambda_{\phi\phi \cO_{\bf0^-}} \\ \lambda_{tt\cO_{\bf0^-}}\end{pmatrix}\\
& +
\sum_{  \cO_{\bf1},\ell^\pm}  \begin{pmatrix} \lambda_{ \phi s\cO_{\bf1}} & \lambda_{ t\phi\cO_{\bf1}} \end{pmatrix} \vec V_{ {\bf1},\Delta,\ell^\pm}  \begin{pmatrix}\lambda_{ \phi s\cO_{\bf1}} \\ \lambda_{ t\phi\cO_{\bf1}} \end{pmatrix} +
\sum_{\cO_{\bf2},\ell^+}  \begin{pmatrix} \lambda_{\bf \phi\phi\cO_{\bf2}} & \lambda_{ ts\cO_{\bf 2}} \end{pmatrix} \vec V_{{\bf 2},\Delta,\ell^+}  \begin{pmatrix} \lambda_{\bf \phi\phi\cO_{\bf2}} \\ \lambda_{ ts\cO_{\bf 2}} \end{pmatrix} \\
  & +
\sum_{\cO_{\bf2},\ell^-}  \lambda_{ ts\cO_{\bf2}}^2  \vec V_{{\bf 2},\Delta,\ell^-} +
\sum_{ \cO_{\bf3} ,\ell^\pm}  \lambda_{t\phi\cO_{\bf 3}}^2  \vec V_{{\bf 3},\Delta,\ell^\pm} +
\sum_{\cO_{\bf4},\ell^+}  \lambda_{tt\cO_{\bf 4}}^2  \vec V_{{\bf 4},\Delta,\ell^+}=0\,,}
 where $\ell^\pm$ denotes which spins appear, and the $V$'s are 22-dimensional vectors of matrix or scalar crossing equations that are ordered as table \ref{table} and written in terms of
\es{Fdefine}{
F^{ij,kl}_{\mp,\Delta,\ell}(u,v)=v^{\frac{\Delta_k+\Delta_j}{2}}g_{\Delta,\ell}^{\Delta_{ij},\Delta_{kl}}(u,v)\mp u^{\frac{\Delta_k+\Delta_j}{2}}g_{\Delta,\ell}^{\Delta_{ij},\Delta_{kl}}(v,u)\,.
}
The explicit form of the $V$'s are given in appendix \ref{Vs}. The same crossing equations were derived and studied independently in~\cite{Go:2019lke}.\footnote{Furthermore, \cite{Go:2019lke} includes a software package {\tt autoboot} that can automatically derive equation~(\ref{crossing}).}

\subsection{Assumptions about the spectrum}
\label{sec:spectrum}

To obtain precise results for the $O(2)$ model, we must input some restrictions on its spectrum and OPE coefficients in order to isolate the theory. Firstly, we impose that $s,\f,t$ are the only relevant scalars in their respective charge sectors. In other words, we impose that $\Delta \geq 3$ for all charge $0,1,2$ scalars after these operators.\footnote{We also forbid any possibility of degenerate scalar contributions at the scaling dimensions $\Delta_{s,\phi,t}$, which would require additional symmetries and by definition place the model outside of the $O(2)$ universality class. While they wouldn't be expected, degenerate contributions at other dimensions are not forbidden by our algorithm.} These assumptions are well-supported by other techniques including Monte Carlo simulations and the $\e$-expansion. 

The dimension of the second charge-0 operator $s'$ is related to the critical exponent $\omega = \Delta_{s'} - 3$, which has been determined to be irrelevant using field theory and numerical techniques \cite{Hasenbusch:2019jkj, Guida:1998bx, Jasch_2001} (see also \cite{Pelissetto:2000ek} for a list of less precise estimates). E.g. \cite{Hasenbusch:2019jkj} gives $\Delta_{s'} = 3.789(4)$. Indeed, irrelevance of this operator is necessary in order to have a critical point rather than a multicritical point in which multiple tunings would be required. 

For the second charge-1 scalar $\phi'$, we are not aware of any direct determination of its scaling dimension. However, in the $\epsilon$-expansion one can show that the na\"ive second charge-1 operator, schematically $(\phi_k)^2 \phi_a$, becomes a descendant of $\phi_a$ \cite{Rychkov:2015naa}. The next charge-1 operators after this are strongly irrelevant close to 4 dimensions, and we are not aware of any evidence that continuation to $\epsilon=1$ could change this property.  Also, Monte Carlo simulations do not show any evidence of a second charge-1 relevant perturbation, which would introduce a new order parameter.

To our knowledge, the dimension of the second charge-2 operator $t'$ has only been determined in the $\epsilon$-expansion \cite{Calabrese:2002bm} to be $\Delta_{t'}\simeq 3.624(10)$, making it squarely irrelevant. Additionally, if this operator corresponded to a relevant perturbation it would have been readily detected in Monte Carlo studies of anisotropic perturbations of the $O(2)$ model~\cite{PhysRevB.84.125136}. 

The lowest-dimension charge-3 scalar in the $O(2)$ model is expected to have dimension $\approx 2.1$ \cite{DePrato:2003yd,PhysRevB.84.125136}.\footnote{We find that this is consistent with estimates based on the extremal functional method \cite{ElShowk:2012hu}.} This value is actually very close to the upper bound imposed by a bootstrap analysis \cite{Nakayama:2016jhq}.\footnote{More precisely the bound requires that given a charge-1 and charge-2 operator of dimension $(\Delta_\phi,\Delta_t)=(0.51905,1.234)$, the OPE $\phi\times t$ must contain a charge-3 operator with dimension smaller than $2.118$. Strictly speaking this bound does not apply to the $O(2)$ model since this choice of dimensions turns out to be excluded. Nevertheless, by continuity, we expect the correct bound to be very close.}  To reflect this, we impose a much weaker bound of $\De \geq 1$ for charge-3 scalars.

For charge-4 scalars, there is strong evidence from the $\e$-expansion \cite{Caselle:1997gf,Carmona:1999rm} and MC \cite{PhysRevB.84.125136,Shao:2019dbi} that there are no relevant charge-4 scalars in the $O(2)$ model. E.g., the recent MC study~\cite{Shao:2019dbi} gives the precise determination $\Delta_{\textrm{charge 4}} = 3.114(2)$. To reflect this, in most of this work we will impose $\De \geq 3$ for charge-4 scalars (following an initial study which imposes the weaker condition $\De \geq 1$). 

\begin{table}
\begin{center}
\begin{tabular}{@{}c|c|c@{}}
\toprule
charge & spin & dimensions \\
\midrule
0 & 0 & $\De_s$ or $\De\geq 3$ \\
1 & 0 & $\De_\f$ or $\De\geq 3$ \\
2 & 0 & $\De_t$ or $\De \geq 3$ \\
3 & 0 & $\De \geq 1$ \\
4 & 0 & $\De \geq 3$ \\
0 & 1 & $\De=2$ or $\De \geq 2+\de_\tau$ \\
0 & 2 & $\De=3$ or $\De\geq 3 + \de_\tau$ \\
$\cR$ & $\ell$ & $\De \geq \ell + 1 + \de_\tau$\\
\bottomrule
\end{tabular}
\end{center}
\caption{\label{tab:spectrumassumptions} Typical assumptions about the spectrum of the $O(2)$ model. In the last line, $\cR,\ell$ represent any choices of representation $\cR$ and spin $\ell$ not already represented in the table. A typical choice of twist gap is $\de_\tau=10^{-6}$.}
\end{table}

For reasons discussed in section~\ref{sec:jumps}, it is useful to impose a small gap $\de \tau$ in twist $\tau=\De-\ell$ above the unitarity bound for the non-scalar operators in the theory. (The unitarity bound for non-scalars is $\tau\geq 1$.) Of course the spectrum must include the $O(2)$ current $J^\mu$ and the stress tensor $T^{\mu\nu}$, so we impose the twist gap only for  operators with dimensions above the current and stress tensor in their respective sectors. (We impose slightly different gaps in these sectors when computing upper bounds on $C_T$ and $C_J$, as discussed in section~\ref{sec:centralcharges}.)

The presence of a small twist gap is expected to be valid in the $O(2)$ model. In the charge-0 sector, Nachtmann's theorem \cite{Nachtmann:1973mr,Komargodski:2012ek,Costa:2017twz}, together with the existence of double-twist operators \cite{Komargodski:2012ek,Fitzpatrick:2012yx}, implies that leading twists $\tau_\ell$ for each even $\ell \geq 4$ satisfy 
\be
1\leq \tau_{4}\leq \tau_\ell \leq 2\De_\f \approx 1.04
\ee
Numerous methods, including the $\e$-expansion, the lightcone bootstrap, and the extremal functional method suggest that $\tau_{4}\approx 1.02$. 
A result from \cite{Meltzer:2018tnm} shows that minimal twists in the charge-2 and charge-4 sectors are equal to or larger than the minimal twist in the charge-0 sector, for each spin. For charges 1 and 3 and odd spins in the $\zMinus$ representation, we can appeal to the $\e$-expansion which shows there are no higher-spin operators with twist near the unitarity bound. Thus, the assumption of a twist gap $\de_\tau < 0.02$ is well-justified. In most of this work, we choose $\de\tau=10^{-6}$. Overall, our assumptions about the spectrum of the $O(2)$ model are listed in table~\ref{tab:spectrumassumptions}.

The OPE coefficients of $J^\mu$ and $T^{\mu\nu}$ are constrained by Ward identities in terms of the two-point coefficients $C_J$ and $C_T$. In our conventions, we have
\be
\l_{\cO \cO T}^2 = \frac{\De_\cO^2}{2C_T/C_T^\mathrm{free}},\quad \l_{\cO\cO J}^2 = \frac{q_\cO^2}{2C_J/C_J^\mathrm{free}},
\ee
where $C_{J,T}^\mathrm{free}$ are the two-point coefficients of $J$ and $T$ in the free $O(2)$ model. Thus, the contribution of these operators to the crossing equation can be parametrized purely in terms of $C_T$ and $C_J$, together with the dimensions and charges of the external scalars $\f,s,t$.

\section{Methods}
\label{sec:methods}

\subsection{Numerical bootstrap bounds}

Given the crossing equations (\ref{crossing}), we compute bounds on CFT quantities in the standard way described in \cite{Rattazzi:2008pe,Kos:2014bka}. Suppose we would like to demonstrate that a hypothetical spectrum is inconsistent. We search for a linear functional $\a$ such that
\be
\label{eq:allpositivity}
\a(\vec V_{{\bf 0}^+,\De,\ell^+}) \succeq 0, \qquad
\a(\vec V_{{\bf 0}^-,\De,\ell^-}) \succeq 0, \qquad
\a(\vec V_{{\bf 1},\De,\ell^\pm}) \succeq 0, \qquad
\a(\vec V_{{\bf 2},\De,\ell^+}) \succeq 0, \nn\\
\a(\vec V_{{\bf 2},\De,\ell^-}) \geq 0, \qquad
\a(\vec V_{{\bf 3},\De,\ell^\pm}) \geq 0, \qquad
\a(\vec V_{{\bf 4},\De,\ell^+}) \geq 0,
\ee
for all combinations of representations, dimensions $\De$, and even or odd spins $\ell^\pm$ in some hypothetical spectrum. Here, ``$M\succeq0$" means ``$M$ is positive-semidefinite." It is conventional to normalize the contribution of the unit operator in the crossing equation to $1$:
\be
\begin{pmatrix} 1&1&1 \end{pmatrix} \a(\vec V_{{\bf 0}^+,0,0}) \begin{pmatrix} 1 \\ 1 \\ 1 \end{pmatrix} &= 1.
\ee
If a functional exists satisfying these conditions, then the hypothetical spectrum is ruled out. We search for a functional using {\tt SDPB} \cite{Landry:2019qug}.

\subsection{Positivity conditions involving the external scalars $s,\phi,t$}

The external operators $s,\phi,t$ appearing in the crossing equations require special treatment when computing bootstrap bounds.\footnote{We use the term ``external" to refer to operators that appear explicitly in the four-point functions being studied, as opposed to ``internal" operators that appear in the conformal block expansion.} 
There are four nonvanishing OPE coefficients involving just $s,\phi,t$. They can be grouped into a vector\footnote{Note that OPE coefficients of scalar operators are symmetric with respect to permutation $\l_{\f_1\f_2\f_3} = \l_{\f_1\f_3\f_2}=\textrm{four other permutations}$.}
\be
\l_\mathrm{ext} &\equiv \begin{pmatrix}
\l_{sss} \\ \l_{\f\f s} \\ \l_{tts} \\ \l_{\f\f t}
\end{pmatrix}.
\ee
We define the $4\x 4$ symmetric matrices $\vec V_\ext$ as the bilinear forms paired with $\l_\ext$ in the crossing equations. $\vec V_\ext$ is given implicitly by
\be
&\l_\mathrm{ext}^T
\vec V_\mathrm{ext}
\l_\mathrm{ext}
\nn\\
&=
\begin{pmatrix} \lambda_{sss} & \lambda_{\phi\phi s} & \lambda_{tt s} \end{pmatrix} \vec V_{{\bf0^+},\Delta_s,0}  \begin{pmatrix}  \lambda_{ss s} \\ \lambda_{\phi\phi s} \\ \lambda_{tt s}  \end{pmatrix}
+
\begin{pmatrix} \lambda_{ \f \f s} & \lambda_{ \phi \f t} \end{pmatrix} \vec V_{ {\bf1},\Delta_\f,0}  \begin{pmatrix}\lambda_{ \phi \f s} \\ \lambda_{\phi \f t} \end{pmatrix}
+
\begin{pmatrix} \lambda_{ \phi\phi t} & \lambda_{ tt s} \end{pmatrix} \vec V_{{\bf 2},\Delta_t,0}  \begin{pmatrix} \lambda_{\phi\phi t} \\ \lambda_{ tts} \end{pmatrix}.
\ee

When computing bounds, we can treat the term $\l_\mathrm{ext}^T\vec V_\mathrm{ext}\l_\mathrm{ext}$ in different ways, depending on our knowledge of $\l_\ext$. If we know nothing about $\l_\mathrm{ext}$, then we can search for a functional $\a$ such that
\be
\label{eq:externalscalarpositivesemidefiniteness}
\a(\vec V_\mathrm{ext}) \succeq 0,
\ee
where ``$\succeq 0$" means ``is positive semidefinite." In this way, we ensure that the contribution of external scalar OPE coefficients to the crossing equation has a definite sign after applying $\a$, independent of the values of those coefficients. Imposing the condition (\ref{eq:externalscalarpositivesemidefiniteness}), we can compute an allowed region $\cD$ for other quantities like operator dimensions.

However, the condition (\ref{eq:externalscalarpositivesemidefiniteness}) is stronger than necessary because it allows the matrix $M_\ext \equiv \l_\ext \l_\ext^T$ to have rank larger than $1$. Specifically, it ensures that $\Tr(M_\ext \a(\vec V_\ext))\geq 0$ for $M_\ext$ of any rank. We would like a procedure that only imposes positivity when $M_\ext$ is a rank-1 matrix.

Such a procedure was described in \cite{Kos:2016ysd,SlavaUnpublished}, and it results in stronger bounds. Suppose first that we know the direction of $\l_\mathrm{ext}$. More precisely, suppose we know the equivalence class $[\l_\mathrm{ext}] \in \mathbb{RP}^3$ of $\l_\ext$ under rescaling by a real number. In this case, the condition (\ref{eq:externalscalarpositivesemidefiniteness}) is too strong, and it suffices to impose the weaker condition\footnote{Here, $\l_\ext$ can be any representative of the equivalence class $[\l_\ext]$.}
\be
\label{eq:weakercondition}
\l_\mathrm{ext}^T \a(\vec V_\mathrm{ext}) \l_\mathrm{ext} \geq 0.
\ee
(Note that $\a(\vec V_\mathrm{ext})$ is a $4\x 4$ matrix, so that $\l_\mathrm{ext}^T \a(\vec V_\mathrm{ext}) \l_\mathrm{ext}$ is a number.) This ensures that the contribution of external scalars to the crossing equation will be positive, independent of the magnitude or sign of $\l_\mathrm{ext}$. If we use the weaker condition (\ref{eq:weakercondition}) to compute bounds on other quantities, we obtain an allowed region $\cD_{[\l_\mathrm{ext}]}$ that is smaller than $\cD$, but depends on the equivalence class $[\l_\mathrm{ext}] \in \RP^3$.

If we don't know $[\l_\mathrm{ext}]$ a-priori, we can scan over its value and compute the regions $\cD_{[\l_\mathrm{ext}]}$ as a function of $[\l_\mathrm{ext}]\in \RP^3$. The union of the resulting allowed regions must be contained inside the original allowed region $\cD$:
\be
\cD' \equiv \bigcup_{[\l_\mathrm{ext}] \in \RP^3} \cD_{[\l_\mathrm{ext}]} \subseteq \cD.
\ee
A key observation of \cite{Kos:2016ysd,SlavaUnpublished} is that this inclusion can be strict --- i.e.\ by scanning over different directions $[\l_\mathrm{ext}]$ in OPE space, and taking the union of the resulting allowed regions, we can obtain a smaller allowed region than if we impose the na\"ive condition (\ref{eq:externalscalarpositivesemidefiniteness}). Scanning over OPE coefficient directions $[\l_\ext]$ allows us to use that $\l_\ext \l_\ext^T$ is rank-1, and get better results. A disadvantage is that we must solve multiple semidefinite programs to compute the new allowed region $\cD'$.

\subsection{An algorithm for scanning over OPE coefficients}
\label{sec:cuttingsurface}

Suppose we would like to determine whether some putative scaling dimensions $(\De_s,\De_\f,\De_t)$ are allowed or not. According to the previous section, we should scan over directions in OPE coefficient space $[\l_\mathrm{ext}]\in \RP^3$. For each direction, we should compute whether a functional $\a$ exists satisfying (\ref{eq:weakercondition}) and (\ref{eq:allpositivity}). If $\a$ does not exist for some $[\l_\mathrm{ext}]$, then the point $(\De_s,\De_\f,\De_t)$ is allowed. If $\a$ exists for all $[\l_\mathrm{ext}]$, then the point $(\De_s,\De_\f,\De_t)$ is disallowed. In this section, we describe an algorithm that makes the scan over $[\l_\ext]\in \RP^3$ very efficient.

Let us choose some initial direction $[\l_\mathrm{1}]\in \RP^3$. Suppose that a functional $\a_1$ exists obeying the condition\footnote{If no such functional exists, then we know $(\De_s,\De_\f,\De_t)$ is an allowed point in dimension space, and we can stop.}
\be
\l_1^T \a_1(\vec V_\mathrm{ext}) \l_1 \geq 0,
\ee
and additionally obeying all other necessary positivity conditions (\ref{eq:allpositivity}) for computing feasibility of the given point $(\De_s,\De_\f,\De_t)$ in dimension space. The key observation is that $Q_1 = \a_1(\vec V_\mathrm{ext})$ defines a bilinear form that is positive not only for $\l_\mathrm{1}$, but also for some neighborhood $U_1\subset \RP^3$ containing $\l_1\in U_1$. That is, $\a_1$ rules out an entire neighborhood $U_1\subset \RP^3$. We can now focus on scanning over the complement $\RP^3 \setminus U_1$. 

\begin{algorithm}[t]
\Begin{
Given a list of functionals $\{\a_1,\dots,\a_n\}$, together with quadratic forms $Q_i = \a_i(\vec V_\ext)$ and regions ruled out by those quadratic forms
\be
\label{eq:quadraticinequalityforU}
U_i &\equiv \{ [\l]\in \RP^3 \textrm{ such that } \l^T Q_i \l \geq 0 \}.
\ee
The allowed region of OPE space is 
\be
\cA_n \equiv \RP^3 \setminus (\cup_{i=1}^n U_i).
\ee
\eIf{$\cA_n$ is empty}{
All directions in OPE space are ruled out.\\
\Return{Disallowed}
}{
Choose some $[\l_{n+1}] \in \cA_n $.\\
Impose the positivity condition $\l_{n+1}^T \a(\vec V_\ext) \l_{n+1} \geq 0$, and
solve the resulting semidefinite program to find a functional $\a_{n+1}$.\\
\eIf{$\a_{n+1}$ exists}{
Append $\a_{n+1}$ to the list $\{\a_1,\dots,\a_{n}\}$ and go to {\bf begin}.
}{
We have failed to rule out all directions in OPE space.\\
\Return{Allowed}
}
}
}
\caption{Cutting surface algorithm for scanning over OPE coefficients.}
\label{alg:cuttingsurface}
\end{algorithm}

This suggests Algorithm~\ref{alg:cuttingsurface} for ruling out a point $(\De_s,\De_\f,\De_t)$ in dimension space.
Algorithm~\ref{alg:cuttingsurface} is similar to so-called ``cutting plane" methods. We have a region $\cA_n$ of allowed OPE directions. We choose a point $[\l_{n+1}]\in \cA_n$ and consult an ``oracle" (the semidefinite program solver) to get a quadratic form $Q_{n+1}$ that rules out that point. This quadratic form cuts away a neighborhood $U_{n+1}$ from $\cA_n$, giving a smaller allowed region $\cA_{n+1} = \cA_n \setminus U_{n+1}$.

In traditional cutting plane methods, an oracle provides {\it linear} forms instead of quadratic forms. If the $U_i$ were half-spaces defined by linear forms, then the above algorithm would exhibit some nice properties. Firstly, the allowed regions $\cA_n$ would be convex. Secondly, if we choose $[\l_{n+1}]\in \cA_n$ to be the center of volume of $\cA_n$ (in some affine coordinates), then the neighborhood $U_{n+1}$ would be guaranteed to cut away half of $\cA_n$. Thus, the volume of $\cA_n$ would decrease exponentially in the number of cuts, and the algorithm would take logarithmic time in the volume of $\cA_n$.\footnote{For example, to search a unit cube in $D$ dimensions, it takes time proportional to $D$. The precise running time depends on how the algorithm terminates. We comment more on this below.}

\begin{figure}[t!]
\begin{center}
\includegraphics[width=\textwidth]{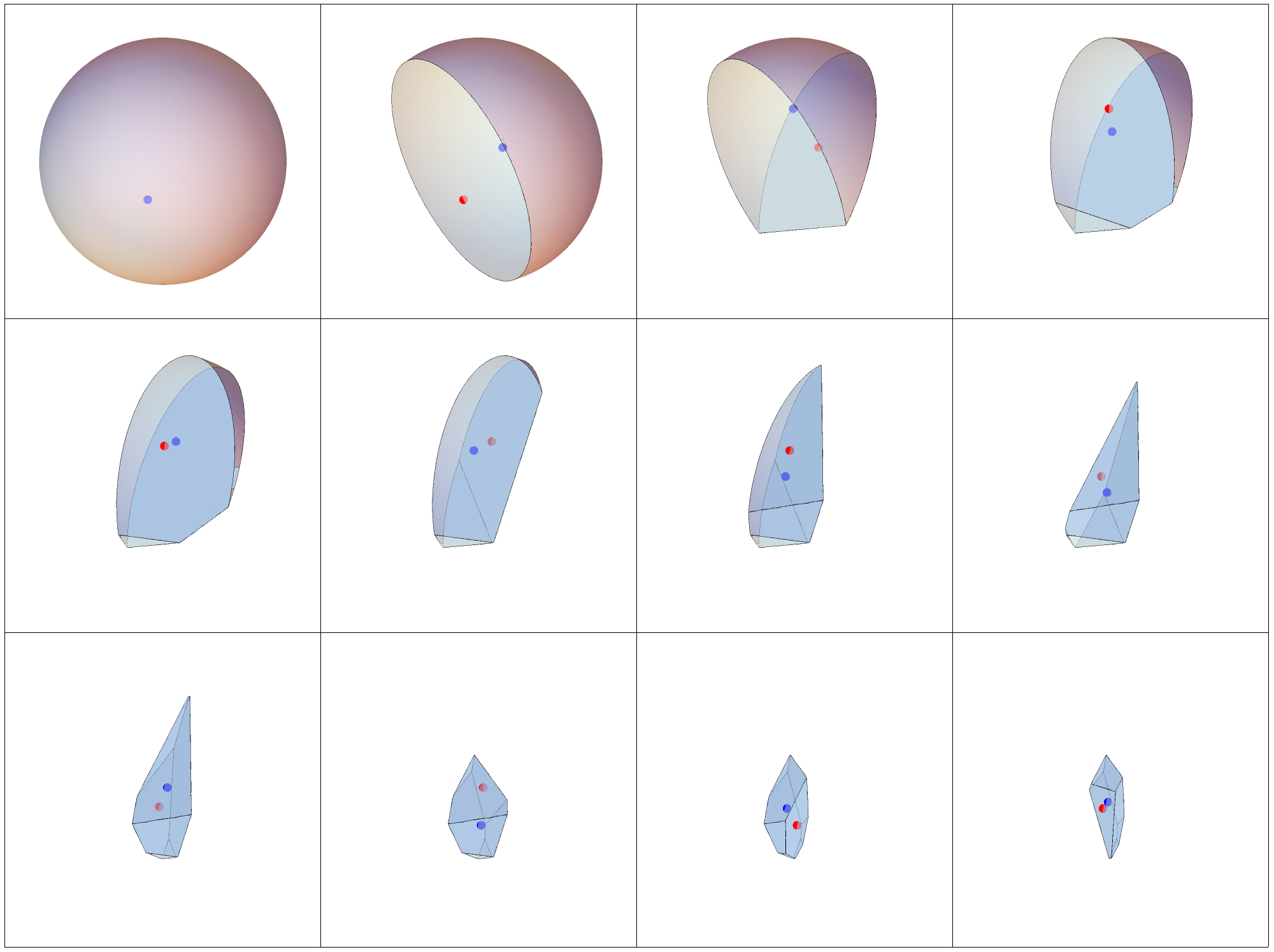}
\caption{Example allowed regions $\cA_1,\dots,\cA_{12}$ of OPE space during the cutting surface algorithm for scanning over OPE coefficients. This example is drawn from our calculation of the $O(2)$ model island with derivative order $\Lambda=43$. We plot OPE space after applying the affine transformation described in figure~\ref{fig:boundingellispoid}, which turns the initial bounding ellipsoid into the unit sphere. For each allowed region $\cA_n$, we show the point $[\l_n]$ most recently ruled out by {\tt SDPB} in red. This point is typically very close to the boundary of the allowed region. We also show the next point to be tested $[\l_{n+1}]$ in blue. We choose the blue point close to the center of $\cA_n$. In the final frame, {\tt SDPB} gives {\it primal feasible \/}for the blue point.
}
\label{fig:cuttingsurface}
\end{center}
\end{figure}

Fortunately, in many examples, we have found that once the allowed region $\cA_n$ becomes sufficiently small, the sets $U_{n+1}$ become very close to half-spaces near the allowed region, see figure~\ref{fig:cuttingsurface}. Recall that $U_{n+1}$ is defined by a quadratic inequality (\ref{eq:quadraticinequalityforU}), and thus generically has curved edges. However, as the algorithm proceeds, the radius of curvature of these edges becomes large relative to the size of the region $\cA_n$ (in some generic affine coordinates on $\RP^3$). Thus, our algorithm approximately inherits many of the nice properties of traditional cutting plane methods. We call our method a ``cutting surface" algorithm.

\subsection{Finding a point $[\l_{n+1}]$}

The most difficult step in the cutting surface algorithm is determining whether $\cA_n$ is non-empty and, if it is non-empty, choosing a point $[\l_{n+1}]\in \cA_n$. For this step, we are given a list of quadratic forms $Q_1,\dots,Q_n \in \R^{m\x m}$, and we wish to find $x=\l_{n+1}\in \R^m$ that is negative with respect to those quadratic forms. (For the computations in this work, $m=4$.) This type of problem is called a {\it quadratically constrained quadratic program} (QCQP), see e.g.\ \cite{Park2017GeneralHF}.

Unfortunately, QCQPs are NP-hard in general.\footnote{A notable exception is $m=2$. In this case, the quadratic forms become quadratic functions of a single variable in an affine patch of $\RP^1$, and the positive and negative regions can be solved for analytically. This case is relevant, for example, in the 3d Ising model problem studied in \cite{Kos:2016ysd}, which involves two OPE coefficients $\l_{\s\s\e}$ and $\l_{\e\e\e}$.} However, we have found several heuristic approaches that work well for the case at hand. Furthermore, these heuristics can be stacked: if one method fails to find a solution, we can try another method. For this work, we applied multiple heuristics, using one to verify the results of another when possible. In the next few subsections, we describe these heuristics.

Because we solve the QCQP using heuristics, our implementation of the cutting surface algorithm is non-rigorous (except when $m=2$). It would be interesting to investigate whether there exists a deterministic algorithm for QCQPs in low dimensions that could be useful in bootstrap calculations.

\subsubsection{Implementation in Mathematica}
\label{sec:mathimplementation}

For low-dimensional cases, where $Q_i\in \R^{m\x m}$ with $m=3,4$, we have implemented the cutting surface algorithm in Mathematica using standard functions. For example, in order to plot the region $\cA_n$ we pass the inequalities $\lambda^{T} Q_1 \lambda < 0, \dots, \lambda^{T} Q_n \lambda < 0$ to the functions {\tt RegionPlot} or {\tt RegionPlot3D}.\footnote{In cases where higher resolution is needed, we could specify a larger set of sample points using the {\tt PlotPoints} option, or we could define a more powerful function {\tt contourRegionPlot3D} which implements an automatic (but sometimes slow) refinement of the boundary, see \url{https://mathematica.stackexchange.com/questions/48486/high-quality-regionplot3d-for-logical-combinations-of-predicates/}.} We then use the {\tt DiscretizeGraphics} function to convert the resulting plot into a {\tt MeshRegion} corresponding to the allowed region. 

If the resulting {\tt MeshRegion} returns as {\tt EmptyRegion[m-1]} then the algorithm terminates. If it is instead nonempty, then there are various approaches one can use to select a point in its interior. One simple and fast option is to take the {\tt RegionCentroid}. This approach works most of the time, but occasionally fails when the allowed region is nonconvex. 

Another simple approach is to select the point which {\tt NMaximize}s the {\tt RegionDistance} to the {\tt RegionBoundary}, subject to the constraint of being inside the allowed region. We found that this approach leads to a working algorithm a majority of the time, but is often slow and sometimes picks suboptimal points. In the next subsection we describe a more robust procedure that we have developed for selecting an optimal point in the interior. 

Another important point is that as the allowed region gets smaller, it is helpful to apply an {\tt AffineTransform} at each iteration of the algorithm to make the allowed region roughly spherical. This for example helps to avoid the problem of missing a very small allowed region. We do this by computing a {\tt BoundingRegion} of the allowed {\tt MeshRegion} (we had good success with the form ``FastOrientedCuboid"), and then constructing an {\tt AffineTransform} which maps it to the unit cube. This transformation then gets applied to all coordinates before iterating.

\subsubsection{Minimizing $Q_n$}
\label{sec:qminimization}

We now describe some heuristics that do not depend on specialized Mathematica features and can in principle be used in general dimensions $m$. One important heuristic takes advantage of allowed regions $\cA_i$ typically becoming close to convex as the cutting surface algorithm proceeds.  Recall that $\cA_{n-1}$ is the region on which all quadratic forms $Q_1,\dots,Q_{n-1}$ are negative. Suppose this region is nonempty. Now let us add an additional quadratic form $Q_n$. We would like to know whether $Q_n$ is positive on $\cA_{n-1}$ (in which case $\cA_n$ is empty). If it is not positive, we would like to find a point $\l_{n+1}\in \cA_{n-1}$ such that $Q_n$ is negative on $\l_{n+1}$.

To do so, consider the function $f(x) = x^T Q_n x / x^T x$, where $x\in \R^m$. Because $f$ is homogeneous of degree zero, $f$ defines a function on $\RP^{m-1}$. We would like to minimize $f$ over $\cA_{n-1}$. If the minimum is negative, then the solution $[x]$ gives a point in $\cA_{n}$.

One possible minimization procedure is gradient descent starting from a point in $\cA_{n-1}$. To ensure that we stay inside $\cA_{n-1}$, we introduce a ``barrier" function
\be
B_\mathrm{\cA_{n-1}}(x) = -\sum_{i=1}^{n-1} \log \frac{x^T Q_i x}{x^T x},
\ee
and minimize the combination
\be
\label{eq:combinedfunction}
f(x) + \g B_\mathrm{\cA_{n-1}}(x),
\ee
where $\g>0$ is a parameter that we choose.
The barrier function is defined so that it is finite inside $\cA_{n-1}$ and diverges to $+\oo$ as one approaches the boundaries of $\cA_{n-1}$ from the interior.
In the limit $\g\to 0$, the minimum of (\ref{eq:combinedfunction}) converges to the minimum of $f(x)$ over $\cA_{n-1}$.

Following standard practice in interior point optimization, we combine gradient descent with decreasing the parameter $\g$. In each iteration, we compute a search direction using Newton's method for the combined function (\ref{eq:combinedfunction}). We then move along this direction and simultaneously decrease $\g$ by a constant factor.

If the region $\cA_{n-1}$ were convex and the function $f$ were convex, then the above algorithm would be guaranteed to find the minimum of $f$. We have found that in practice, convexity holds approximately for both the region $\cA_{n-1}$ and the function $f(x)$. Thus, typically this algorithm finds a suitable minimum after a single run. To increase its likelihood of success, we attempt the descent algorithm from many different randomly chosen starting points inside $\cA_{n-1}$. We sample random starting points using the hessian line search method detailed in section~\ref{sec:hessianlinesearch}.

We can make some shortcuts to the standard interior point method. First we observe that $Q_n$ is usually very small for $\lambda_n$. This means $\lambda_n$ is in fact already quite close to the $Q_n=0$ surface. One shortcut is that we can draw a line starting from $\lambda_n$ along the gradient of the function defined by $Q_n$, then test whether there is a feasible point on this line. Another shortcut is that we can simply sample some random points around $\lambda_n$. Both shortcuts have a very good chance to succeed, and are very cheap compared to the interior point method described above. Therefore we perform the shortcuts before the standard interior point method.

For the computations in this work, the simple method of minimizing $Q_n$ over $\cA_{n-1}$ works most of the time. It will be interesting to explore its applicability to higher-dimensional spaces of OPE coefficients and other bootstrap problems.

\subsubsection{Semidefinite relaxation and rank minimization}
\label{sec:semidefiniterelaxation}

Another heuristic uses the method of semidefinite relaxation, which is standard in the literature on QCQPs \cite{Park2017GeneralHF}. Recall that we would like to solve the QCQP: find $x$ such that $x^T Q_i x \leq 0$ for all $i=1,\dots,n$ (which is equivalent to $[x]\in \cA_n$). This can also be written as:
\be
\label{eq:almostansdp}
\textrm{Find $X\succeq 0$ such that $\Tr(X Q_i) \leq 0$ for all $i=1,\dots,n$, and $\mathrm{rank}(X)=1$}.
\ee
Here, $X$ is an $m\x m$ matrix and ``$\succeq$" means ``is positive semidefinite". If such an $X$ exists, then it can be written $X=xx^T$, and $x$ provides the required solution to the QCQP.

Equation (\ref{eq:almostansdp}) {\it almost} defines a semidefinite program. The only difference is the condition $\mathrm{rank}(X)=1$. Removing the rank-1 condition, we obtain the {\it semidefinite relaxation} of the original QCQP. Solving the semidefinite relaxation gives two possible outcomes:
\begin{itemize}
\item The semidefinite relaxation is infeasible (i.e.\ $X$ does not exist satisfying the conditions $\Tr(X Q_i)\leq 0$ and $X\succeq 0$). In this case, the original QCQP is necessarily infeasible. Thus, we can rigorously conclude that $\cA_{n}$ is empty.

\item The semidefinite relaxation is feasible. Typically, the resulting matrix $X$ is not particularly close to rank 1, so we must perform some additional work to find whether a solution of the QCQP exists.
\end{itemize}

In the case where the semidefinite relaxation is feasible, we use the method described in \cite{iterativerankpenalty} for finding low-rank solutions of semidefinite programs. This method involves solving a sequence of semidefinite programs with objective functions designed to successively decrease the $m-1$ smallest eigenvalues of $X$. We solve the semidefinite relaxation and the subsequent rank-minimization SDPs using {\tt SDPB}.

If rank minimization succeeds, we are left with a positive semidefinite matrix $X$ with one large eigenvalue and several small eigenvalues. To find a rank-1 solution $x x^T$, we apply the random sampling method described in \cite{Park2017GeneralHF}. We take random samples $x \in \R^m$ with covariance matrix $X=\<x x^T\>$. By construction, each inequality in the QCQP is true in expectation:
\be
\< x^T Q_i x\> &= \Tr(Q_i \<x x^T\>)=\Tr(Q_i X) \leq 0.
\ee
Thus, there is a reasonable probability of finding a sample $x$ for which all inequalities in the QCQP are true. If such a sample exists, we have solved the QCQP. If we do not find such a sample, then we cannot conclude anything about the QCQP.

An implementation of the algorithm described in this section is available online.\footnote{\url{https://gitlab.com/davidsd/quadratic-net/}} In our testing, it worked consistently in cases where OPE space is relatively low-dimensional $m\leq 4$. Indeed, this algorithm is capable of finding solutions to the QCQP in cases where the $Q_n$-minimization of section~\ref{sec:qminimization} fails (for example because $\cA_{n-1}$ has a complicated or elongated shape).  Although it takes only a few minutes to run, SDP relaxation methods are more computationally intensive than the $Q_n$-minimization. Thus, we use them as a final heuristic, which we run only when other heuristics have failed to solve the QCQP.

\subsubsection{Choosing $[\l_{n+1}]$}
\label{sec:hessianlinesearch}

When $\cA_n$ is non-empty, the heuristics in sections~\ref{sec:mathimplementation}, \ref{sec:qminimization}, and \ref{sec:semidefiniterelaxation} will usually find a point $[x]\in \cA_n$. However, to make the cutting surface algorithm as efficient as possible, we would like to choose $[\l_{n+1}]$ roughly in the ``center" of $\cA_n$. In the approach using standard Mathematica functions, one possibility is to choose $[\l_{n+1}]$ to be the {\tt RegionCentroid} of the allowed OPE region. However, for the other approaches it is important to have methods that don't require detailed knowledge of the shape of $\cA_n$ (which can be expensive to compute).

One simple approach is to minimize the barrier function $B_{\cA_n}(x)$ over $\cA_n$ (using $[x]$ as an initial point). However, for very elongated regions $\cA_n$, the minimum of the barrier function is sometimes not particularly close to the center of volume.

Note that in the case $m=2$, where OPE space $\RP^{m-1}$ is 1-dimensional, it is trivial to find a suitable $[\l_{n+1}]$. The allowed region is a union of line segments that we can solve for analytically. We can then choose the midpoint of the longest line segment (in some affine coordinates).

We can use this observation in higher dimensions. Let us start with a point $[x_0]\in \RP^{m-1}$ and choose a random line $\ell_0\subset \RP^{m-1}$ containing $[x_0]$. The intersection of the line $\ell_0$ with the region $\cA_n$ is a union of line segments (typically a single segment), and we can choose $[x_1]$ to be the midpoint of one of these segments. Repeating in this way, we obtain a sequence of points $[x_k]$ that are at the midpoints of random lines intersecting $\cA_n$. This sequence does not typically converge to a single point. However, later points in the sequence are good candidates for $[\l_{n+1}]$.\footnote{In practice, we take the last $10$ points in a long sequence and average them in some affine coordinates.} To randomly sample the line $\ell_i$, we choose coordinates around $[x_i]$ in which the Hessian of the barrier function $B_{\cA_n}(x)$ at $[x_i]$ becomes a diagonal matrix with entries $\pm 1$. In these coordinates, the region $\cA_n$ typically looks roughly spherical around $[x_i]$. We then use a uniform distribution on an infinitesimal sphere around $x_i$ in these coordinates. We call this method a ``hessian line search."

The hessian line search can be modified to randomly sample points inside $\cA_n$, with applications to the $Q_n$-minimization method of section~\ref{sec:qminimization}. Instead of choosing $x_{i+1}$ to be the midpoint of a line segment in $\ell_i \cap \cA_n$, we can choose it randomly along a segment.

\subsubsection{Bounding ellipsoids}

The cutting surface method becomes most efficient when the radius of curvature of the surface defined by the quadratic form $Q_n$ is small compared to the size of the region $\cA_{n-1}$. If we start with the allowed region $\cA_0 = \RP^{m-1}$, then it might take several iterations of the algorithm before this happens. Indeed, in our testing, the cutting surface algorithm often spent significant time cutting away parts of $\RP^{m-1}$ that are known to be far from the correct values of OPE coefficients. To avoid this problem, it is useful to impose a ``bounding box" in OPE space. An efficient way to do this is to pick a bounding ellipsoid, and choose $Q_1$ to be the quadratic form that rules out the exterior of the ellipsoid.

\begin{figure}[t!]
\begin{center}
\includegraphics[width=.9\textwidth]{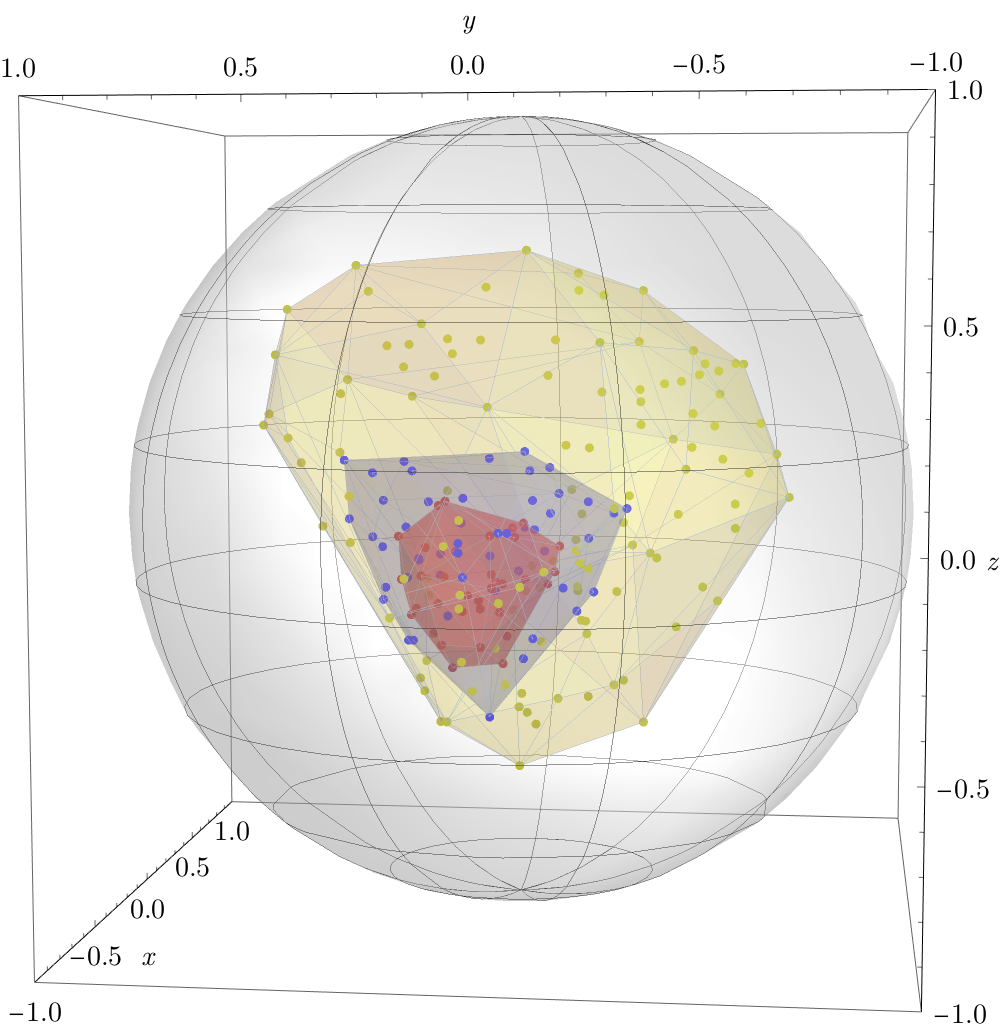}
\caption{Allowed points in external scalar OPE coefficient space, found while computing the allowed island in dimension space, for $\Lambda=27$ (yellow), $\Lambda=35$ (blue), and $\Lambda=43$ (red), together with a choice of bounding ellipsoid (gray).  For each set of points, we also show their convex hull in the same color. To plot the points, we applied an affine transformation to make the $\Lambda=27$ region roughly spherical. The relationship between the displayed coordinates $x,y,z$ and the OPE coefficients is $\l_\ext=(751.0591846177696 - 362.65959721052656x - 131.334377405401y - 41.46952958591952z, 1, 3383.753238900843 + 695.8131625006117x - 1729.4094085965235y - 
  607.9744222068027z, -12562.290081255807 + 123.88628689820867x - 3799.4579787849975y + 10949.506824631871z)$. After finding the $\Lambda=27$ points, we chose the gray sphere as a bounding ellipsoid for the computation with $\Lambda=35$. No $\Lambda=35$ (blue) points are near the edge of the bounding ellipsoid, which justifies this choice. We used the same bounding ellipsoid for the computation with $\Lambda=43$. Again, no $\Lambda=43$ (red) points are near the edge of the bounding ellipsoid.
}
\label{fig:boundingellispoid}
\end{center}
\end{figure}

Imposing a bounding ellipsoid is a non-rigorous optimization and should be done with care. As we worked our way up in the number of derivatives of the crossing equations, we used the following strategy. At an initial derivative order $\Lambda$, we keep track of all values of OPE coefficients of allowed points. We choose an ellipsoid $\cE$ that contains these values and is also enlarged by an $O(1)$ factor. We then increase $\Lambda \to \Lambda'$ and use $\cE$ as a bounding ellipsoid for the cutting surface algorithm. As a check on this method, we can inspect the set of allowed OPE coefficients found at derivative order $\Lambda'$ and see if any of them are close to the boundary of $\cE$. In practice, they never are, see figure~\ref{fig:boundingellispoid}. (In fact, they are almost never outside the cloud of points computed at derivative order $\Lambda$, so the enlargement by an $O(1)$ factor is unnecessary.) We can now find a new ellipsoid $\cE'$ and continue.

\subsection{Hot-starting}

The cutting surface algorithm requires solving multiple SDPs to rule out a single point $(\De_\f,\De_s,\De_t)$ in dimension space. For example, for the computation described in section~\ref{sec:opescanresults}, each point in dimension space required solving an average of $\sim35$ SDPs (not including the tiny SDPs encountered in the semidefinite relaxation method of section~\ref{sec:semidefiniterelaxation}). Fortunately, many of these SDPs can be solved extremely quickly using {\it hot-starting} \cite{Go:2019lke}: we reuse the final state of the semidefinite program solver from a previous calculation as the initial state in a new calculation. In practice, hot-starting means passing an old checkpoint file as an argument to {\tt SDPB}.

\begin{figure}[t]
\begin{center}
\includegraphics[scale=.45]{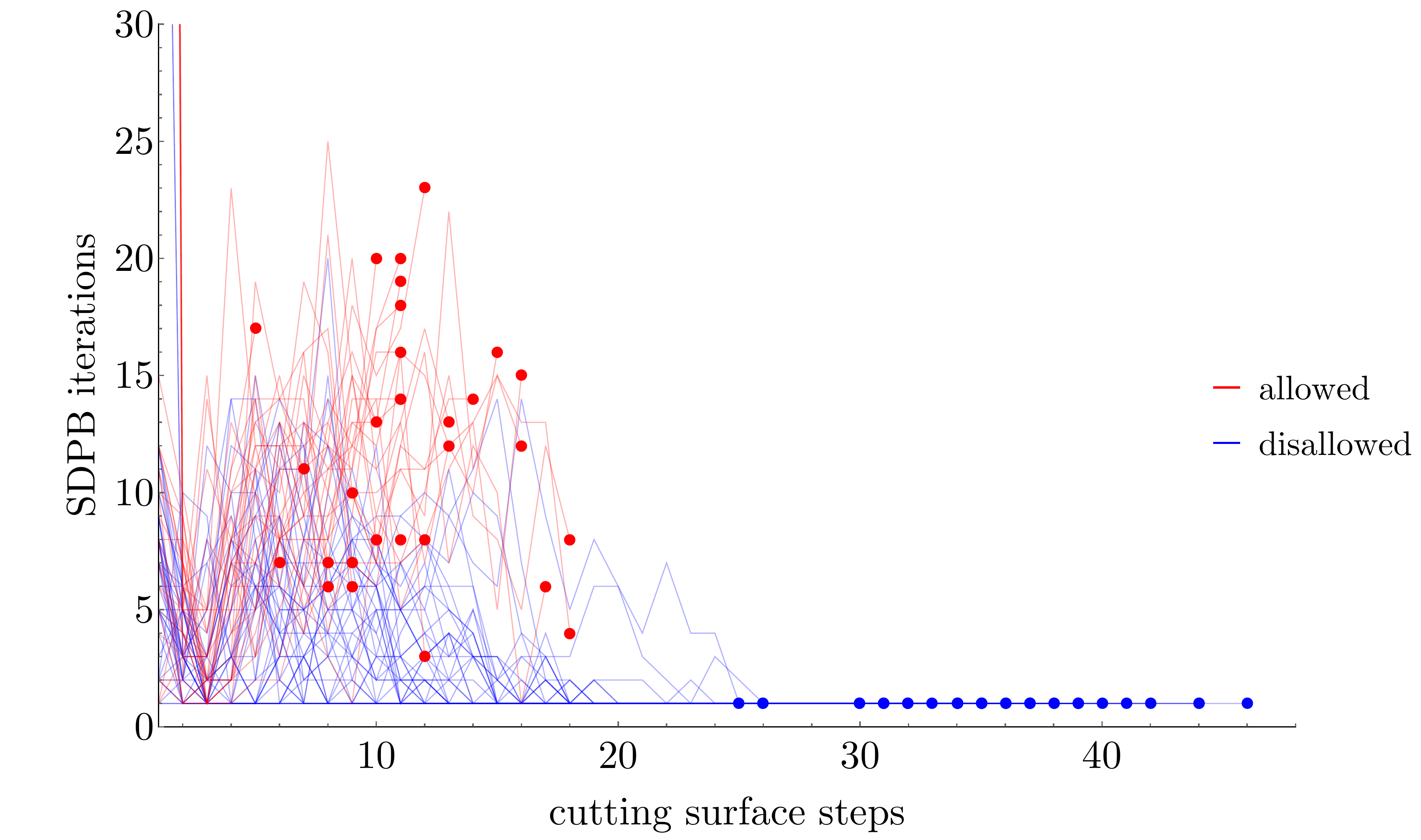}
\caption{
Number of iterations of {\tt SDPB} in each step of the cutting surface algorithm, for our computation of the $O(2)$ model island with $\Lambda=43$. Hot-starting drastically reduces the number of iterations throughout the computation. The blue paths represent OPE scans that eventually terminate by ruling out a point in dimension space. The red paths represent scans that eventually terminate by finding an allowed (primal) point. We mark the end of each path with a dot. At the beginning of the computation, a small number of points require $\sim 200$ {\tt SDPB} iterations during the first step of the cutting surface algorithm. Once the checkpoints from those {\tt SDPB} runs have been generated, hot-starting ensures that most subsequent runs take $\lesssim 20$ {\tt SDPB} iterations. The first 10-20 steps of the cutting surface algorithm typically require 1-15 {\tt SDPB} iterations each. If the point is allowed, the algorithm typically finds it within 20 steps. If the point is disallowed, subsequent steps of the cutting surface algorithm take fewer iterations, with the last several steps requiring 1 iteration each.
}
\label{fig:iterationcount}
\end{center}
\end{figure}

Hot-starting is particularly advantageous in the cutting surface algorithm because SDPs only change by a small amount with each new run. Specifically, the only difference between subsequent SDPs is the replacement of the positivity condition $\l_n^T \a(V_\ext) \l_n\geq 0$ by the new condition $\l_{n+1}^T \a(V_\ext) \l_{n+1}\geq 0$. Thus, the previous checkpoint contains a functional that already satisfies all other positivity conditions in the semidefinite program. In practice, the new condition $\l_{n+1}^T \a(V_\ext) \l_{n+1}\geq 0$ is satisfied after a small number of iterations of {\tt SDPB}. Furthermore, the number of iterations typically decreases over the course of the cutting surface algorithm, see figure~\ref{fig:iterationcount}.

Hot-starting is useful also for different points in dimension space. In practice, we keep a list of checkpoint files from all runs of {\tt SDPB} over the course of a computation. For each new point in dimension space, we find the newest checkpoint file corresponding to the closest point in dimension space, and use it to initiate the cutting surface algorithm.

\begin{figure}[t!]
\begin{center}
\subfigure[]{\includegraphics[width=0.45\textwidth]{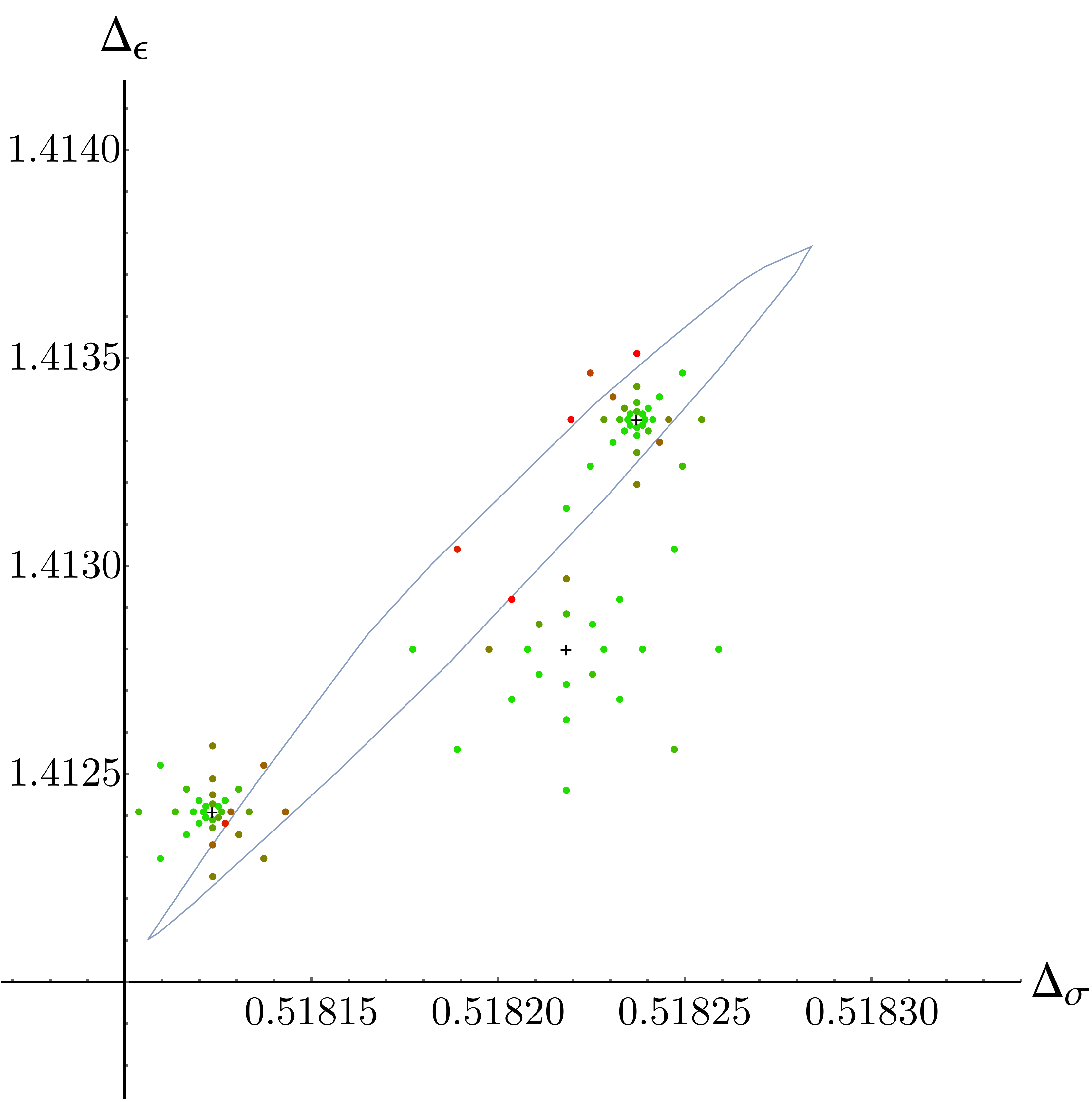}\label{fig:hotstart1plot}}
\subfigure[]{\includegraphics[width=0.45\textwidth]{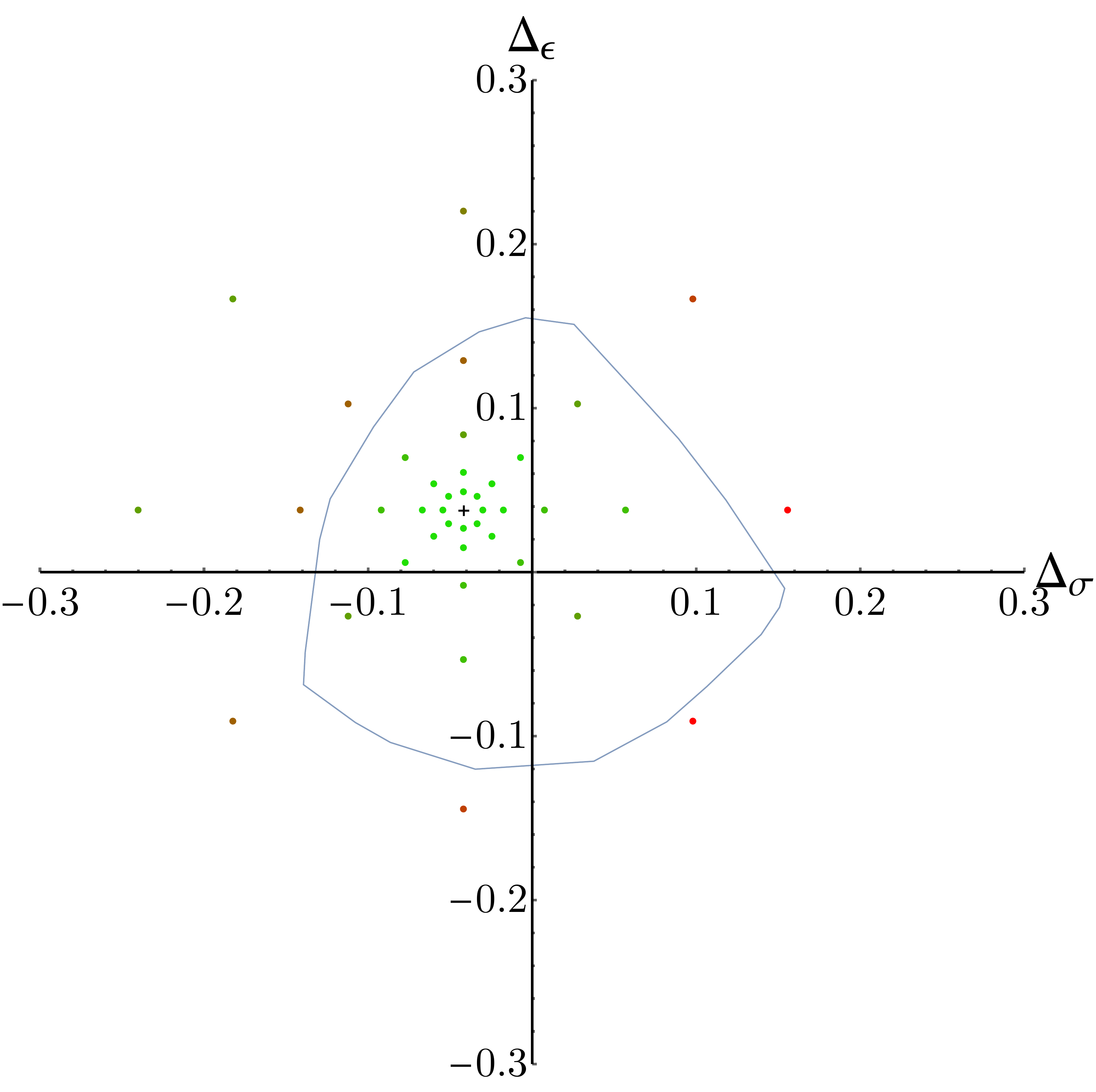}\label{fig:hotstart2plot}}
\caption{
Hot-starting effectiveness for different nearby checkpoints. The blue line is the boundary of the 3d Ising from $\sigma,\epsilon$ mixed correlator bootstrap. The setup is the same as that of the dark blue region of figure 3 in \cite{Kos:2014bka} except $n_{max}=10$ (i.e.\ $\Lambda=19$). We fix $P_0$ (indicated by a cross) to be various points and hot-start the $P_0$ computation with checkpoints taken from nearby points $P_i$ around $P_0$. The color of $P_i$ indicates how many {\tt SDPB} iterations is needed for the hot-started computation. Red corresponds to 8 iterations, while green corresponds to 1 iteration. Without hot-starting, the typical number of iterations is about 80.
On the left: we fix $P_0$ to be $(0.518123, 1.412409)$, $(0.518237, 1.413352)$, $(0.518218, 1.412800)$. On the right: the $(\Delta_\sigma,\Delta_\epsilon)$ space is transformed such that the island is roughly spherical. We fix $P_0$ to be $(0.518217, 1.413221)$.} 
\end{center}
\end{figure}

To demonstrate the effectiveness of hot-starting in dimension space, we study the 3d Ising model $\sigma,\epsilon$ mixed correlator bootstrap described in \cite{Kos:2014bka}. We choose a fixed point $P_0$ in dimension space and hot-start $P_0$ with several checkpoints from nearby points $P_i$, see figure~\ref{fig:hotstart1plot}. We observed that in general when $P_i$ is close to $P_0$, the number of iterations is smaller. In figure~\ref{fig:hotstart2plot}, we show the effectiveness of hot-starting in a transformed space, where the Ising island is roughly a spherical shape. We see that the concept of ``nearest" is better behaved in this transformed space. 

Let us mention one additional practical optimization. In each step of the cutting surface algorithm for scanning OPE coefficients, we must solve semidefinite programs that are nearly identical: they differ only in the positivity conditions associated to the external scalars $\f,s,t$. Consequently, we can avoid re-generating the entire SDP and only re-generate the conditions for the external scalars.

\subsection{Primal/dual jumps}
\label{sec:jumps}

When testing feasibility of an SDP (as opposed to optimizing an objective function), {\tt SDPB} includes some features that allow the solver to terminate more quickly. Internally, {\tt SDPB} uses a modified Newton's method to simultaneously solve three types of equations: primal feasibility equations, dual feasibility equations, and an equation relating the two. For our purposes, the dual feasibility equations are the most important. A functional $\a$ exists if and only if the dual feasibility equations are satisfied. If {\tt SDPB} detects that it is possible to solve either the primal or dual feasibility equations during an iteration, then it does so immediately. We call such events primal/dual jumps.

When testing feasibility, a dual jump means that a functional $\a$ has been found (and the solver can terminate). In practice, a primal jump means a functional will not be found (so the solver can terminate in this case as well). The observation that we can stop after a primal jump was made in \cite{Simmons-Duffin:2015qma}. As far as we are aware, it has not been rigorously established. However, this does not affect the validity of the resulting bootstrap bounds, which depends only on the existence of functionals.

To make {\tt SDPB} terminate in the event of primal/dual jumps, we supply the options {\tt --detectPrimalFeasibleJump} and {\tt --detectDualFeasibleJump}. We have found that it is important to disallow {\tt SDPB} from terminating for other reasons. For example, over the course of the cutting surface algorithm, the primal error can get quite small, and often goes below reasonable values of {\tt primalErrorThreshold}. However, in practice only primal/dual jumps are good reasons to terminate. Thus, we recommend turning off the options {\tt --findPrimalFeasible} and {\tt --findDualFeasible}, and setting {\tt primalErrorThreshold} and {\tt dualErrorThreshold} extremely small (e.g.\ $10^{-200}$). Our precise parameters are listed in appendix~\ref{app:software}.

The existence of dual feasible jumps is sensitive to the precise bootstrap problem being solved. In our initial bootstrap implementation for correlators of $\f,s,t$, we did not observe any dual feasible jumps. In these cases, {\tt SDPB} would run for many iterations, with the {\tt dualError} (which indicates failure of the dual feasibility equations to be satisfied) steadily decreasing but never jumping to zero. We observed that during these iterations, {\tt SDPB} was working hard to find functionals that were positive when acting on operators close to the unitarity bound. We alleviate this problem by imposing a small gap in twist $\tau=\De-J$. Specifically, we impose
\be
\tau \geq \tau_\textrm{unitarity} + \de\tau,
\ee
(where $\tau_\textrm{unitarity}$ is the unitarity bound)
in all spin/symmetry sectors not containing conserved currents. (This condition is in addition to other gaps.) The extremely conservative choice $\de\tau=10^{-6}$ is sufficient to restore dual feasible jumps. Imposing this small twist gap dramatically increases the efficiency of our methods.

\subsection{Delaunay triangulation in dimension space}

Given the above methods for determining whether a point $(\De_\f,\De_s,\De_t)$ in dimension space is allowed, we would like to search for the full allowed region. For simplicity, first consider the one-dimensional case, where we have a single parameter $\De$. We can map $\De$-space efficiently using binary search between known points.  Suppose we have a list of values $\De_1 < \De_2 < \dots < \De_n$, that are known to be either allowed or disallowed. We define 
\be
\label{eq:pivalues}
p_i = \begin{cases}
0 & \textrm{if $\De_i$ is disallowed,}\\
1 & \textrm{if $\De_i$ is allowed.}
\end{cases}
\ee
For each case where $p_i\neq p_{i+1}$, we perform a binary search between $\De_i$ and $\De_{i+1}$ to find the precise threshold between allowed and disallowed.

We can reinterpret this method as follows. We can define a ``probability" $p(\De)$ that a given point is allowed. Our eventual goal will be to make $p(\De)$ as close to $0$ or $1$ as possible for all $\De$. A reasonable approximation for $p(\De)$ is via linear interpolation between the values $p(\De_i)=p_i$. To improve our knowledge of the allowed region as quickly as possible, the next test point $\De_\textrm{test}$ should have probability $p(\De_\textrm{test})=1/2$. If there are multiple such points, we should choose the one with the smallest slope $|p'(\De_\textrm{test})|$.\footnote{If we are testing points in parallel, then we can order the points in order of increasing slope and test the first few.} We then test whether $\De_\mathrm{test}$ is allowed, add it to the list of known values, and repeat the algorithm.

The above method generalizes to higher dimensions. Consider a vector of dimensions $\vec \De\in \R^k$. Suppose that we have a list of points $\vec \De_{1},\vec \De_{2},\dots,\vec \De_{n}\in \R^k$ and values $p_i$ defined as in (\ref{eq:pivalues}). To define a probability function $p(\vec \De)$, we perform a Delaunay triangulation of the set of known points.\footnote{Delaunay triangulations in 2 or 3 dimensions can be computed in Mathematica. In general, they can be computed efficiently using the software package {\tt qhull} \cite{Barber96thequickhull}.} Within each simplex of the triangulation, we define $p(\vec\De)$ via linear interpolation between its values $p_i$ at the vertices. Within each simplex, the points satisfying $p(\vec \De)=1/2$ are either empty or form a codimension-1 polyhedron. For every nonempty polyhedron, we define a candidate point as the mean of the vertices of the polyhedron. We choose $\vec \De_\mathrm{test}$ as the candidate point inside the simplex with the largest ``crossing distance", which is defined as the minimum distance between two vertices of the simplex with different values of $p_i$. After testing $\vec \De_\mathrm{test}$, we add it to the list of known points and repeat the algorithm.

\begin{figure}[t!]
\begin{center}
\includegraphics[scale=1.]{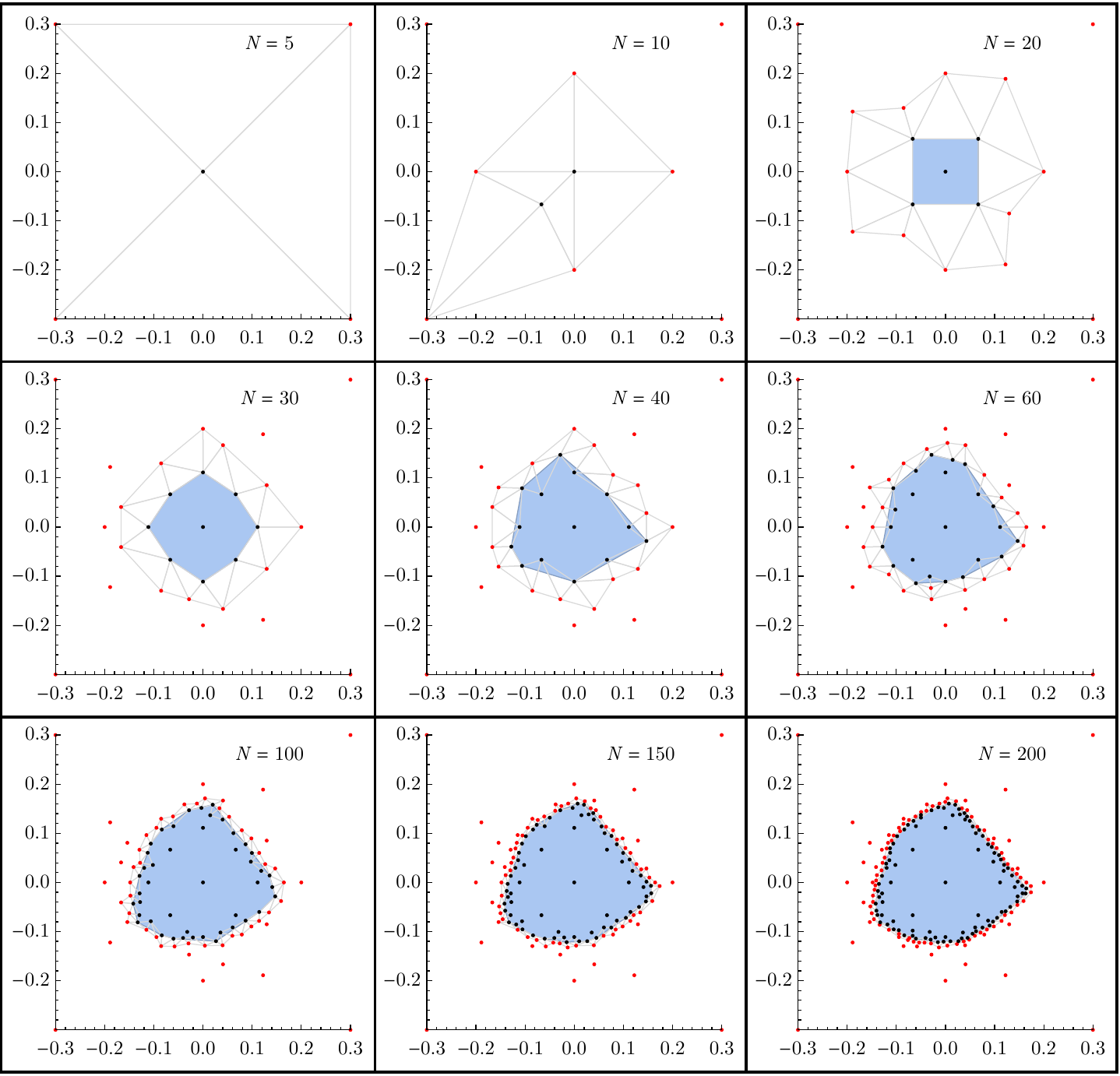}
\caption{
A series of images show intermediate states of the Delaunay triangulation algorithm for 3d Ising $\sigma$,$\epsilon$ mixed correlator bootstrap. The setup is same as the dark blue region of figure 3 in \cite{Kos:2014bka} except $n_{max}=10$ ($\Lambda=19$). We transformed the $(\Delta_\sigma,\Delta_\epsilon)$ space such that the 3d Ising island is roughly spherical. The red points are disallowed, while black points are allowed. The blue region is the convex hull of the black points. The $N$ in each plot is the total number of sampled points. 
}
\label{fig:delaunayani}
\end{center}
\end{figure}

We illustrate this algorithm in 2 dimensions in figure~\ref{fig:delaunayani}.

To work properly, Delaunay triangulation search requires sufficiently good initial conditions. For example, in the 1-dimensional case (binary search), we only obtain a correct picture of the allowed region if each connected allowed component and each connected disallowed component contains at least one initial point. Similarly, in higher dimensions, we only find an allowed region if we start with at least one point inside that region.

For this work, we found suitable initial conditions by first studying low derivative order $\Lambda$, and then working our way up in $\Lambda$. Our typical workflow is as follows: Based on computations at $\Lambda=15,19,23$, we found that the allowed region is a nearly convex island, and it can be made approximately spherical by a particular affine transformation. For each subsequent computation, we applied an affine transformation determined by the previous computation before performing the Delaunay search. This increases the efficiency of the search and makes it easier to correctly resolve corners sharp corners and other features in the boundary of the island.

Because the shape of the island is so simple, Delaunay triangulation works properly given a single allowed point, together with enough disallowed points that the island does not extend outside the convex hull of the  disallowed points. When increasing $\Lambda$, we can reuse all disallowed points from lower values of $\Lambda$. What remains is to determine an allowed point at the new value of $\Lambda$. We guess the allowed point in dimension space by extrapolating the way that the island shrinks with $\Lambda$, and choosing a point in the center of the extrapolated island. We test this point, and if the result is {\it primal feasible}, we can initiate a Delaunay search for the island. If the point is ruled out, we must make a different guess.

\section{Results}
\label{sec:results}

\subsection{Dimension bounds without OPE scans}

In this section, we show bounds on the dimensions $\De_\f,\De_s$ computed {\it without} the algorithm described in section~\ref{sec:cuttingsurface} for scanning over OPE coefficients. We also explore effects of imposing a more or less conservative gap in the charge-4 scalar sector.

\begin{figure}[h]
\begin{center}
\includegraphics[scale=.45]{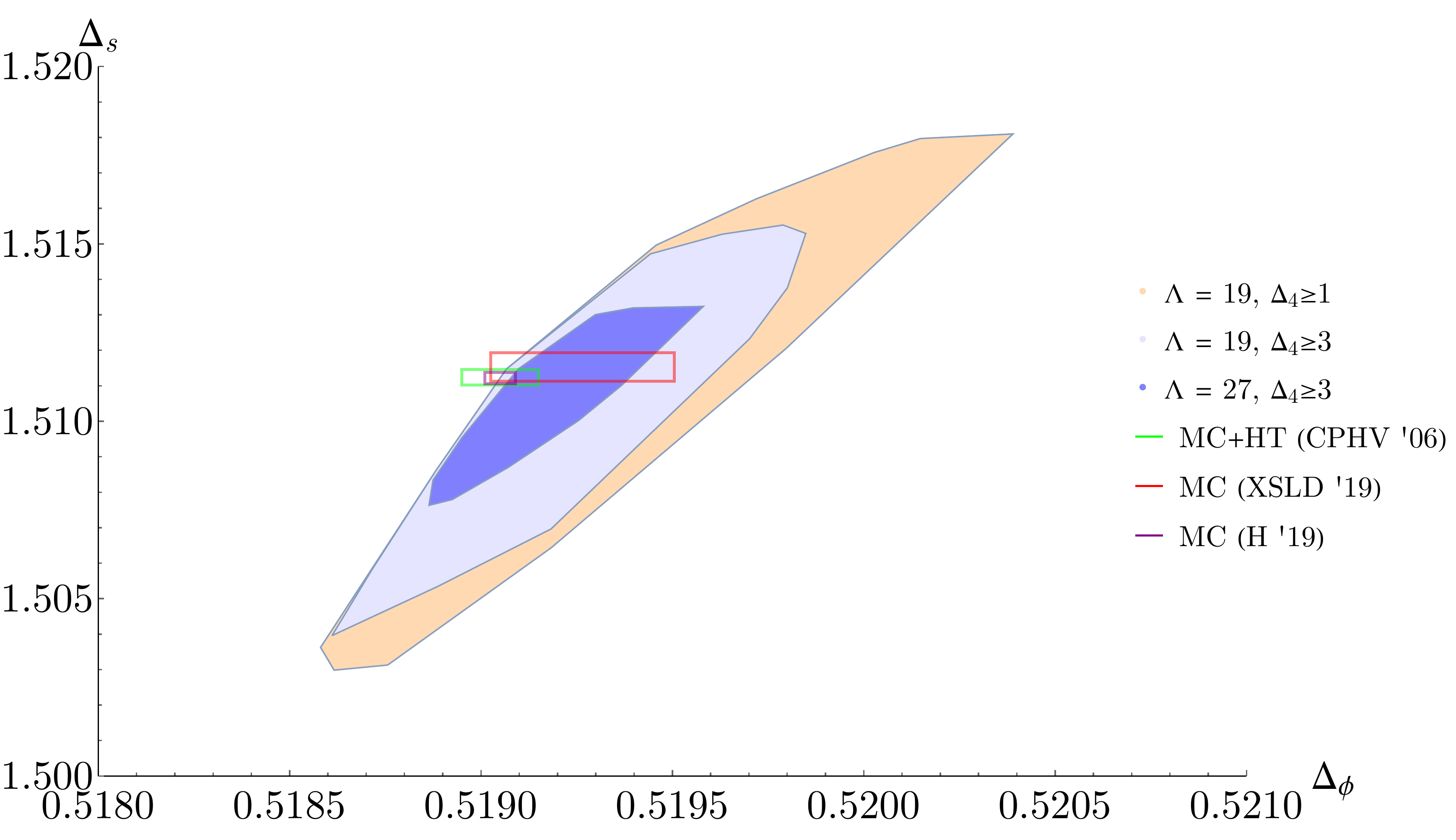}
\caption{Bounds on the scaling dimensions $\De_\f,\De_s$ computed {\it without} the cutting surface algorithm described in section~\ref{sec:cuttingsurface}. The light orange region shows the bound computed with $\Lambda=19$ and a conservative gap assumption in the charge-4 scalar sector $\De_4\geq 1$. The light blue region shows the bound at $\Lambda=19$ with a stronger gap assumption $\De_4\geq 3$. The dark blue region shows the bound at $\Lambda=27$ with the stronger gap assumption. The results are compared with the recent Monte Carlo studies~\cite{Xu:2019mvy,Hasenbusch:2019jkj} and an earlier study combining Monte Carlo simulations with high temperature expansion calculations in~\cite{Campostrini:2006ms}.
These bounds were computed at relatively low resolution, so the edges of the island show some artifacts.
}
\label{fig:2dislandvarious}
\end{center}
\end{figure}

Figure~\ref{fig:2dislandvarious} shows bounds with different gap assumptions and different values of $\Lambda$, all computed without scanning over OPE coefficients. The light orange region shows a bound with $\Lambda=19$ and the conservative assumption that the lowest dimension charge-4 scalar operator has dimension $\De_4\geq 1$. Evidence from other techniques supports the hypothesis that in fact $\De_4\geq 3$. The light blue region shows the resulting bound after imposing this stronger gap assumption. Finally, the dark blue region shows the result of imposing the stronger gap assumption and increasing the derivative order to $\Lambda=27$.

We see that the stronger gap assumption reduces the size of the island by approximately 30\% in both dimensions. Furthermore, imposing the gap assumption causes the island to shrink relatively quickly with $\Lambda$. Here, we see that increasing $\Lambda$ from $19$ to $27$ causes the island to shrink by an additional factor of 2. Because the stronger gap is well-motivated and significantly improves the results, we include it in our computations. For comparison in figure~\ref{fig:2dislandvarious}, we show the Monte Carlo and high temperature expansion result from \cite{Campostrini:2006ms} and more recent Monte Carlo result from \cite{Xu:2019mvy}. Without scanning over OPE coefficients, the bootstrap results are less precise.

\subsection{Dimension bounds with OPE scans}
\label{sec:opescanresults}

Now we show our results obtained from scanning over OPE coefficients using the cutting surface algorithm described above. The plots in this section compute the allowed values of $\{\Delta_{\phi}, \Delta_s, \Delta_t\}$ assuming irrelevance of the second charge 0,1,2 operators and first charge 4 operator. The stress tensor and conserved current are assumed in the spectrum with coefficients constrained by Ward identities. All other operators are allowed to exist at any scaling dimension above $\ell + 1 + \delta \tau$ with $\delta \tau = 10^{-6}$. 

\begin{figure}[ht]
\begin{center}
\includegraphics[scale=.45]{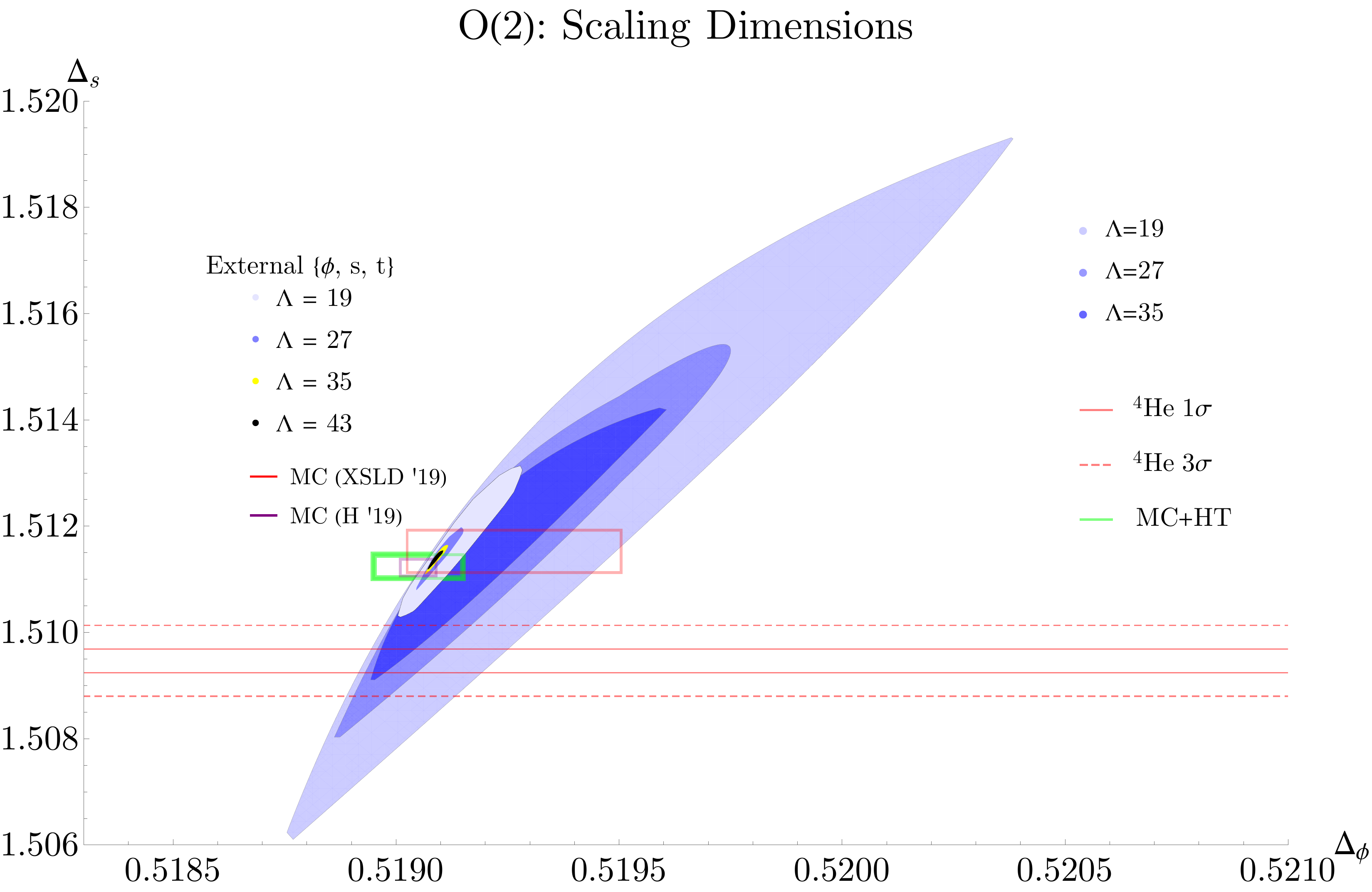}
\caption{Superposition of the new $O(2)$ islands using the $\{\phi_i, s, t_{ij}\}$ system and OPE scans over the earlier bootstrap results from~\cite{Kos:2016ysd} which used the $\{\phi_i, s\}$ system.}
\label{fig:2dIsland-old-superposition}
\end{center}
\end{figure}

\begin{figure}[ht]
\begin{center}
\includegraphics[scale=.45]{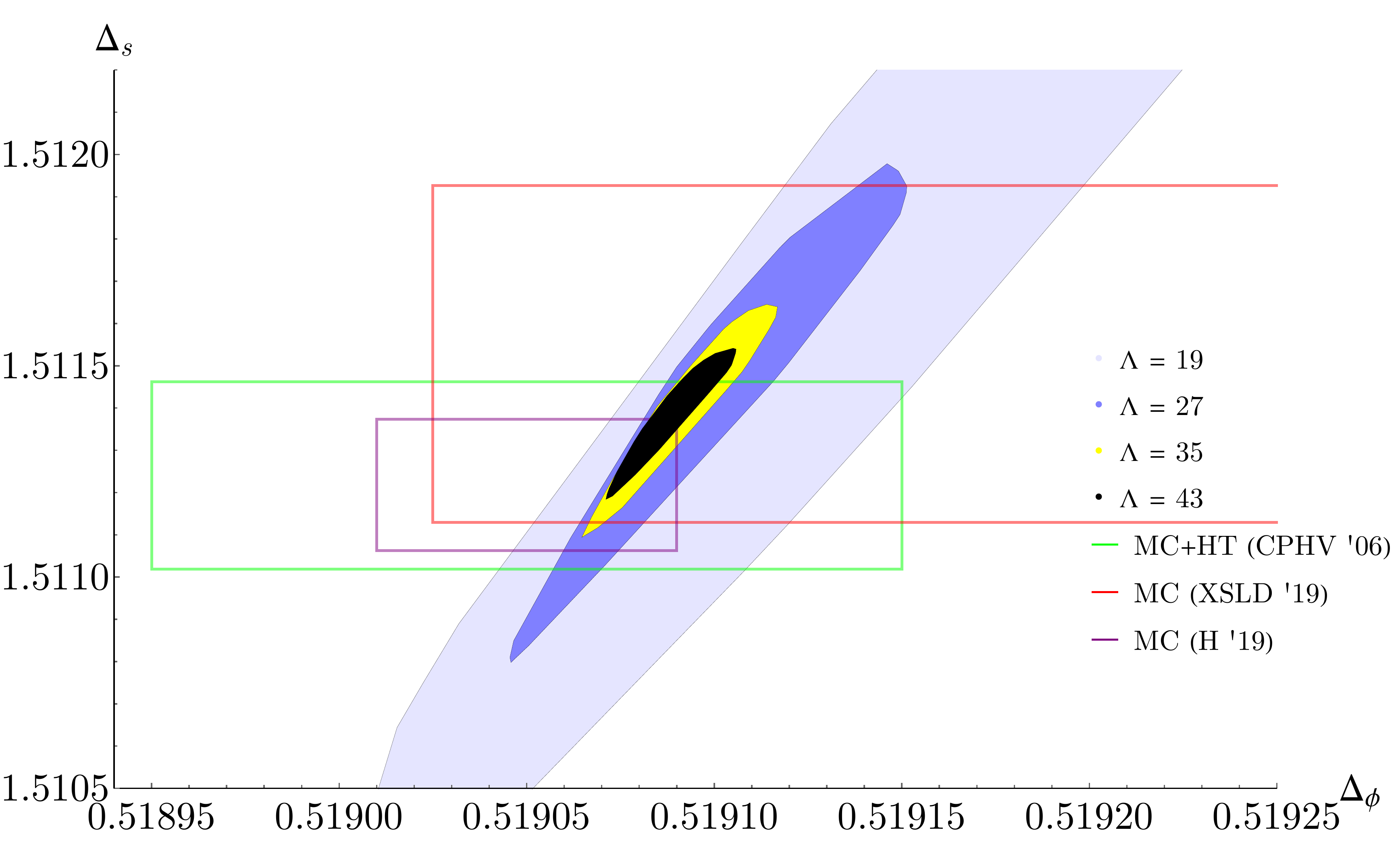}
\caption{New $O(2)$ islands using the $\{\phi_i, s, t_{ij}\}$ system and OPE scans at $\Lambda = 19, 27, 35, 43$. This plot shows the projection to the $\{\Delta_\phi, \Delta_s\}$ plane. The results are compared with the recent Monte Carlo studies~\cite{Xu:2019mvy,Hasenbusch:2019jkj} and an earlier study combining Monte Carlo simulations with high temperature expansion calculations~\cite{Campostrini:2006ms}.}
\label{fig:2dIsland-with-OPE}
\end{center}
\end{figure}

\begin{figure}[ht]
\begin{center}
\includegraphics[scale=.45]{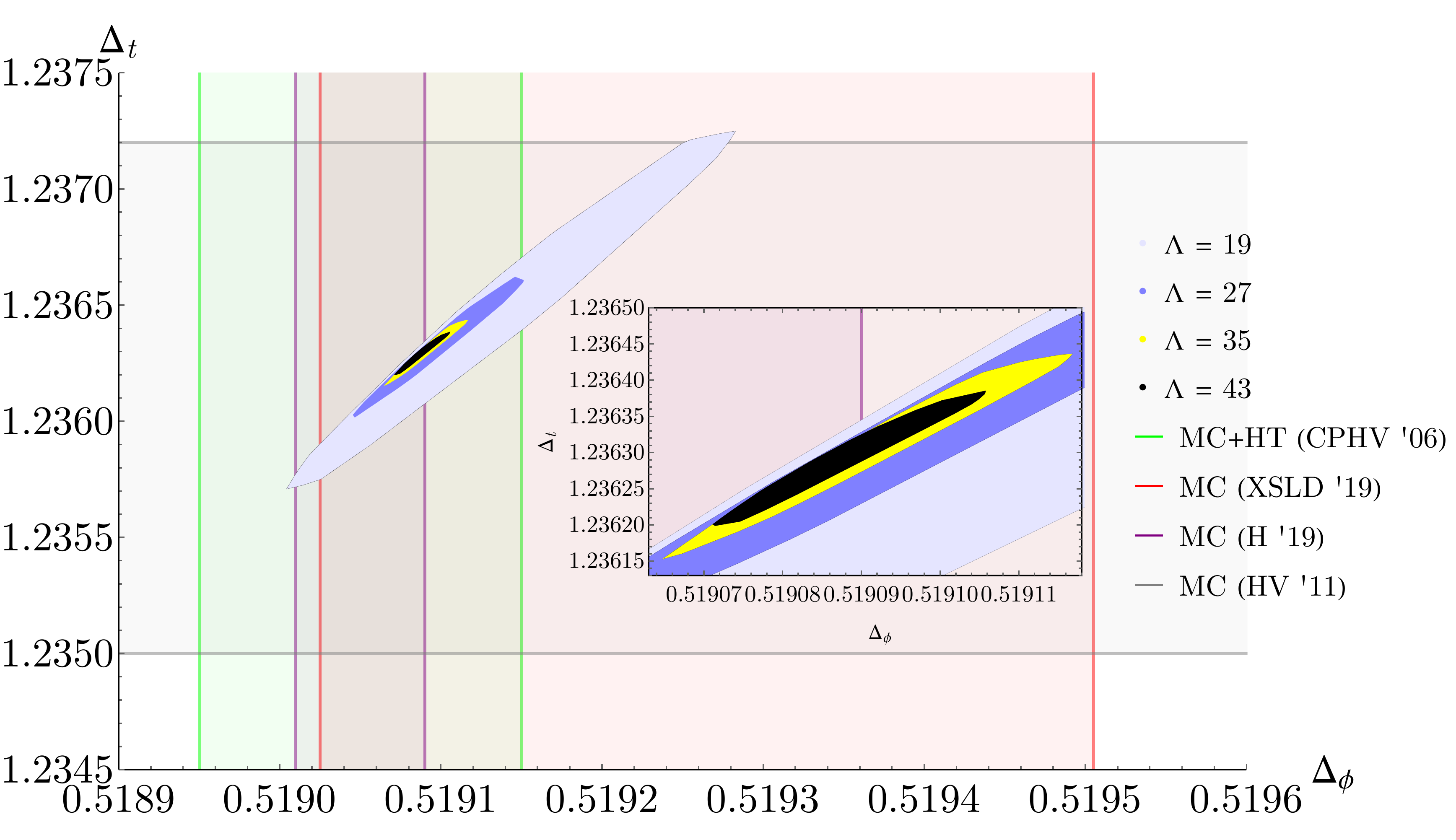}
\caption{New $O(2)$ islands using the $\{\phi_i, s, t_{ij}\}$ system and OPE scans at $\Lambda = 19, 27, 35, 43$. This plot shows the projection to the $\{\Delta_{\phi}, \Delta_{t}\}$ plane. The results for $\Delta_{\phi}$ are compared with the recent Monte Carlo studies~\cite{Xu:2019mvy,Hasenbusch:2019jkj} and an earlier study combining Monte Carlo simulations with high temperature expansion calculations~\cite{Campostrini:2006ms}, while the results for $\Delta_{t}$ are compared with the Monte Carlo study~\cite{PhysRevB.84.125136}. The latter is also compatible with the earlier pseudo-$\epsilon$ expansion estimate $\Delta_{t}=1.237(8)$~\cite{Calabrese:2004ca}.}
\label{fig:2dIsland-phit-with-OPE}
\end{center}
\end{figure}

Figures~\ref{fig:2dIsland-old-superposition} and~\ref{fig:2dIsland-with-OPE} shows our determinations of the allowed regions at derivative order $\Lambda = 19,27,35,43$, projected to the $\{\Delta_{\phi}, \Delta_s\}$ plane. Figure~\ref{fig:2dIsland-phit-with-OPE} also shows the projection to the $\{\Delta_{\phi}, \Delta_t\}$ plane and figure~\ref{fig:3dIsland-with-OPE-untransformed} in the introduction shows a view of the 3d region at $\Lambda=43$. The improvement relative to figure~\ref{fig:2dislandvarious} is readily apparent. In particular the conformal bootstrap results exclude the values of $\Delta_s$ extracted from ${}^4$He measurements~\cite{Lipa:2003zz} and improve upon but appear compatible with both earlier~\cite{Campostrini:2006ms} and recent results from Monte Carlo simulations~\cite{Xu:2019mvy,Hasenbusch:2019jkj}. 

The plotted regions are obtained by constructing the Delaunay triangulation of our tested points, selecting the triangles that contain both allowed and disallowed points, and plotting the convex hull of the points in the interior of these triangles that are midway between the allowed and disallowed vertices. This represents our best determination of the allowed region at a given $\Lambda$, but has a small error associated with the distance between the boundary and the nearest disallowed point. This ``best-fit" region gives the determinations
\be
\Delta_{\phi} &= 0.519088(17^*), \\
\Delta_{s} &= 1.51136(18^*),\\
\Delta_{t} &= 1.23629(9^*).
\ee  
More conservatively we can consider the convex hull of the disallowed points in the Delaunay triangles straddling the boundary of the allowed region. We believe that every point outside of this more conservative region is excluded by the conformal bootstrap, giving the rigorous error bars
\be
\Delta_{\phi} &= 0.519088(\bf{22}), \\
\Delta_{s} &= 1.51136(\bf{22}),\\
\Delta_{t} &= 1.23629(\bf{11}).
\ee  

Each allowed point in dimension space comes paired with an allowed point in the space of OPE coefficient ratios. At $\Lambda=43$ these allowed OPE coefficient ratios live in the ranges
\be
\label{eq:operatios}
\frac{\lambda_{sss}}{\lambda_{\f\f s}} &= 1.20926(46^*), \\
\frac{\lambda_{tts}}{\lambda_{\f\f s}} &= 1.82227(19^*),\\
\frac{\lambda_{\f\f t}}{\lambda_{\f\f s}} &= 1.765918(64^*).
\ee  
The full allowed region in OPE coefficient space may be slightly larger.\footnote{Using the scaling dimension region as a guide we would estimate that the range of allowed values may increase in size by $\sim 20\%$ when going from the computed allowed points at $\Lambda=43$ to the ``best-fit" allowed region.} The full set of computed points at $\Lambda=43$ are shown in figure~\ref{fig:OPEratiosL43-untransformed} and listed in appendix~\ref{app:points}.

\begin{figure}[h]
\begin{center}
\includegraphics[scale=.35]{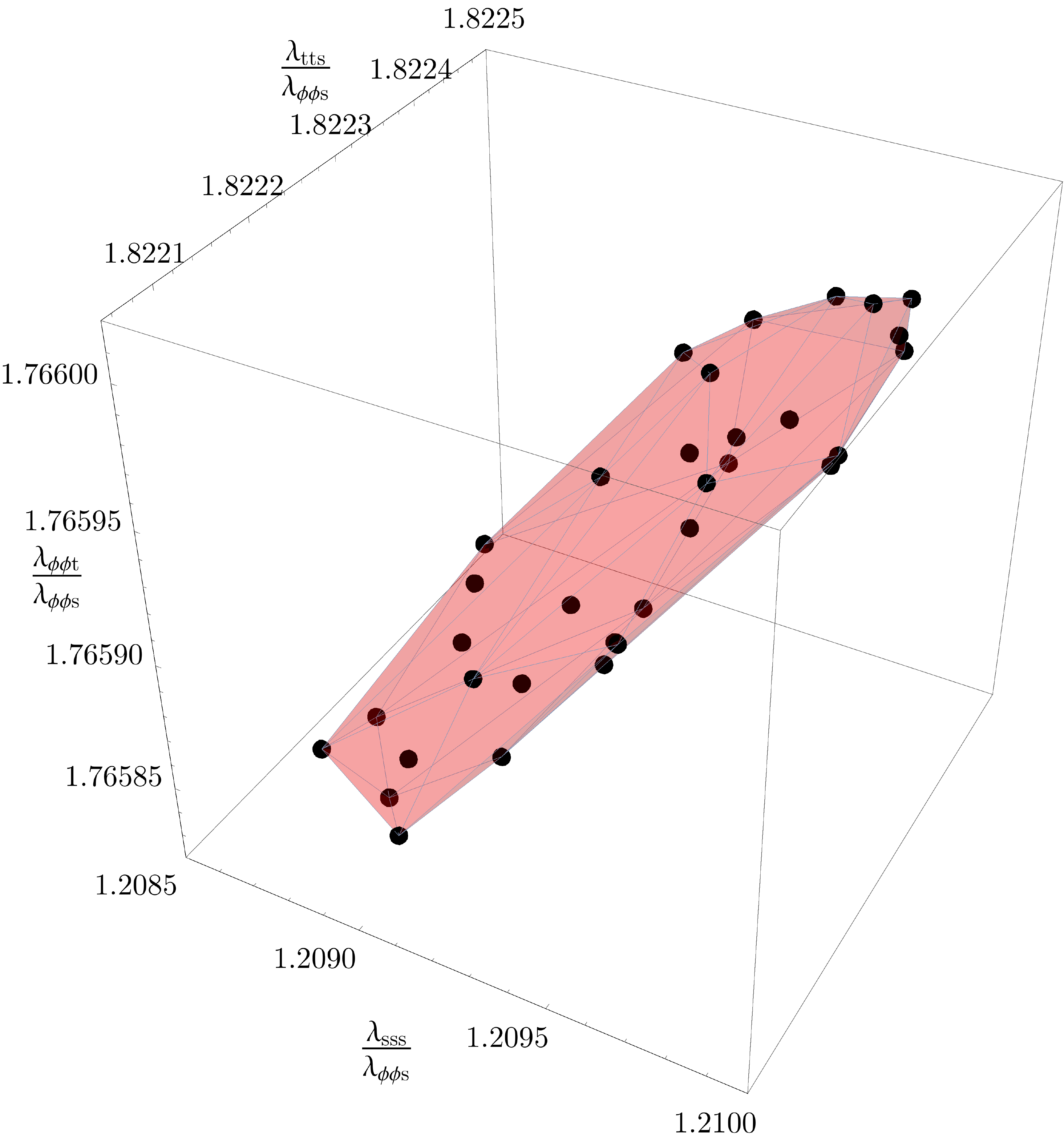}
\caption{Allowed points in the space of OPE coefficient ratios computed using the $\{\phi_i, s, t_{ij}\}$ system at $\Lambda = 43$. The convex hull of these points (red) gives an estimate for the allowed values of these coefficients. The projection of the full 6d allowed region will be slightly larger so the shown region is non-rigorous.}
\label{fig:OPEratiosL43-untransformed}
\end{center}
\end{figure}

\subsection{Central charges and $\lambda_{\f\f s}$}
\label{sec:centralcharges}

As stated in section~\ref{sec:spectrum}, the two-point coefficient $C_T$ for stress tensors  and the two-point coefficient $C_J$ for the $O(2)$ current  appear in the crossing equations. These coefficients are interesting for several reasons. For example they are related to transport in quantum critical systems, giving the leading term in the high frequency expansion at finite temperature \cite{Katz:2014rla,Iliesiu:2018fao}. In particular, the zero temperature conductivity of the $O(2)$ model is given by \cite{Katz:2014rla}
\be
2\pi \s_\infty &= \frac{2\pi C_J}{16 C_J^\mathrm{free}}.
\ee 

It should be possible to produce an island in the combined space of scaling dimensions $\De_\f,\De_s,\De_t$, OPE coefficient ${\lambda_{\phi \phi s}}$, and coefficients $C_T,C_J$. In particular, this would give a determination of $C_T$ and $C_J$ with rigorous error bars. Due to limits on computational resources, we have not yet attempted this computation. Instead, we will content ourselves with non-rigorous estimates of $C_T$, $C_J$ and ${\lambda_{\phi \phi s}}$. We chose 7 allowed points (shown in table~\ref{tab:thesevenpoints} of appendix~\ref{app:points}) in our island computed with $\Lambda=43$ derivatives. For each point, we computed upper and lower bounds on $C_J, C_T$, and the OPE coefficient ${\lambda_{\phi \phi s}}$ with $\Lambda=35$ derivatives. The largest upper bound and smallest lower bound give an estimate for these quantities. 

In order to compute upper bounds on $C_J$ ($C_T$), we must assume a gap between the conserved current (stress tensor) and other operators in the same spin and global symmetry sector. When computing upper bounds on $C_J$, we assume all other spin-1 charge-0 operators have dimension $\De\geq 3$. When computing upper bounds on $C_T$, we assume all other spin-2 charge-0 operators have dimension $\De\geq 4$. These assumptions are well-supported by estimates from the $\e$-expansion and from the extremal functional method.

We find
\be
C_J/C_J^\mathrm{free}&=0.904395(28^*),\\
C_T/C_T^\mathrm{free}&=0.944056(15^*),
\ee
where in both cases the error bars are non-rigorous. Our result for $C_J$ gives a new determination of the zero-temperature conductivity
\be
2\pi \s_\infty &= 0.355155(11^*).
\ee
We also find
\begin{align}
\lambda_{\phi \phi s}=0.687126(27^*).
\end{align}
Combining this result with the OPE ratios (\ref{eq:operatios}) and adding errors in quadrature leads to the values quoted in table~\ref{tab:results}.

\subsection{Estimates from the extremal functional method}\label{functional}

The extremal functional method \cite{Poland:2010wg,ElShowk:2012hu} is a non-rigorous method for estimating a large amount of CFT data from a small number of computations. We hope to present a more detailed analysis of our extremal functionals for the $O(2)$ model in future work. For now, we give estimates of the dimensions of a few important low-lying scalars in table~\ref{tab:resultsefm}. To obtain extremal functionals, we chose 20 allowed points in the $\Lambda=43$ island and computed lower and upper bounds on the norm of the external OPE vector $|\l_\ext|$ with derivative order $\Lambda=27$ (shown in table~\ref{tab:the20points} of appendix~\ref{app:points}) . Comparing the zeros of the resulting functionals, we identified \emph{stable} zeros whose positions did not vary significantly as we changed the point in the island \cite{Simmons-Duffin:2016wlq}. Thus, for 20 points, we have 40 different values of $\Delta_{s'}$, $\Delta_{t'}$, $\Delta_{\textrm{charge 3}}$, and $\Delta_{\textrm{charge 4}}$ (half of them are from the lower bound computations, while another half are from the upper bound computations). The gaps we impose are the same as in the OPE scan discussed before, except that we set the twist gap $\delta_\tau$ to $10^{-4}$.

 \begin{table*}[h!]
\centering
\begin{tabular}{@{}cc|cc@{}}
	\toprule
Dim & Method & value & ref \\
	\midrule
 $\Delta_{s'}$ & MC & 3.789(4) & \cite{Hasenbusch:2019jkj} \\
  & CB & $3.794(8^*)$ & \\
  	 \midrule
 $\Delta_{t'}$ & FT & 3.624(10) &\cite{Calabrese:2002bm} \\
  & CB & $3.650(2^*)$ & \\
  	 \midrule
 $\Delta_{\textrm{charge 3}}$ & MC & 2.1085(20) &  \cite{PhysRevB.84.125136} \\
 & CB & $2.1086(3^*)$& \\
  	 \midrule
 $\Delta_{\textrm{charge 4}}$ & MC & 3.114(2)  &  \cite{Shao:2019dbi} \\
 & CB & $3.14(2^*) $ & \\
	\midrule
\end{tabular}
	\caption{Comparison of conformal bootstrap (CB) estimates using the extremal functional method with previous Monte Carlo (MC) and $\e$-expansion (FT) determinations of operator dimensions. The values for the extremal functional determinations are means across the 40 different extremal spectra, and the errors are the standard deviations. We mark the errors with a $^*$ to emphasize that they are non-rigorous. Here, ``charge-3" and ``charge-4" refer to the lowest-dimension scalars with the given charges, which in field theory language are $\f_{(i}\f_{j} \f_{k)}$ and $\f_{(i}\f_j\f_k\f_{l)}$. 
	\label{tab:resultsefm}}
\end{table*}

\section*{Acknowledgements}

We thank David Meltzer, Slava Rychkov and Ettore Vicari for discussions. AV and SMC thank Filip Kos for collaboration at an early stage of this project. NS thanks Junchen Rong for discussions. DSD and JL thank Brad Filippone for discussions on the Lambda Point Experiment and for his excellent demonstration of the superfluid $^4$He phase transition, performed annually (on Earth) for Caltech undergrads in Physics 2c/12c. WL, JL, and DSD are supported by Simons Foundation grant 488657 (Simons Collaboration on the Nonperturbative Bootstrap). DSD and JL are also supported by a Sloan Research Fellowship, and a DOE Early Career Award under grant no.\ DE-SC0019085. DP is supported by Simons Foundation grant 488651 (Simons Collaboration on the Nonperturbative Bootstrap) and DOE grant no.\ DE-SC0020318. NS and AV are supported by the
European Research Council (ERC) Starting Grant no.\ 758903. AV is also supported by the Swiss National Science Foundation (SNSF) under grant no.\ PP00P2-163670. SMC is supported by a Zuckerman STEM Leadership Fellowship.

This work used the Extreme Science and Engineering Discovery Environment (XSEDE) Comet Cluster at the San Diego Supercomputing Center (SDSC) through allocation PHY190023, which is supported by National Science Foundation grant number ACI-1548562. This work also used the EPFL SCITAS cluster, which is supported by the SNSF grant PP00P2-163670, the Caltech High Performance Cluster, partially supported by a grant from the Gordon and Betty Moore Foundation, and the Grace computing cluster, supported by the facilities and staff of the Yale University Faculty of Sciences High Performance Computing Center.

\appendix

\section{Code availability}
\label{app:code}

All code used in this work is available online. This includes
\begin{itemize}
\item The semidefinite program solver {\tt SDPB}:\\ \href{https://github.com/davidsd/sdpb}{\tt https://github.com/davidsd/sdpb} 
\item Code for generating tables of scalar conformal blocks:\\ \href{https://gitlab.com/bootstrapcollaboration/scalar_blocks}{\tt https://gitlab.com/bootstrapcollaboration/scalar\_blocks}
\item A Mathematica framework for bootstrap calculations, including implementations of the cutting surface and Delaunay triangulation algorithms described in section~\ref{sec:methods}:\\ \href{https://gitlab.com/bootstrapcollaboration/simpleboot}{\tt https://gitlab.com/bootstrapcollaboration/simpleboot}.
\item A Haskell framework for concurrent computations on an HPC cluster:\\ \href{https://github.com/davidsd/hyperion}{\tt https://github.com/davidsd/hyperion}
\item Haskell libraries for bootstrap computations, including implementations of the cutting surface and Delaunay triangulation algorithms described in section~\ref{sec:methods}:\\
\href{https://gitlab.com/davidsd/sdpb-haskell}{\tt https://gitlab.com/davidsd/sdpb-haskell} \\
\href{https://gitlab.com/davidsd/hyperion-bootstrap}{\tt https://gitlab.com/davidsd/hyperion-bootstrap} \\
\href{https://gitlab.com/davidsd/hyperion-projects}{\tt https://gitlab.com/davidsd/hyperion-projects}
\item A Haskell library and standalone executable for solving quadratically constrained problems by a combination of semidefinite relaxation and other heuristics\\
\href{https://gitlab.com/davidsd/quadratic-net/}{\tt https://gitlab.com/davidsd/quadratic-net/}
\end{itemize}

\section{Software setup and parameters}
\label{app:software}
The computations of the $O(2)$ model islands described in section~\ref{sec:opescanresults} with $\Lambda=19,27$ were performed on the Caltech HPC Cluster and the Yale Grace Cluster. For the computations with $\Lambda=35$ and $\Lambda=43$, we tested possible primal points using the Caltech and Yale clusters. In each case, after finding a few initial primal points, the main Delaunay triangulation search was performed on the XSEDE  \cite{6866038} Comet Cluster at the San Diego Supercomputing Center through allocation PHY190023. The computation of the $\Lambda=35$ island took 192K core-hours and was completed in 4 days. The computation of the $\Lambda=43$ island took 1.03M core-hours and was completed in 2 weeks.  

In table~\ref{islandparameter}, we list the {\tt SDPB} and {\tt scalar\_blocks} parameters for the $\Lambda=35,43$ island computations. (Parameters for other values of $\Lambda$ are available upon request.) In table~\ref{extremalparameter}, we list the parameters for the extremal functional computations with $\Lambda=27$ section~\ref{functional}. Note that for the island computation, the parameters {\tt findPrimalFeasible}, {\tt findDualFeasible}, {\tt detectPrimalFeasibleJump}, and {\tt detectDualFeasibleJump} are set in accordance with the discussion in section~\ref{sec:jumps}.

\begin{table}
\begin{center}
\begin{tabular}{@{}c|c|c@{}}
	\toprule
$\Lambda$ &  35 & 43\\
{\small\texttt{keptPoleOrder}}&  32 & 40\\
{\small\texttt{order}}&  80 & 90 \\
{\small\texttt{spins}} & $S_{35}$ & $S_{43}$ \\
{\small\texttt{precision}} &  960 & 1024\\
{\small\texttt{dualityGapThreshold}} & $10^{-30}$ & $10^{-75}$ \\
{\small\texttt{primalErrorThreshold}}& $10^{-200}$ & $10^{-200}$\\
{\small\texttt{dualErrorThreshold}} & $10^{-200}$ & $10^{-200}$\\ 
{\small\texttt{findPrimalFeasible}} & false & false \\
{\small\texttt{findDualFeasible}} & false & false \\
{\small\texttt{detectPrimalFeasibleJump}} & true & true \\
{\small\texttt{detectDualFeasibleJump}} & true & true \\
{\small\texttt{initialMatrixScalePrimal}} & $10^{50}$ & $10^{60}$ \\
{\small\texttt{initialMatrixScaleDual}} & $10^{50}$ & $10^{60}$\\
{\small\texttt{feasibleCenteringParameter}} & 0.1 & 0.1 \\
{\small\texttt{infeasibleCenteringParameter}} & 0.3 & 0.3 \\
{\small\texttt{stepLengthReduction}} & 0.7 & 0.7 \\
{\small\texttt{maxComplementarity}} & $10^{160}$ & $10^{200}$ \\
 \bottomrule
\end{tabular}
\caption{Parameters for the computations in section~\ref{sec:opescanresults}. The sets $S_{35,43}$ are defined in (\ref{eq:spinsets}).}
\label{islandparameter}
\end{center}
\end{table}

\begin{table}
\begin{center}
\begin{tabular}{@{}c|c@{}}
	\toprule
$\Lambda$ &  27 \\
{\small\texttt{keptPoleOrder}}&  12 \\
{\small\texttt{order}}&  60 \\
{\small\texttt{spins}} & $S_{27}$ \\
{\small\texttt{precision}} &  900 \\
{\small\texttt{dualityGapThreshold}} & $10^{-80}$ \\
{\small\texttt{primalErrorThreshold}}& $10^{-200}$ \\
{\small\texttt{dualErrorThreshold}} & $10^{-100}$\\ 
{\small\texttt{initialMatrixScalePrimal}} & $10^{20}$\\
{\small\texttt{initialMatrixScaleDual}} & $10^{20}$\\
{\small\texttt{feasibleCenteringParameter}} & 0.1\\
{\small\texttt{infeasibleCenteringParameter}} & 0.3\\
{\small\texttt{stepLengthReduction}} & 0.7\\
{\small\texttt{maxComplementarity}} & $10^{200}$\\
 \bottomrule
\end{tabular}
\caption{Parameters for the computations in section~\ref{functional}. The set $S_{27}$ is defined in (\ref{eq:spinsets}).}
\label{extremalparameter}
\end{center}
\end{table}

The sets of spins used for each value of $\Lambda$ were
\be
\label{eq:spinsets}
S_{27} &= \{0,\dots,31\}\cup \{49,50\}, \nn\\
S_{35} &= \{0,\dots,44\}\cup \{47, 48, 51, 52, 55, 56, 59, 60, 63, 64, 67, 68\},\nn\\
S_{43} &= \{0,\dots,64\}\cup \{67, 68, 71, 72, 75, 76, 79, 80, 83, 84, 87, 88\}.
\ee

\section{Tensor structures}\label{Ts}

In this appendix we compute all the $O(2)$ structures $T^R_{\mathcal{R}_1\mathcal{R}_2\mathcal{R}_3\mathcal{R}_4}(y_i)$ that appear in the block expansion \eqref{4point} for the 4-point functions we consider as listed in table \ref{table}.  The block expansion in the $s$-channel is derived by inserting a complete set of states
\es{complete}{
& \sum_{\alpha=\cO,P \cO,PP\cO} \frac{\left\langle { \varphi^1_{\mathcal{R}_1}(x_1,y_1)  \varphi^2_{\mathcal{R}_2}(x_2,y_2) |\alpha\rangle\langle\alpha| \varphi^3_{\mathcal{R}_3}(x_3,y_3)  \varphi^4_{\mathcal{R}_4}(x_4,y_4) } \right\rangle   }{\langle\alpha|\alpha\rangle}\,,
}
where $\alpha$ runs over an orthogonal basis of operators $\cO$ (and descendents) in irrep $R$ that appear in the OPEs $\varphi^1\times\varphi^2$ and $\varphi^3\times\varphi^4$. The 4-point structure $T^R_{\mathcal{R}_1\mathcal{R}_2\mathcal{R}_3\mathcal{R}_4}(y_i)$ can then be written in terms of the $O(2)$ structures $T_{\mathcal{R}_i\mathcal{R}_j }^R(y_i,y_j,y)$ of each of the pair of 3-point functions as
\es{3sto4}{
T^R_{\mathcal{R}_1\mathcal{R}_2\mathcal{R}_3\mathcal{R}_4}(y_i) = (T_{\mathcal{R}_1\mathcal{R}_2 R}(y_1,y_2,y),T_{\mathcal{R}_3\mathcal{R}_4 R}(y_3,y_4,y))\,,
}
where $(f(y),g(y))$ denotes the contraction over $y$ in index free notation. When $R$ is ${\bf 0}^\pm$, the contraction is just multiplication of the three-point structures. When $R$ has nonzero charge $n$, this contraction can be derived by expanding each rank $n$ $O(2)$ tensor in the basis
\es{basis}{
e=\frac{1}{\sqrt{2}}
\begin{pmatrix}
 1
\\ i 
\end{pmatrix}
\,,\qquad 
e=\frac{1}{\sqrt{2}}
\begin{pmatrix} 
1
\\-i
\end{pmatrix}\,,
}
as
\es{tensExp}{
f(y)=f(e)(y\cdot \bar e)^n + f(\bar e)(y\cdot  e)^n\,,
}
and similarly for $g(y)$. This basis has the convenient properties $e\cdot e=\bar e\cdot\bar e=0$ and $e\cdot \bar e=1$, so that the contraction of the tensors in index free notation is
\es{contract}{
(f(y),g(y))=f(e)g(\bar e)+f(\bar e)g(e)\,.
}
The result of these contractions can then be written in terms of the quantities
\es{ws}{
w_i\equiv y_i\cdot e\,,\qquad \bar w_i\equiv y_i\cdot \bar e\,,
} 
which have the properties
\es{w2}{
y_i\cdot y_i=w_i\bar w_i=0\,,\qquad y_i\cdot y_j=w_i\bar w_j+\bar w_iw_j\,,\qquad y_i\wedge y_j=i(w_i\bar w_j-\bar w_iw_j)\,,
}
which imply that $w_i=0$ or $\bar w_i=0$ since $y_i^2=0$ by definition.

The utility of this derivation is that each 3-point structure establishes a convention for the OPE coefficient $\lambda_{\varphi_i\varphi_j\cO}$ of the associated 3-point function, so computing the 4-point structures in terms of these 3-point structures ensures that the coefficients $\lambda_{\varphi_1\varphi_2\cO}\lambda_{\varphi_3\varphi_4\cO}$ that appear in \eqref{4point} can be consistently identified with these OPE coefficients. For each $s$- and $t$-channel configuration in table \ref{table} with an independent $O(2)$ structure, the resulting four-point structures are:
\es{1111}{
\langle \phi \phi \phi \phi \rangle:\qquad  &T^{\bf0^+}_{{\bf1}_i{\bf1}_j{\bf1}_k{\bf1}_l}=(w_i\bar w_j+\bar w_iw_j)(w_k\bar w_l+\bar w_kw_l)\,,\\
&T^{\bf0^-}_{{\bf1}_i{\bf1}_j{\bf1}_k{\bf1}_l}=-(w_i\bar w_j-\bar w_iw_j)(w_k\bar w_l-\bar w_kw_l)\,,\\
&T^{\bf2}_{{\bf1}_i{\bf1}_j{\bf1}_k{\bf1}_l}=w_iw_j\bar w_k\bar w_l+\bar w_i\bar w_jw_kw_l\,,\\
\langle tttt\rangle:\qquad  &T^{\bf0^+}_{{\bf2}_i{\bf2}_j{\bf2}_k{\bf2}_l}=(w_i\bar w_j+\bar w_iw_j)^2(w_k\bar w_l+\bar w_kw_l)^2\,,\\
&T^{\bf0^-}_{{\bf2}_i{\bf2}_j{\bf2}_k{\bf2}_l}=-(w_i^2\bar w_j^2-\bar w_i^2w_j^2)(w_k^2\bar w_l^2-\bar w^2_kw_l^2)\,,\\
&T^{\bf4}_{{\bf2}_i{\bf2}_j{\bf2}_k{\bf2}_l}=(w_iw_j\bar w_k\bar w_l+\bar w_i\bar w_jw_kw_l)^2\,,\\
\langle t\phi t\phi\rangle\,, \langle \phi tt\phi\rangle:\qquad &T^{\bf1}_{{\bf2}_i{\bf1}_j{\bf2}_k{\bf1}_l}=(w_i\bar w_j+\bar w_iw_j)(w_k\bar w_l+\bar w_kw_l)(w_i\bar w_k+\bar w_i w_k)\,,\\
&T^{\bf3}_{{\bf2}_i{\bf1}_j{\bf2}_k{\bf1}_l}=w_i^2w_j\bar w_k^2\bar w_l+\bar w_i^2\bar w_j w_k^2 w_l\,,\\
\langle tt\phi \phi\rangle :\qquad &T^{\bf0^+}_{{\bf2}_i{\bf2}_j{\bf1}_k{\bf1}_l}=(w_i\bar w_j+\bar w_iw_j)^2(w_k\bar w_l+\bar w_kw_l)\,,\\
&T^{\bf0^-}_{{\bf2}_i{\bf1}_j{\bf2}_k{\bf1}_l}=-(w_i^2\bar w_j^2-\bar w_i^2w_j^2)(w_k\bar w_l-\bar w_kw_l)\,,\\
\langle ssss\rangle:\qquad& T^{\bf0^+}_{{\bf0^+}{\bf0^+}{\bf0^+}{\bf0^+}}=1\,,\\
\langle\phi s\phi s\rangle\,,\langle s\phi\phi s\rangle:\qquad &T^{\bf1}_{{\bf1}_i{\bf0^+}{\bf1}_k{\bf0^+}}=w_i\bar w_k+\bar w_iw_k\,,\\
\langle t st s\rangle\,,\langle  stt s\rangle:\qquad &T^{\bf2}_{{\bf2}_i{\bf0^+}{\bf2}_k{\bf0^+}}=w_i^2\bar w_k^2+\bar w_i^2w_k^2\,,\\
\langle t ts s\rangle:\qquad &T^{\bf0^+}_{{\bf2}_i{\bf2}_j{\bf0^+}{\bf0^+}}=(w_i\bar w_j+\bar w_iw_j)^2\,,\\
\langle \phi \phi s s\rangle:\qquad & T^{\bf0^+}_{{\bf1}_i{\bf1}_j{\bf0^+}{\bf0^+}}=w_i\bar w_j+\bar w_iw_j\,,\\
\langle\phi s\phi t\rangle\,,\langle s\phi\phi t\rangle:\qquad &T^{\bf1}_{{\bf1}_i{\bf0^+}{\bf1}_k{\bf2}_l}=(w_k\bar w_l+\bar w_kw_l)(w_i\bar w_l+\bar w_iw_l)\,,\\
\langle \phi \phi st\rangle:\qquad &T^{\bf2}_{{\bf1}_i{\bf1}_j{\bf 0^+}{\bf2}_l}=w_l^2\bar w_j\bar w_i+\bar w_l^2 w_j w_i\,.\\
}

\section{Crossing vectors}\label{Vs}

Here we write the explicit vectors of crossing equations. In the following, an entry of $0$ will denote either a scalar or matrix of scalars depending on if the crossing equation is a scalar or a matrix.

\begin{footnotesize}
\es{V0p}{
\vec V_{{\bf 0^+},\Delta,\ell^+} =\begin{pmatrix}
\begin{pmatrix}
0&0&0\\
0&{2F_{ - ,\Delta ,\ell}^{\phi \phi ,\phi \phi }}&0\\
0&0&0
\end{pmatrix}
\\
0\\
\begin{pmatrix}
0&0&0\\
0&{ - 2F_{ + ,\Delta ,\ell}^{\phi \phi ,\phi \phi }}&0\\
0&0&0
\end{pmatrix}
\\
\begin{pmatrix}
0&0&0\\
0&0&0\\
0&0&{2F_{ - ,\Delta ,\ell}^{tt,tt}}
\end{pmatrix}
\\
0\\
\begin{pmatrix}
0&0&0\\
0&0&0\\
0&0&{ - 2F_{ + ,\Delta ,\ell}^{tt,tt}}
\end{pmatrix}
\\
0\\
0\\
\begin{pmatrix}
0&0&0\\
0&0&{F_{ - ,\Delta ,\ell}^{tt,\phi \phi }}\\
0&{F_{ - ,\Delta ,\ell}^{tt,\phi \phi }}&0
\end{pmatrix}
\\
\begin{pmatrix}
0&0&0\\
0&0&{F_{ - ,\Delta ,\ell}^{tt,\phi \phi }}\\
0&{F_{ - ,\Delta ,\ell}^{tt,\phi \phi }}&0
\end{pmatrix}
\\
\begin{pmatrix}
0&0&0\\
0&0&{ - F_{ + ,\Delta ,\ell}^{tt,\phi \phi }}\\
0&{ - F_{ + ,\Delta ,\ell}^{tt,\phi \phi }}&0
\end{pmatrix}
\\
\begin{pmatrix}
0&0&0\\
0&0&{ - F_{ + ,\Delta ,\ell}^{tt,\phi \phi }}\\
0&{ - F_{ + ,\Delta ,\ell}^{tt,\phi \phi }}&0
\end{pmatrix}
\\
\begin{pmatrix}
{2F_{ - ,\Delta ,\ell}^{ss,ss}}&0&0\\
0&0&0\\
0&0&0
\end{pmatrix}
\\
0\\
0\\
\begin{pmatrix}
0&0&{F_{ - ,\Delta ,\ell}^{tt,ss}}\\
0&0&0\\
{F_{ - ,\Delta ,\ell}^{tt,ss}}&0&0
\end{pmatrix}
\\
\begin{pmatrix}
0&0&{F_{ + ,\Delta ,\ell}^{tt,ss}}\\
0&0&0\\
{F_{ + ,\Delta ,\ell}^{tt,ss}}&0&0
\end{pmatrix}
\\
\begin{pmatrix}
0&{F_{ - ,\Delta ,\ell}^{\phi \phi ,ss}}&0\\
{F_{ - ,\Delta ,\ell}^{\phi \phi ,ss}}&0&0\\
0&0&0
\end{pmatrix}
\\
\begin{pmatrix}
0&{F_{ + ,\Delta ,\ell}^{\phi \phi ,ss}}&0\\
{F_{ + ,\Delta ,\ell}^{\phi \phi ,ss}}&0&0\\
0&0&0
\end{pmatrix}
\\
0\\
0\\
0
\end{pmatrix}\,,
\qquad\qquad
\vec{V}_{{\bf 0^-},\Delta,\ell^-}  = 
\begin{pmatrix}
\begin{pmatrix}
{  2F_{ - ,\Delta ,\ell}^{\phi \phi ,\phi \phi }}&0\\
0&0
\end{pmatrix}
\\
\begin{pmatrix}
{-4F_{ - ,\Delta ,\ell}^{\phi \phi ,\phi \phi }}&0\\
0&0
\end{pmatrix}
\\
\begin{pmatrix}
{ 2F_{ + ,\Delta ,\ell}^{\phi \phi ,\phi \phi }}&0\\
0&0
\end{pmatrix}
\\
\begin{pmatrix}
0&0\\
0&{  2F_{ - ,\Delta ,\ell}^{tt,tt}}
\end{pmatrix}
\\
\begin{pmatrix}
0&0\\
0&{-4F_{ - ,\Delta ,\ell}^{tt,tt}}
\end{pmatrix}
\\
\begin{pmatrix}
0&0\\
0&{2F_{ + ,\Delta ,\ell}^{tt,tt}}
\end{pmatrix}
\\
0\\
0\\
\begin{pmatrix}
0&{-F_{ - ,\Delta ,\ell}^{tt,\phi \phi }}\\
{-F_{ - ,\Delta ,\ell}^{tt,\phi \phi }}&0
\end{pmatrix}
\\
\begin{pmatrix}
0&{  F_{ - ,\Delta ,\ell}^{tt,\phi \phi }}\\
{  F_{ - ,\Delta ,\ell}^{tt,\phi \phi }}&0
\end{pmatrix}
\\
\begin{pmatrix}
0&{  F_{ + ,\Delta ,\ell}^{tt,\phi \phi }}\\
{  F_{ + ,\Delta ,\ell}^{tt,\phi \phi }}&0
\end{pmatrix}
\\
\begin{pmatrix}
0&{-F_{ + ,\Delta ,\ell}^{tt,\phi \phi }}\\
{-F_{ + ,\Delta ,\ell}^{tt,\phi \phi }}&0
\end{pmatrix}
\\
0\\
0\\
0\\
0\\
0\\
0\\
0\\
0\\
0\\
0
\end{pmatrix}\,,
}
\end{footnotesize}

\begin{footnotesize}

\es{V1}{
{\vec V_{{\bf1},\Delta,\ell^\pm}} = \begin{pmatrix}
0\\
0\\
0\\
0\\
0\\
0\\
 \begin{pmatrix}
0&0\\
0&{2(-1)^\ell F_{ - ,\Delta ,\ell}^{t\phi ,t\phi }}
 \end{pmatrix}
 \\
 \begin{pmatrix}
0&0\\
0&{2(-1)^\ell  F_{ + ,\Delta ,\ell}^{t\phi ,t\phi }}
 \end{pmatrix}
 \\
0\\
 \begin{pmatrix}
0&0\\
0&{2F_{ - ,\Delta ,\ell}^{\phi t,t\phi }}
 \end{pmatrix}
 \\
0\\
 \begin{pmatrix}
0&0\\
0&{2F_{ + ,\Delta ,\ell}^{\phi t,t\phi }}
 \end{pmatrix}
 \\
0\\
 \begin{pmatrix}
{2(-1)^\ell F_{ - ,\Delta ,\ell}^{\phi s,\phi s}}&0\\
0&0
 \end{pmatrix}
 \\
0\\
0\\
0\\
 \begin{pmatrix}
{ 2 F_{ - ,\Delta ,\ell}^{s\phi ,\phi s}}&0\\
0&0
 \end{pmatrix}
 \\
 \begin{pmatrix}
{ - 2 F_{ + ,\Delta ,\ell}^{s\phi ,\phi s}}&0\\
0&0
 \end{pmatrix}
 \\
 \begin{pmatrix}
0&{ F_{ - ,\Delta ,\ell}^{\phi s,\phi t}}\\
{  F_{ - ,\Delta ,\ell}^{\phi s,\phi t}}&0
 \end{pmatrix}
 \\
 \begin{pmatrix}
0&{  (-1)^\ell F_{ - ,\Delta ,\ell}^{s\phi ,\phi t}}\\
{ (-1)^\ell F_{ - ,\Delta ,\ell}^{s\phi ,\phi t}}&0
 \end{pmatrix}
 \\
 \begin{pmatrix}
0&{ (-1)^\ell F_{ + ,\Delta ,\ell}^{s\phi ,\phi t}}\\
{ (-1)^\ell F_{ + ,\Delta ,\ell}^{s\phi ,\phi t}}&0
 \end{pmatrix}
 \end{pmatrix}\,,\qquad
{\vec V_{{\bf2},\Delta,\ell^+}} = \begin{pmatrix}
0\\
 \begin{pmatrix}
{2F_{ - ,\Delta ,\ell}^{\phi \phi ,\phi \phi }}&0\\
0&0
 \end{pmatrix}
 \\
 \begin{pmatrix}
{2F_{ + ,\Delta ,J\ell}^{\phi \phi ,\phi \phi }}&0\\
0&0
 \end{pmatrix}
 \\
0\\
0\\
0\\
0\\
0\\
0\\
0\\
0\\
0\\
0\\
0\\
 \begin{pmatrix}
0&0\\
0&{2F_{ - ,\Delta ,\ell}^{ts,ts}}
 \end{pmatrix}
 \\
 \begin{pmatrix}
0&0\\
0&{2F_{ - ,\Delta ,\ell}^{st,ts}}
 \end{pmatrix}
 \\
 \begin{pmatrix}
0&0\\
0&{ - 2F_{ + ,\Delta ,\ell}^{st,ts}}
 \end{pmatrix}
 \\
0\\
0\\
0\\
 \begin{pmatrix}
0&{F_{ - ,\Delta ,\ell}^{\phi \phi ,st}}\\
{F_{ - ,\Delta ,\ell}^{\phi \phi ,st}}&0
 \end{pmatrix}
 \\
 \begin{pmatrix}
0&{ - F_{ + ,\Delta ,\ell}^{\phi \phi ,st}}\\
{ - F_{ + ,\Delta ,\ell}^{\phi \phi ,st}}&0
 \end{pmatrix}
 \end{pmatrix}\,,
}
\end{footnotesize}

\begin{footnotesize}
\es{V2m}{
\vec{V}_{{{\bf 2},\Delta,\ell^-}} =  \begin{pmatrix}
0\\
0\\
0\\
0\\
0\\
0\\
0\\
0\\
0\\
0\\
0\\
0\\
0\\
0\\
{-2F_{ - ,\Delta ,\ell}^{ts,ts}}\\
{  2F_{ - ,\Delta ,\ell}^{st,ts}}\\
{-2F_{ + ,\Delta ,\ell}^{st,ts}}\\
0\\
0\\
0\\
0\\
0
\end{pmatrix}\,,\quad
\vec{V}_{{\bf3},\Delta,\ell^\pm} =  \begin{pmatrix}
0\\
0\\
0\\
0\\
0\\
0\\
{2{{( - 1)}^\ell} F_{ - ,\Delta ,\ell}^{t\phi ,t\phi }}\\
{ -2 {{( - 1)}^\ell} F_{ + ,\Delta ,\ell}^{t\phi ,t\phi }}\\
{2 F_{ - ,\Delta ,\ell}^{\phi t,t\phi }}\\
0\\
{2F_{ + ,\Delta ,\ell}^{\phi t,t\phi }}\\
0\\
0\\
0\\
0\\
0\\
0\\
0\\
0\\
0\\
0\\
0
\end{pmatrix}\,,\quad
\vec{V}_{{\bf4},\Delta,\ell^+} = \begin{pmatrix}
0\\
0\\
0\\
0\\
{2F_{ - ,\Delta ,\ell}^{tt,tt}}\\
{2F_{ + ,\Delta ,\ell}^{tt,tt}}\\
0\\
0\\
0\\
0\\
0\\
0\\
0\\
0\\
0\\
0\\
0\\
0\\
0\\
0\\
0\\
0
\end{pmatrix}\,.
}
\end{footnotesize}

\section{Computed points}
\label{app:points}

\begin{table}
\centering
\begin{tabular}{@{}c|c|c|c|c|c@{}}
\toprule
$\De_\f$ & $\De_s$ & $\De_t$ & $\frac{\lambda_{sss}}{\lambda_{\f\f s}}$ & $\frac{\lambda_{tts}}{\lambda_{\f\f s}}$ & $\frac{\lambda_{\f\f t}}{\lambda_{\f\f s}}$ \\
\midrule
 0.519091478 & 1.51141697 & 1.23631316 & 1.23631316 & 1.23631316 & 1.23631316 \\
 0.519088325 & 1.51139275 & 1.23629816 & 1.23629816 & 1.23629816 & 1.23629816 \\
 0.519085258 & 1.51131148 & 1.23626768 & 1.23626768 & 1.23626768 & 1.23626768 \\
 0.519083027 & 1.51130787 & 1.23626810 & 1.23626810 & 1.23626810 & 1.23626810 \\
 0.519084900 & 1.51132513 & 1.23626125 & 1.23626125 & 1.23626125 & 1.23626125 \\
 0.519101167 & 1.51147622 & 1.23635261 & 1.23635261 & 1.23635261 & 1.23635261 \\
 0.519079494 & 1.51130889 & 1.23625139 & 1.23625139 & 1.23625139 & 1.23625139 \\
 0.519088780 & 1.51141601 & 1.23631255 & 1.23631255 & 1.23631255 & 1.23631255 \\
 0.519099104 & 1.51149674 & 1.23636042 & 1.23636042 & 1.23636042 & 1.23636042 \\
 0.519074036 & 1.51122813 & 1.23622003 & 1.23622003 & 1.23622003 & 1.23622003 \\
 0.519075834 & 1.51124069 & 1.23621228 & 1.23621228 & 1.23621228 & 1.23621228 \\
 0.519086133 & 1.51140646 & 1.23629626 & 1.23629626 & 1.23629626 & 1.23629626 \\
 0.519091591 & 1.51144378 & 1.23631466 & 1.23631466 & 1.23631466 & 1.23631466 \\
 0.519101492 & 1.51147122 & 1.23635764 & 1.23635764 & 1.23635764 & 1.23635764 \\
 0.519095922 & 1.51143997 & 1.23632426 & 1.23632426 & 1.23632426 & 1.23632426 \\
 0.519089922 & 1.51145388 & 1.23632418 & 1.23632418 & 1.23632418 & 1.23632418 \\
 0.519096569 & 1.51145694 & 1.23634757 & 1.23634757 & 1.23634757 & 1.23634757 \\
 0.519078927 & 1.51129693 & 1.23625413 & 1.23625413 & 1.23625413 & 1.23625413 \\
 0.519085163 & 1.51135762 & 1.23627100 & 1.23627100 & 1.23627100 & 1.23627100 \\
 0.519095326 & 1.51148380 & 1.23634189 & 1.23634189 & 1.23634189 & 1.23634189 \\
 0.519081546 & 1.51129674 & 1.23625401 & 1.23625401 & 1.23625401 & 1.23625401 \\
 0.519078552 & 1.51131491 & 1.23624987 & 1.23624987 & 1.23624987 & 1.23624987 \\
 0.519104279 & 1.51152063 & 1.23637609 & 1.23637609 & 1.23637609 & 1.23637609 \\
 0.519077715 & 1.51124447 & 1.23623187 & 1.23623187 & 1.23623187 & 1.23623187 \\
 0.519074849 & 1.51125858 & 1.23622291 & 1.23622291 & 1.23622291 & 1.23622291 \\
 0.519081236 & 1.51134317 & 1.23627346 & 1.23627346 & 1.23627346 & 1.23627346 \\
 0.519087675 & 1.51137648 & 1.23630209 & 1.23630209 & 1.23630209 & 1.23630209 \\
 0.519092708 & 1.51139697 & 1.23631672 & 1.23631672 & 1.23631672 & 1.23631672 \\
 0.519080005 & 1.51131897 & 1.23624739 & 1.23624739 & 1.23624739 & 1.23624739 \\
 0.519096168 & 1.51149661 & 1.23635270 & 1.23635270 & 1.23635270 & 1.23635270 \\
 0.519073619 & 1.51122331 & 1.23621261 & 1.23621261 & 1.23621261 & 1.23621261 \\
 0.519085778 & 1.51132384 & 1.23628076 & 1.23628076 & 1.23628076 & 1.23628076 \\
 0.519075030 & 1.51121405 & 1.23622082 & 1.23622082 & 1.23622082 & 1.23622082 \\
\bottomrule
\end{tabular}
\caption{\label{tab:allowedpointsL43}Allowed points in the $\Lambda=43$ island.}
\end{table}

\begin{table}
\centering
\begin{tabular}{@{}c|c|c@{}}
\toprule
$\De_\f$ & $\De_s$ & $\De_t$  \\
\midrule
 0.519102918 & 1.51155239 & 1.23637912 \\
 0.519108668 & 1.51153259 & 1.23638777 \\
 0.519084234 & 1.51130123 & 1.23627522 \\
 0.519086029 & 1.51135180 & 1.23626619 \\
 0.519093006 & 1.51136316 & 1.23630104 \\
 0.519074320 & 1.51118490 & 1.23620389 \\
 0.519102521 & 1.51148465 & 1.23635216 \\
 0.519109629 & 1.51158377 & 1.23640397 \\
 0.519077293 & 1.51123239 & 1.23621599 \\
 0.519086333 & 1.51131583 & 1.23626238 \\
 0.519088681 & 1.51132415 & 1.23628443 \\
 0.519103625 & 1.51156649 & 1.23639258 \\
 0.519097531 & 1.51152436 & 1.23635506 \\
 0.519104829 & 1.51155133 & 1.23639016 \\
 0.519106540 & 1.51151848 & 1.23638657 \\
 0.519099641 & 1.51149324 & 1.23634449 \\
 0.519091345 & 1.51145618 & 1.23631607 \\
 0.519099918 & 1.51143846 & 1.23634527 \\
 0.519081611 & 1.51130758 & 1.23626964 \\
 0.519093364 & 1.51150044 & 1.23634838 \\
 0.519090250 & 1.51142367 & 1.23630035 \\
 0.519095800 & 1.51148226 & 1.23635217 \\
 0.519079641 & 1.51134752 & 1.23626571 \\
 0.519066632 & 1.51113867 & 1.23617714 \\
 0.519089008 & 1.51142764 & 1.23631975 \\
 0.519081958 & 1.51136634 & 1.23626970 \\
 0.519073136 & 1.51120099 & 1.23619139 \\
 0.519079477 & 1.51125698 & 1.23623547 \\
 0.519092469 & 1.51139541 & 1.23629967 \\
 0.519091772 & 1.51141270 & 1.23632592 \\
 0.519092673 & 1.51139406 & 1.23632062 \\
 0.519069909 & 1.51118566 & 1.23618371 \\
 0.519101366 & 1.51149948 & 1.23637079 \\
 0.519090130 & 1.51135761 & 1.23628884 \\
 0.519082450 & 1.51133857 & 1.23628007 \\
 0.519107673 & 1.51153927 & 1.23639093 \\
 0.519096449 & 1.51141705 & 1.23632503 \\
 0.519074457 & 1.51126505 & 1.23622784 \\
 0.519089400 & 1.51140721 & 1.23631629 \\
 0.519080527 & 1.51132179 & 1.23624620 \\
 0.519075418 & 1.51126954 & 1.23622125 \\
 0.519071183 & 1.51118343 & 1.23619298 \\
\bottomrule
\end{tabular}
\caption{\label{tab:disallowedpointsL43}Disallowed points computed at $\Lambda=43$.}
\end{table}

\begin{table}
\centering
\begin{tabular}{@{}c|c|c|c|c|c@{}}
\toprule
$\De_\f$ & $\De_s$ & $\De_t$ & $\frac{\lambda_{sss}}{\lambda_{\f\f s}}$ & $\frac{\lambda_{tts}}{\lambda_{\f\f s}}$ & $\frac{\lambda_{\f\f t}}{\lambda_{\f\f s}}$ \\
\midrule
0.519101167 & 1.51147622 & 1.23635261 & 1.20936871 & 1.82235941 & 1.76596240\\
0.519079494 & 1.51130889 & 1.23625139 & 1.20934084 & 1.82223619 & 1.76589343\\
0.519089922 & 1.51145388 & 1.23632418 & 1.20972662 & 1.82245009 & 1.76596250\\
0.519075834 & 1.51124069 & 1.23621228 & 1.20906418 & 1.82207926 & 1.76585335\\
0.519075030 & 1.51121405 & 1.23622082 & 1.20879917 & 1.82210575 & 1.76586410\\
0.519091591 & 1.51144378 & 1.23631466 & 1.20970116 & 1.82235729 & 1.76594661\\
0.519086715 & 1.51136546 & 1.23628759 & 1.20932228 & 1.82228021 & 1.76592047\\
\bottomrule
\end{tabular}
\caption{\label{tab:thesevenpoints}Allowed points in the $\Lambda=43$ island used for computing upper and lower bounds on $C_T$, $C_J$, and $\l_{\f\f s}$.}
\end{table}

\begin{table}
\centering
\begin{tabular}{@{}c|c|c|c|c|c@{}}
\toprule
$\De_\f$ & $\De_s$ & $\De_t$ & $\frac{\lambda_{sss}}{\lambda_{\f\f s}}$ & $\frac{\lambda_{tts}}{\lambda_{\f\f s}}$ & $\frac{\lambda_{\f\f t}}{\lambda_{\f\f s}}$ \\
\midrule
0.519130434&1.51173444&1.23648971&1.20977354&1.82254374&1.76606470\\
0.519135171&1.51172427&1.23649356&1.20947477&1.82245370&1.76605159\\
0.519076518&1.51110487&1.23620503&1.20766586&1.82191247&1.76584197\\
0.519115548&1.51167580&1.23642873&1.21014420&1.82257643&1.76603227\\
0.519113909&1.51170936&1.23646025&1.21013097&1.82272756&1.76607582\\
0.519096732&1.51147972&1.23636344&1.20944426&1.82251617&1.76600087\\
0.519128801&1.51168098&1.23648846&1.20929738&1.82252856&1.76605495\\
0.519119255&1.51170685&1.23646324&1.21007964&1.82275976&1.76606055\\
0.519109342&1.51150256&1.23640031&1.20891847&1.82236481&1.76600112\\
0.519087647&1.51141667&1.23630721&1.20963450&1.82247476&1.76594440\\
0.519105802&1.51141826&1.23635621&1.20856734&1.82219520&1.76595563\\
0.519125142&1.51173460&1.23646472&1.21012577&1.82250871&1.76605236\\
0.519107610&1.51164424&1.23640715&1.21022297&1.82258938&1.76603036\\
0.519115226&1.51174173&1.23647414&1.21033291&1.82281805&1.76609054\\
0.519084390&1.51137895&1.23628833&1.20979136&1.82229748&1.76593252\\
0.519096529&1.51153244&1.23635748&1.20995866&1.82250999&1.76599060\\
0.519122718&1.51168123&1.23647847&1.20940108&1.82261368&1.76607344\\
0.519138689&1.51177044&1.23653770&1.20947377&1.82262309&1.76609008\\
0.519057668&1.51097950&1.23611240&1.20794762&1.82181966&1.76576836\\
0.519074424&1.51116298&1.23616082&1.20864157&1.82181577&1.76579563\\
\bottomrule
\end{tabular}
\caption{\label{tab:the20points}Allowed points in the $\Lambda=35$ island used for obtaining low-lying scalar operator dimensions via the extremal functional method.}
\end{table}

\clearpage
\bibliography{Biblio}

\providecommand{\href}[2]{#2}\begingroup\raggedright\begin{thebibliography}{100}

\bibitem{Rattazzi:2008pe}
R.~Rattazzi, V.~S. Rychkov, E.~Tonni, and A.~Vichi, ``{Bounding scalar operator
  dimensions in 4D CFT},''
  \href{http://dx.doi.org/10.1088/1126-6708/2008/12/031}{{\em JHEP} {\bfseries
  0812} (2008) 031},
\href{http://arxiv.org/abs/0807.0004}{{\ttfamily arXiv:0807.0004 [hep-th]}}.

\bibitem{Rychkov:2009ij}
V.~S. Rychkov and A.~Vichi, ``{Universal Constraints on Conformal Operator
  Dimensions},'' \href{http://dx.doi.org/10.1103/PhysRevD.80.045006}{{\em
  Phys.Rev.} {\bfseries D80} (2009) 045006},
\href{http://arxiv.org/abs/0905.2211}{{\ttfamily arXiv:0905.2211 [hep-th]}}.

\bibitem{Poland:2018epd}
D.~Poland, S.~Rychkov, and A.~Vichi, ``{The Conformal Bootstrap: Theory,
  Numerical Techniques, and Applications},''
  \href{http://dx.doi.org/10.1103/RevModPhys.91.015002}{{\em Rev. Mod. Phys.}
  {\bfseries 91} no.~1, (2019) 15002},
  \href{http://arxiv.org/abs/1805.04405}{{\ttfamily arXiv:1805.04405
  [hep-th]}}.
[Rev. Mod. Phys.91,015002(2019)].

\bibitem{Chester:2019wfx}
S.~M. Chester, ``{Weizmann Lectures on the Numerical Conformal Bootstrap},''
\href{http://arxiv.org/abs/1907.05147}{{\ttfamily arXiv:1907.05147 [hep-th]}}.

\bibitem{Kos:2014bka}
F.~Kos, D.~Poland, and D.~Simmons-Duffin, ``{Bootstrapping Mixed Correlators in
  the 3D Ising Model},'' \href{http://dx.doi.org/10.1007/JHEP11(2014)109}{{\em
  JHEP} {\bfseries 11} (2014) 109},
\href{http://arxiv.org/abs/1406.4858}{{\ttfamily arXiv:1406.4858 [hep-th]}}.

\bibitem{Kos:2015mba}
F.~Kos, D.~Poland, D.~Simmons-Duffin, and A.~Vichi, ``{Bootstrapping the O(N)
  Archipelago},'' \href{http://dx.doi.org/10.1007/JHEP11(2015)106}{{\em JHEP}
  {\bfseries 11} (2015) 106},
\href{http://arxiv.org/abs/1504.07997}{{\ttfamily arXiv:1504.07997 [hep-th]}}.

\bibitem{Kos:2016ysd}
F.~Kos, D.~Poland, D.~Simmons-Duffin, and A.~Vichi, ``{Precision Islands in the
  Ising and $O(N)$ Models},''
  \href{http://dx.doi.org/10.1007/JHEP08(2016)036}{{\em JHEP} {\bfseries 08}
  (2016) 036},
\href{http://arxiv.org/abs/1603.04436}{{\ttfamily arXiv:1603.04436 [hep-th]}}.

\bibitem{Rong:2018okz}
J.~Rong and N.~Su, ``{Bootstrapping minimal $\mathcal{N}=1$ superconformal
  field theory in three dimensions},''
\href{http://arxiv.org/abs/1807.04434}{{\ttfamily arXiv:1807.04434 [hep-th]}}.

\bibitem{Agmon:2019imm}
N.~B. Agmon, S.~M. Chester, and S.~S. Pufu, ``{The M-theory Archipelago},''
  \href{http://dx.doi.org/10.1007/JHEP02(2020)010}{{\em JHEP} {\bfseries 02}
  (2020) 010}, \href{http://arxiv.org/abs/1907.13222}{{\ttfamily
  arXiv:1907.13222 [hep-th]}}.

\bibitem{Li:2016wdp}
Z.~Li and N.~Su, ``{Bootstrapping Mixed Correlators in the Five Dimensional
  Critical O(N) Models},''
  \href{http://dx.doi.org/10.1007/JHEP04(2017)098}{{\em JHEP} {\bfseries 04}
  (2017) 098},
\href{http://arxiv.org/abs/1607.07077}{{\ttfamily arXiv:1607.07077 [hep-th]}}.

\bibitem{Nakayama:2016jhq}
Y.~Nakayama and T.~Ohtsuki, ``{Conformal Bootstrap Dashing Hopes of Emergent
  Symmetry},'' \href{http://dx.doi.org/10.1103/PhysRevLett.117.131601}{{\em
  Phys. Rev. Lett.} {\bfseries 117} no.~13, (2016) 131601},
\href{http://arxiv.org/abs/1602.07295}{{\ttfamily arXiv:1602.07295
  [cond-mat.str-el]}}.

\bibitem{Li:2017ddj}
D.~Li, D.~Meltzer, and A.~Stergiou, ``{Bootstrapping mixed correlators in 4D $
  \mathcal{N} $ = 1 SCFTs},''
  \href{http://dx.doi.org/10.1007/JHEP07(2017)029}{{\em JHEP} {\bfseries 07}
  (2017) 029},
\href{http://arxiv.org/abs/1702.00404}{{\ttfamily arXiv:1702.00404 [hep-th]}}.

\bibitem{Behan:2018hfx}
C.~Behan, ``{Bootstrapping the long-range Ising model in three dimensions},''
  \href{http://dx.doi.org/10.1088/1751-8121/aafd1b}{{\em J. Phys.} {\bfseries
  A52} no.~7, (2019) 075401},
\href{http://arxiv.org/abs/1810.07199}{{\ttfamily arXiv:1810.07199 [hep-th]}}.

\bibitem{Kousvos:2018rhl}
S.~R. Kousvos and A.~Stergiou, ``{Bootstrapping Mixed Correlators in
  Three-Dimensional Cubic Theories},''
  \href{http://dx.doi.org/10.21468/SciPostPhys.6.3.035}{{\em SciPost Phys.}
  {\bfseries 6} no.~3, (2019) 035},
\href{http://arxiv.org/abs/1810.10015}{{\ttfamily arXiv:1810.10015 [hep-th]}}.

\bibitem{Kousvos:2019hgc}
S.~R. Kousvos and A.~Stergiou, ``{Bootstrapping Mixed Correlators in
  Three-Dimensional Cubic Theories II},''
  \href{http://dx.doi.org/10.21468/SciPostPhys.8.6.085}{{\em SciPost Phys.}
  {\bfseries 8} no.~6, (2020) 085},
  \href{http://arxiv.org/abs/1911.00522}{{\ttfamily arXiv:1911.00522
  [hep-th]}}.

\bibitem{Iliesiu:2015qra}
L.~Iliesiu, F.~Kos, D.~Poland, S.~S. Pufu, D.~Simmons-Duffin, and R.~Yacoby,
  ``{Bootstrapping 3D Fermions},''
  \href{http://dx.doi.org/10.1007/JHEP03(2016)120}{{\em JHEP} {\bfseries 03}
  (2016) 120},
\href{http://arxiv.org/abs/1508.00012}{{\ttfamily arXiv:1508.00012 [hep-th]}}.

\bibitem{Iliesiu:2017nrv}
L.~Iliesiu, F.~Kos, D.~Poland, S.~S. Pufu, and D.~Simmons-Duffin,
  ``{Bootstrapping 3D Fermions with Global Symmetries},''
  \href{http://dx.doi.org/10.1007/JHEP01(2018)036}{{\em JHEP} {\bfseries 01}
  (2018) 036},
\href{http://arxiv.org/abs/1705.03484}{{\ttfamily arXiv:1705.03484 [hep-th]}}.

\bibitem{Karateev:2019pvw}
D.~Karateev, P.~Kravchuk, M.~Serone, and A.~Vichi, ``{Fermion Conformal
  Bootstrap in 4d},'' \href{http://dx.doi.org/10.1007/JHEP06(2019)088}{{\em
  JHEP} {\bfseries 06} (2019) 088},
\href{http://arxiv.org/abs/1902.05969}{{\ttfamily arXiv:1902.05969 [hep-th]}}.

\bibitem{Dymarsky:2017xzb}
A.~Dymarsky, J.~Penedones, E.~Trevisani, and A.~Vichi, ``{Charting the space of
  3D CFTs with a continuous global symmetry},''
  \href{http://dx.doi.org/10.1007/JHEP05(2019)098}{{\em JHEP} {\bfseries 05}
  (2019) 098},
\href{http://arxiv.org/abs/1705.04278}{{\ttfamily arXiv:1705.04278 [hep-th]}}.

\bibitem{Reehorst:2019pzi}
M.~Reehorst, E.~Trevisani, and A.~Vichi, ``{Mixed Scalar-Current bootstrap in
  three dimensions},''
\href{http://arxiv.org/abs/1911.05747}{{\ttfamily arXiv:1911.05747 [hep-th]}}.

\bibitem{Dymarsky:2017yzx}
A.~Dymarsky, F.~Kos, P.~Kravchuk, D.~Poland, and D.~Simmons-Duffin, ``{The 3d
  Stress-Tensor Bootstrap},''
  \href{http://dx.doi.org/10.1007/JHEP02(2018)164}{{\em JHEP} {\bfseries 02}
  (2018) 164},
\href{http://arxiv.org/abs/1708.05718}{{\ttfamily arXiv:1708.05718 [hep-th]}}.

\bibitem{Rattazzi:2010yc}
R.~Rattazzi, S.~Rychkov, and A.~Vichi, ``{Bounds in 4D Conformal Field Theories
  with Global Symmetry},''
  \href{http://dx.doi.org/10.1088/1751-8113/44/3/035402}{{\em J.Phys.}
  {\bfseries A44} (2011) 035402},
\href{http://arxiv.org/abs/1009.5985}{{\ttfamily arXiv:1009.5985 [hep-th]}}.

\bibitem{Vichi:2011ux}
A.~Vichi, ``{Improved bounds for CFT's with global symmetries},''
  \href{http://dx.doi.org/10.1007/JHEP01(2012)162}{{\em JHEP} {\bfseries 1201}
  (2012) 162},
\href{http://arxiv.org/abs/1106.4037}{{\ttfamily arXiv:1106.4037 [hep-th]}}.

\bibitem{Poland:2011ey}
D.~Poland, D.~Simmons-Duffin, and A.~Vichi, ``{Carving Out the Space of 4D
  CFTs},'' \href{http://dx.doi.org/10.1007/JHEP05(2012)110}{{\em JHEP}
  {\bfseries 1205} (2012) 110},
\href{http://arxiv.org/abs/1109.5176}{{\ttfamily arXiv:1109.5176 [hep-th]}}.

\bibitem{Kos:2013tga}
F.~Kos, D.~Poland, and D.~Simmons-Duffin, ``{Bootstrapping the $O(N)$ vector
  models},'' \href{http://dx.doi.org/10.1007/JHEP06(2014)091}{{\em JHEP}
  {\bfseries 06} (2014) 091},
\href{http://arxiv.org/abs/1307.6856}{{\ttfamily arXiv:1307.6856 [hep-th]}}.

\bibitem{Berkooz:2014yda}
M.~Berkooz, R.~Yacoby, and A.~Zait, ``{Bounds on $\mathcal{N} = 1$
  superconformal theories with global symmetries},''
  \href{http://dx.doi.org/10.1007/JHEP01(2015)132,
  10.1007/JHEP08(2014)008}{{\em JHEP} {\bfseries 08} (2014) 008},
  \href{http://arxiv.org/abs/1402.6068}{{\ttfamily arXiv:1402.6068 [hep-th]}}.
[Erratum: JHEP01,132(2015)].

\bibitem{Nakayama:2014lva}
Y.~Nakayama and T.~Ohtsuki, ``{Approaching the conformal window of $O(n)\times
  O(m)$ symmetric Landau-Ginzburg models using the conformal bootstrap},''
  \href{http://dx.doi.org/10.1103/PhysRevD.89.126009}{{\em Phys.Rev.}
  {\bfseries D89} no.~12, (2014) 126009},
\href{http://arxiv.org/abs/1404.0489}{{\ttfamily arXiv:1404.0489 [hep-th]}}.

\bibitem{Caracciolo:2014cxa}
F.~Caracciolo, A.~Castedo~Echeverri, B.~von Harling, and M.~Serone, ``{Bounds
  on OPE Coefficients in 4D Conformal Field Theories},''
  \href{http://dx.doi.org/10.1007/JHEP10(2014)020}{{\em JHEP} {\bfseries 10}
  (2014) 020},
\href{http://arxiv.org/abs/1406.7845}{{\ttfamily arXiv:1406.7845 [hep-th]}}.

\bibitem{Nakayama:2014sba}
Y.~Nakayama and T.~Ohtsuki, ``{Bootstrapping phase transitions in QCD and
  frustrated spin systems},''
  \href{http://dx.doi.org/10.1103/PhysRevD.91.021901}{{\em Phys. Rev.}
  {\bfseries D91} no.~2, (2015) 021901},
\href{http://arxiv.org/abs/1407.6195}{{\ttfamily arXiv:1407.6195 [hep-th]}}.

\bibitem{Chester:2014gqa}
S.~M. Chester, S.~S. Pufu, and R.~Yacoby, ``{Bootstrapping $O(N)$ vector models
  in 4 $< d <$ 6},'' \href{http://dx.doi.org/10.1103/PhysRevD.91.086014}{{\em
  Phys. Rev.} {\bfseries D91} no.~8, (2015) 086014},
\href{http://arxiv.org/abs/1412.7746}{{\ttfamily arXiv:1412.7746 [hep-th]}}.

\bibitem{Nakayama:2014yia}
Y.~Nakayama and T.~Ohtsuki, ``{Five dimensional $O(N)$-symmetric CFTs from
  conformal bootstrap},''
  \href{http://dx.doi.org/10.1016/j.physletb.2014.05.058}{{\em Phys. Lett.}
  {\bfseries B734} (2014) 193--197},
\href{http://arxiv.org/abs/1404.5201}{{\ttfamily arXiv:1404.5201 [hep-th]}}.

\bibitem{Chester:2015qca}
S.~M. Chester, S.~Giombi, L.~V. Iliesiu, I.~R. Klebanov, S.~S. Pufu, and
  R.~Yacoby, ``{Accidental Symmetries and the Conformal Bootstrap},''
  \href{http://dx.doi.org/10.1007/JHEP01(2016)110}{{\em JHEP} {\bfseries 01}
  (2016) 110},
\href{http://arxiv.org/abs/1507.04424}{{\ttfamily arXiv:1507.04424 [hep-th]}}.

\bibitem{Chester:2015lej}
S.~M. Chester, L.~V. Iliesiu, S.~S. Pufu, and R.~Yacoby, ``{Bootstrapping
  $O(N)$ Vector Models with Four Supercharges in $3 \leq d \leq4$},''
  \href{http://dx.doi.org/10.1007/JHEP05(2016)103}{{\em JHEP} {\bfseries 05}
  (2016) 103},
\href{http://arxiv.org/abs/1511.07552}{{\ttfamily arXiv:1511.07552 [hep-th]}}.

\bibitem{Chester:2016wrc}
S.~M. Chester and S.~S. Pufu, ``{Towards bootstrapping QED$_{3}$},''
  \href{http://dx.doi.org/10.1007/JHEP08(2016)019}{{\em JHEP} {\bfseries 08}
  (2016) 019},
\href{http://arxiv.org/abs/1601.03476}{{\ttfamily arXiv:1601.03476 [hep-th]}}.

\bibitem{Nakayama:2016knq}
Y.~Nakayama, ``{Bootstrap bound for conformal multi-flavor QCD on lattice},''
  \href{http://dx.doi.org/10.1007/JHEP07(2016)038}{{\em JHEP} {\bfseries 07}
  (2016) 038},
\href{http://arxiv.org/abs/1605.04052}{{\ttfamily arXiv:1605.04052 [hep-th]}}.

\bibitem{Iha:2016ppj}
H.~Iha, H.~Makino, and H.~Suzuki, ``{Upper bound on the mass anomalous
  dimension in many-flavor gauge theories: a conformal bootstrap approach},''
  \href{http://dx.doi.org/10.1093/ptep/ptw046}{{\em PTEP} {\bfseries 2016}
  no.~5, (2016) 053B03},
\href{http://arxiv.org/abs/1603.01995}{{\ttfamily arXiv:1603.01995 [hep-th]}}.

\bibitem{Nakayama:2017vdd}
Y.~Nakayama, ``{Bootstrap experiments on higher dimensional CFTs},''
  \href{http://dx.doi.org/10.1142/S0217751X18500367}{{\em Int. J. Mod. Phys.}
  {\bfseries A33} no.~07, (2018) 1850036},
\href{http://arxiv.org/abs/1705.02744}{{\ttfamily arXiv:1705.02744 [hep-th]}}.

\bibitem{Rong:2017cow}
J.~Rong and N.~Su, ``{Scalar CFTs and Their Large N Limits},''
  \href{http://dx.doi.org/10.1007/JHEP09(2018)103}{{\em JHEP} {\bfseries 09}
  (2018) 103},
\href{http://arxiv.org/abs/1712.00985}{{\ttfamily arXiv:1712.00985 [hep-th]}}.

\bibitem{Chester:2017vdh}
S.~M. Chester, L.~V. Iliesiu, M.~Mezei, and S.~S. Pufu, ``{Monopole Operators
  in $U(1)$ Chern-Simons-Matter Theories},''
  \href{http://dx.doi.org/10.1007/JHEP05(2018)157}{{\em JHEP} {\bfseries 05}
  (2018) 157},
\href{http://arxiv.org/abs/1710.00654}{{\ttfamily arXiv:1710.00654 [hep-th]}}.

\bibitem{Stergiou:2018gjj}
A.~Stergiou, ``{Bootstrapping hypercubic and hypertetrahedral theories in three
  dimensions},'' \href{http://dx.doi.org/10.1007/JHEP05(2018)035}{{\em JHEP}
  {\bfseries 05} (2018) 035},
\href{http://arxiv.org/abs/1801.07127}{{\ttfamily arXiv:1801.07127 [hep-th]}}.

\bibitem{Li:2018lyb}
Z.~Li, ``{Solving QED$_3$ with Conformal Bootstrap},''
\href{http://arxiv.org/abs/1812.09281}{{\ttfamily arXiv:1812.09281 [hep-th]}}.

\bibitem{Rong:2019qer}
J.~Rong and N.~Su, ``{Bootstrapping the $\mathcal{N}=1$ Wess-Zumino models in
  three dimensions},''
\href{http://arxiv.org/abs/1910.08578}{{\ttfamily arXiv:1910.08578 [hep-th]}}.

\bibitem{Rychkov:2011et}
S.~Rychkov, ``{Conformal Bootstrap in Three Dimensions?},''
\href{http://arxiv.org/abs/1111.2115}{{\ttfamily arXiv:1111.2115 [hep-th]}}.

\bibitem{ElShowk:2012ht}
S.~El-Showk, M.~F. Paulos, D.~Poland, S.~Rychkov, D.~Simmons-Duffin, and
  A.~Vichi, ``{Solving the 3D Ising Model with the Conformal Bootstrap},''
  \href{http://dx.doi.org/10.1103/PhysRevD.86.025022}{{\em Phys. Rev.}
  {\bfseries D86} (2012) 025022},
\href{http://arxiv.org/abs/1203.6064}{{\ttfamily arXiv:1203.6064 [hep-th]}}.

\bibitem{Gaiotto:2013nva}
D.~Gaiotto, D.~Mazac, and M.~F. Paulos, ``{Bootstrapping the 3d Ising twist
  defect},'' \href{http://dx.doi.org/10.1007/JHEP03(2014)100}{{\em JHEP}
  {\bfseries 03} (2014) 100},
\href{http://arxiv.org/abs/1310.5078}{{\ttfamily arXiv:1310.5078 [hep-th]}}.

\bibitem{El-Showk:2014dwa}
S.~El-Showk, M.~F. Paulos, D.~Poland, S.~Rychkov, D.~Simmons-Duffin, and
  A.~Vichi, ``{Solving the 3d Ising Model with the Conformal Bootstrap II.
  c-Minimization and Precise Critical Exponents},''
  \href{http://dx.doi.org/10.1007/s10955-014-1042-7}{{\em J. Stat. Phys.}
  {\bfseries 157} (2014) 869},
\href{http://arxiv.org/abs/1403.4545}{{\ttfamily arXiv:1403.4545 [hep-th]}}.

\bibitem{Chang:2017cdx}
C.-M. Chang, M.~Fluder, Y.-H. Lin, and Y.~Wang, ``{Spheres, Charges,
  Instantons, and Bootstrap: A Five-Dimensional Odyssey},''
  \href{http://dx.doi.org/10.1007/JHEP03(2018)123}{{\em JHEP} {\bfseries 03}
  (2018) 123},
\href{http://arxiv.org/abs/1710.08418}{{\ttfamily arXiv:1710.08418 [hep-th]}}.

\bibitem{Li:2017kck}
Z.~Li and N.~Su, ``{3D CFT Archipelago from Single Correlator Bootstrap},''
\href{http://arxiv.org/abs/1706.06960}{{\ttfamily arXiv:1706.06960 [hep-th]}}.

\bibitem{Hasegawa:2018yqg}
C.~Hasegawa and Y.~Nakayama, ``{Three ways to solve critical $\phi^4$ theory on
  $4-\epsilon$ dimensional real projective space: perturbation, bootstrap, and
  Schwinger-Dyson equation},''
  \href{http://dx.doi.org/10.1142/S0217751X18500495}{{\em Int. J. Mod. Phys.}
  {\bfseries A33} no.~08, (2018) 1850049},
\href{http://arxiv.org/abs/1801.09107}{{\ttfamily arXiv:1801.09107 [hep-th]}}.

\bibitem{Gowdigere:2018lxz}
C.~N. Gowdigere, J.~Santara, and Sumedha, ``{Conformal Bootstrap Signatures of
  the Tricritical Ising Universality Class},''
\href{http://arxiv.org/abs/1811.11442}{{\ttfamily arXiv:1811.11442 [hep-th]}}.

\bibitem{Stergiou:2019dcv}
A.~Stergiou, ``{Bootstrapping MN and Tetragonal CFTs in Three Dimensions},''
  \href{http://dx.doi.org/10.21468/SciPostPhys.7.1.010}{{\em SciPost Phys.}
  {\bfseries 7} (2019) 010},
\href{http://arxiv.org/abs/1904.00017}{{\ttfamily arXiv:1904.00017 [hep-th]}}.

\bibitem{Poland:2010wg}
D.~Poland and D.~Simmons-Duffin, ``{Bounds on 4D Conformal and Superconformal
  Field Theories},'' \href{http://dx.doi.org/10.1007/JHEP05(2011)017}{{\em
  JHEP} {\bfseries 1105} (2011) 017},
\href{http://arxiv.org/abs/1009.2087}{{\ttfamily arXiv:1009.2087 [hep-th]}}.

\bibitem{Beem:2013qxa}
C.~Beem, L.~Rastelli, and B.~C. van Rees, ``{The $\mathcal{N}=4$ Superconformal
  Bootstrap},'' \href{http://dx.doi.org/10.1103/PhysRevLett.111.071601}{{\em
  Phys.Rev.Lett.} {\bfseries 111} (2013) 071601},
\href{http://arxiv.org/abs/1304.1803}{{\ttfamily arXiv:1304.1803 [hep-th]}}.

\bibitem{Alday:2013opa}
L.~F. Alday and A.~Bissi, ``{The superconformal bootstrap for structure
  constants},'' \href{http://dx.doi.org/10.1007/JHEP09(2014)144}{{\em JHEP}
  {\bfseries 09} (2014) 144},
\href{http://arxiv.org/abs/1310.3757}{{\ttfamily arXiv:1310.3757 [hep-th]}}.

\bibitem{Alday:2014qfa}
L.~F. Alday and A.~Bissi, ``{Generalized bootstrap equations for $
  \mathcal{N}=4 $ SCFT},''
  \href{http://dx.doi.org/10.1007/JHEP02(2015)101}{{\em JHEP} {\bfseries 02}
  (2015) 101},
\href{http://arxiv.org/abs/1404.5864}{{\ttfamily arXiv:1404.5864 [hep-th]}}.

\bibitem{Chester:2014fya}
S.~M. Chester, J.~Lee, S.~S. Pufu, and R.~Yacoby, ``{The $ \mathcal{N}=8 $
  superconformal bootstrap in three dimensions},''
  \href{http://dx.doi.org/10.1007/JHEP09(2014)143}{{\em JHEP} {\bfseries 09}
  (2014) 143},
\href{http://arxiv.org/abs/1406.4814}{{\ttfamily arXiv:1406.4814 [hep-th]}}.

\bibitem{Beem:2014zpa}
C.~Beem, M.~Lemos, P.~Liendo, L.~Rastelli, and B.~C. van Rees, ``{The $
  \mathcal{N}=2 $ superconformal bootstrap},''
  \href{http://dx.doi.org/10.1007/JHEP03(2016)183}{{\em JHEP} {\bfseries 03}
  (2016) 183},
\href{http://arxiv.org/abs/1412.7541}{{\ttfamily arXiv:1412.7541 [hep-th]}}.

\bibitem{Bobev:2015jxa}
N.~Bobev, S.~El-Showk, D.~Mazac, and M.~F. Paulos, ``{Bootstrapping SCFTs with
  Four Supercharges},'' \href{http://dx.doi.org/10.1007/JHEP08(2015)142}{{\em
  JHEP} {\bfseries 08} (2015) 142},
\href{http://arxiv.org/abs/1503.02081}{{\ttfamily arXiv:1503.02081 [hep-th]}}.

\bibitem{Beem:2015aoa}
C.~Beem, M.~Lemos, L.~Rastelli, and B.~C. van Rees, ``{The (2, 0)
  superconformal bootstrap},''
  \href{http://dx.doi.org/10.1103/PhysRevD.93.025016}{{\em Phys. Rev.}
  {\bfseries D93} no.~2, (2016) 025016},
\href{http://arxiv.org/abs/1507.05637}{{\ttfamily arXiv:1507.05637 [hep-th]}}.

\bibitem{Poland:2015mta}
D.~Poland and A.~Stergiou, ``{Exploring the Minimal 4D $\mathcal{N}=1$ SCFT},''
  \href{http://dx.doi.org/10.1007/JHEP12(2015)121}{{\em JHEP} {\bfseries 12}
  (2015) 121},
\href{http://arxiv.org/abs/1509.06368}{{\ttfamily arXiv:1509.06368 [hep-th]}}.

\bibitem{Lemos:2015awa}
M.~Lemos and P.~Liendo, ``{Bootstrapping $ \mathcal{N}=2 $ chiral
  correlators},'' \href{http://dx.doi.org/10.1007/JHEP01(2016)025}{{\em JHEP}
  {\bfseries 01} (2016) 025},
\href{http://arxiv.org/abs/1510.03866}{{\ttfamily arXiv:1510.03866 [hep-th]}}.

\bibitem{Lin:2015wcg}
Y.-H. Lin, S.-H. Shao, D.~Simmons-Duffin, Y.~Wang, and X.~Yin, ``{$ \mathcal{N}
  $ = 4 superconformal bootstrap of the K3 CFT},''
  \href{http://dx.doi.org/10.1007/JHEP05(2017)126}{{\em JHEP} {\bfseries 05}
  (2017) 126},
\href{http://arxiv.org/abs/1511.04065}{{\ttfamily arXiv:1511.04065 [hep-th]}}.

\bibitem{Lin:2016gcl}
Y.-H. Lin, S.-H. Shao, Y.~Wang, and X.~Yin, ``{(2, 2) superconformal bootstrap
  in two dimensions},'' \href{http://dx.doi.org/10.1007/JHEP05(2017)112}{{\em
  JHEP} {\bfseries 05} (2017) 112},
\href{http://arxiv.org/abs/1610.05371}{{\ttfamily arXiv:1610.05371 [hep-th]}}.

\bibitem{Bae:2016jpi}
J.-B. Bae, D.~Gang, and J.~Lee, ``{3d $\mathcal{N}=2$ minimal SCFTs from
  Wrapped M5-branes},'' \href{http://dx.doi.org/10.1007/JHEP08(2017)118}{{\em
  JHEP} {\bfseries 08} (2017) 118},
\href{http://arxiv.org/abs/1610.09259}{{\ttfamily arXiv:1610.09259 [hep-th]}}.

\bibitem{Lemos:2016xke}
M.~Lemos, P.~Liendo, C.~Meneghelli, and V.~Mitev, ``{Bootstrapping
  $\mathcal{N}=3$ superconformal theories},''
  \href{http://dx.doi.org/10.1007/JHEP04(2017)032}{{\em JHEP} {\bfseries 04}
  (2017) 032},
\href{http://arxiv.org/abs/1612.01536}{{\ttfamily arXiv:1612.01536 [hep-th]}}.

\bibitem{Beem:2016wfs}
C.~Beem, L.~Rastelli, and B.~C. van Rees, ``{More ${\mathcal N}=4$
  superconformal bootstrap},''
  \href{http://dx.doi.org/10.1103/PhysRevD.96.046014}{{\em Phys. Rev.}
  {\bfseries D96} no.~4, (2017) 046014},
\href{http://arxiv.org/abs/1612.02363}{{\ttfamily arXiv:1612.02363 [hep-th]}}.

\bibitem{Cornagliotto:2017dup}
M.~Cornagliotto, M.~Lemos, and V.~Schomerus, ``{Long Multiplet Bootstrap},''
  \href{http://dx.doi.org/10.1007/JHEP10(2017)119}{{\em JHEP} {\bfseries 10}
  (2017) 119},
\href{http://arxiv.org/abs/1702.05101}{{\ttfamily arXiv:1702.05101 [hep-th]}}.

\bibitem{Chang:2017xmr}
C.-M. Chang and Y.-H. Lin, ``{Carving Out the End of the World or
  (Superconformal Bootstrap in Six Dimensions)},''
  \href{http://dx.doi.org/10.1007/JHEP08(2017)128}{{\em JHEP} {\bfseries 08}
  (2017) 128},
\href{http://arxiv.org/abs/1705.05392}{{\ttfamily arXiv:1705.05392 [hep-th]}}.

\bibitem{Cornagliotto:2017snu}
M.~Cornagliotto, M.~Lemos, and P.~Liendo, ``{Bootstrapping the $(A_1,A_2)$
  Argyres-Douglas theory},''
  \href{http://dx.doi.org/10.1007/JHEP03(2018)033}{{\em JHEP} {\bfseries 03}
  (2018) 033},
\href{http://arxiv.org/abs/1711.00016}{{\ttfamily arXiv:1711.00016 [hep-th]}}.

\bibitem{Agmon:2017xes}
N.~B. Agmon, S.~M. Chester, and S.~S. Pufu, ``{Solving M-theory with the
  Conformal Bootstrap},'' \href{http://dx.doi.org/10.1007/JHEP06(2018)159}{{\em
  JHEP} {\bfseries 06} (2018) 159},
\href{http://arxiv.org/abs/1711.07343}{{\ttfamily arXiv:1711.07343 [hep-th]}}.

\bibitem{Baggio:2017mas}
M.~Baggio, N.~Bobev, S.~M. Chester, E.~Lauria, and S.~S. Pufu, ``{Decoding a
  Three-Dimensional Conformal Manifold},''
  \href{http://dx.doi.org/10.1007/JHEP02(2018)062}{{\em JHEP} {\bfseries 02}
  (2018) 062},
\href{http://arxiv.org/abs/1712.02698}{{\ttfamily arXiv:1712.02698 [hep-th]}}.

\bibitem{Liendo:2018ukf}
P.~Liendo, C.~Meneghelli, and V.~Mitev, ``{Bootstrapping the half-BPS line
  defect},'' \href{http://dx.doi.org/10.1007/JHEP10(2018)077}{{\em JHEP}
  {\bfseries 10} (2018) 077},
\href{http://arxiv.org/abs/1806.01862}{{\ttfamily arXiv:1806.01862 [hep-th]}}.

\bibitem{Atanasov:2018kqw}
A.~Atanasov, A.~Hillman, and D.~Poland, ``{Bootstrapping the Minimal 3D
  SCFT},'' \href{http://dx.doi.org/10.1007/JHEP11(2018)140}{{\em JHEP}
  {\bfseries 11} (2018) 140},
\href{http://arxiv.org/abs/1807.05702}{{\ttfamily arXiv:1807.05702 [hep-th]}}.

\bibitem{Chang:2019dzt}
C.-M. Chang, M.~Fluder, Y.-H. Lin, S.-H. Shao, and Y.~Wang, ``{3d N=4 Bootstrap
  and Mirror Symmetry},''
\href{http://arxiv.org/abs/1910.03600}{{\ttfamily arXiv:1910.03600 [hep-th]}}.

\bibitem{Rattazzi:2010gj}
R.~Rattazzi, S.~Rychkov, and A.~Vichi, ``{Central Charge Bounds in 4D Conformal
  Field Theory},'' \href{http://dx.doi.org/10.1103/PhysRevD.83.046011}{{\em
  Phys.Rev.} {\bfseries D83} (2011) 046011},
\href{http://arxiv.org/abs/1009.2725}{{\ttfamily arXiv:1009.2725 [hep-th]}}.

\bibitem{Caracciolo:2009bx}
F.~Caracciolo and V.~S. Rychkov, ``{Rigorous Limits on the Interaction Strength
  in Quantum Field Theory},''
  \href{http://dx.doi.org/10.1103/PhysRevD.81.085037}{{\em Phys. Rev.}
  {\bfseries D81} (2010) 085037},
\href{http://arxiv.org/abs/0912.2726}{{\ttfamily arXiv:0912.2726 [hep-th]}}.

\bibitem{Liendo:2012hy}
P.~Liendo, L.~Rastelli, and B.~C. van Rees, ``{The Bootstrap Program for
  Boundary CFT$_d$},'' \href{http://dx.doi.org/10.1007/JHEP07(2013)113}{{\em
  JHEP} {\bfseries 1307} (2013) 113},
\href{http://arxiv.org/abs/1210.4258}{{\ttfamily arXiv:1210.4258 [hep-th]}}.

\bibitem{ElShowk:2012hu}
S.~El-Showk and M.~F. Paulos, ``{Bootstrapping Conformal Field Theories with
  the Extremal Functional Method},''
  \href{http://dx.doi.org/10.1103/PhysRevLett.111.241601}{{\em Phys. Rev.
  Lett.} {\bfseries 111} no.~24, (2013) 241601},
\href{http://arxiv.org/abs/1211.2810}{{\ttfamily arXiv:1211.2810 [hep-th]}}.

\bibitem{Nakayama:2016cim}
Y.~Nakayama, ``{Bootstrapping critical Ising model on three-dimensional real
  projective space},''
  \href{http://dx.doi.org/10.1103/PhysRevLett.116.141602}{{\em Phys. Rev.
  Lett.} {\bfseries 116} no.~14, (2016) 141602},
\href{http://arxiv.org/abs/1601.06851}{{\ttfamily arXiv:1601.06851 [hep-th]}}.

\bibitem{Echeverri:2016ztu}
A.~Castedo~Echeverri, B.~von Harling, and M.~Serone, ``{The Effective
  Bootstrap},'' \href{http://dx.doi.org/10.1007/JHEP09(2016)097}{{\em JHEP}
  {\bfseries 09} (2016) 097},
\href{http://arxiv.org/abs/1606.02771}{{\ttfamily arXiv:1606.02771 [hep-th]}}.

\bibitem{Cappelli:2018vir}
A.~Cappelli, L.~Maffi, and S.~Okuda, ``{Critical Ising Model in Varying
  Dimension by Conformal Bootstrap},''
  \href{http://dx.doi.org/10.1007/JHEP01(2019)161}{{\em JHEP} {\bfseries 01}
  (2019) 161},
\href{http://arxiv.org/abs/1811.07751}{{\ttfamily arXiv:1811.07751 [hep-th]}}.

\bibitem{Simmons-Duffin:2015qma}
D.~Simmons-Duffin, ``{A Semidefinite Program Solver for the Conformal
  Bootstrap},'' \href{http://dx.doi.org/10.1007/JHEP06(2015)174}{{\em JHEP}
  {\bfseries 06} (2015) 174},
\href{http://arxiv.org/abs/1502.02033}{{\ttfamily arXiv:1502.02033 [hep-th]}}.

\bibitem{Landry:2019qug}
W.~Landry and D.~Simmons-Duffin, ``{Scaling the semidefinite program solver
  SDPB},''
\href{http://arxiv.org/abs/1909.09745}{{\ttfamily arXiv:1909.09745 [hep-th]}}.

\bibitem{Go:2019lke}
M.~Go and Y.~Tachikawa, ``{autoboot: A generator of bootstrap equations with
  global symmetry},'' \href{http://dx.doi.org/10.1007/JHEP06(2019)084}{{\em
  JHEP} {\bfseries 06} (2019) 084},
\href{http://arxiv.org/abs/1903.10522}{{\ttfamily arXiv:1903.10522 [hep-th]}}.

\bibitem{Lipa:2003zz}
J.~Lipa, J.~Nissen, D.~Stricker, D.~Swanson, and T.~Chui, ``{Specific heat of
  liquid helium in zero gravity very near the lambda point},''
\href{http://dx.doi.org/10.1103/PhysRevB.68.174518}{{\em Phys.Rev.} {\bfseries
  B68} (2003) 174518}.

\bibitem{Hasenbusch:2019jkj}
M.~Hasenbusch, ``{Monte Carlo study of an improved clock model in three
  dimensions},'' \href{http://dx.doi.org/10.1103/PhysRevB.100.224517}{{\em
  Phys. Rev. B} {\bfseries 100} no.~22, (2019) 224517},
  \href{http://arxiv.org/abs/1910.05916}{{\ttfamily arXiv:1910.05916
  [cond-mat.stat-mech]}}.

\bibitem{PhysRevB.84.125136}
M.~Hasenbusch and E.~Vicari, ``Anisotropic perturbations in three-dimensional
  o($n$)-symmetric vector models,''
  \href{http://dx.doi.org/10.1103/PhysRevB.84.125136}{{\em Phys. Rev. B}
  {\bfseries 84} (Sep, 2011) 125136},
  \href{http://arxiv.org/abs/1108.0491}{{\ttfamily arXiv:1108.0491
  [cond-mat.stat-mech]}}.

\bibitem{tilley1990superfluidity}
D.~Tilley and J.~Tilley, {\em Superfluidity and Superconductivity}.
\newblock Graduate Student Series in Physics. Taylor \& Francis, 1990.
\newblock \url{https://books.google.it/books?id=I6JtWd3J8MIC}.

\bibitem{Moldover:1979zz}
M.~R. Moldover, J.~V. Sengers, R.~W. Gammon, and R.~J. Hocken, ``{Gravity
  effects in fluids near the gas-liquid critical point},''
\href{http://dx.doi.org/10.1103/RevModPhys.51.79}{{\em Rev. Mod. Phys.}
  {\bfseries 51} (1979) 79--99}.

\bibitem{Lipa:1996zz}
J.~A. Lipa, D.~R. Swanson, J.~A. Nissen, T.~C.~P. Chui, and U.~E. Israelsson,
  ``{Heat Capacity and Thermal Relaxation of Bulk Helium very near the Lambda
  Point},''
\href{http://dx.doi.org/10.1103/PhysRevLett.76.944}{{\em Phys. Rev. Lett.}
  {\bfseries 76} (1996) 944--947}.

\bibitem{Lipa:2000zz}
J.~A. Lipa, D.~R. Swanson, J.~A. Nissen, Z.~K. Geng, P.~R. Williamson, D.~A.
  Stricker, T.~C.~P. Chui, U.~E. Israelsson, and M.~Larson, ``{Specific Heat of
  Helium Confined to a 57- mum Planar Geometry near the Lambda Point},''
\href{http://dx.doi.org/10.1103/PhysRevLett.84.4894}{{\em Phys. Rev. Lett.}
  {\bfseries 84} (2000) 4894--4897}.

\bibitem{Pelissetto:2000ek}
A.~Pelissetto and E.~Vicari, ``{Critical phenomena and renormalization-group
  theory},'' \href{http://dx.doi.org/10.1016/S0370-1573(02)00219-3}{{\em Phys.
  Rept.} {\bfseries 368} (2002) 549--727},
\href{http://arxiv.org/abs/cond-mat/0012164}{{\ttfamily
  arXiv:cond-mat/0012164}}.

\bibitem{Burovski_2006}
E.~Burovski, J.~Machta, N.~Prokof’ev, and B.~Svistunov, ``High-precision
  measurement of the thermal exponent for the three-dimensional xy universality
  class,'' \href{http://dx.doi.org/10.1103/physrevb.74.132502}{{\em Physical
  Review B} {\bfseries 74} no.~13, (Oct, 2006) }.
  \url{http://dx.doi.org/10.1103/PhysRevB.74.132502}.

\bibitem{Sokolov:2014mfa}
A.~I. Sokolov and M.~A. Nikitina, ``{Critical Exponents of Superfluid Helium
  and Pseudo-$\epsilon$ Expansion},''
  \href{http://dx.doi.org/10.1016/j.physa.2015.10.036}{{\em Physica} {\bfseries
  A444} (2016) 177},
\href{http://arxiv.org/abs/1402.4318}{{\ttfamily arXiv:1402.4318
  [cond-mat.stat-mech]}}.

\bibitem{Xu:2019mvy}
W.~Xu, Y.~Sun, J.-P. Lv, and Y.~Deng, ``{High-precision Monte Carlo study of
  several models in the three-dimensional U(1) universality class},''
  \href{http://dx.doi.org/10.1103/PhysRevB.100.064525}{{\em Phys. Rev.}
  {\bfseries B100} no.~6, (2019) 064525},
\href{http://arxiv.org/abs/1908.10990}{{\ttfamily arXiv:1908.10990
  [cond-mat.stat-mech]}}.

\bibitem{Campostrini:2006ms}
M.~Campostrini, M.~Hasenbusch, A.~Pelissetto, and E.~Vicari, ``{The Critical
  exponents of the superfluid transition in He-4},''
  \href{http://dx.doi.org/10.1103/PhysRevB.74.144506}{{\em Phys. Rev.}
  {\bfseries B74} (2006) 144506},
\href{http://arxiv.org/abs/cond-mat/0605083}{{\ttfamily arXiv:cond-mat/0605083
  [cond-mat]}}.

\bibitem{Simmons-Duffin:2016wlq}
D.~Simmons-Duffin, ``{The Lightcone Bootstrap and the Spectrum of the 3d Ising
  CFT},'' \href{http://dx.doi.org/10.1007/JHEP03(2017)086}{{\em JHEP}
  {\bfseries 03} (2017) 086},
\href{http://arxiv.org/abs/1612.08471}{{\ttfamily arXiv:1612.08471 [hep-th]}}.

\bibitem{Albayrak:2019gnz}
S.~Albayrak, D.~Meltzer, and D.~Poland, ``{More Analytic Bootstrap:
  Nonperturbative Effects and Fermions},''
  \href{http://dx.doi.org/10.1007/JHEP08(2019)040}{{\em JHEP} {\bfseries 08}
  (2019) 040},
\href{http://arxiv.org/abs/1904.00032}{{\ttfamily arXiv:1904.00032 [hep-th]}}.

\bibitem{Guida:1998bx}
R.~Guida and J.~Zinn-Justin, ``{Critical exponents of the N vector model},''
  \href{http://dx.doi.org/10.1088/0305-4470/31/40/006}{{\em J. Phys.}
  {\bfseries A31} (1998) 8103--8121},
\href{http://arxiv.org/abs/cond-mat/9803240}{{\ttfamily arXiv:cond-mat/9803240
  [cond-mat]}}.

\bibitem{Jasch_2001}
F.~Jasch and H.~Kleinert, ``Fast-convergent resummation algorithm and critical
  exponents of phi4-theory in three dimensions,''
  \href{http://dx.doi.org/10.1063/1.1289377}{{\em Journal of Mathematical
  Physics} {\bfseries 42} no.~1, (Jan, 2001) 52--73}.
  \url{http://dx.doi.org/10.1063/1.1289377}.

\bibitem{Rychkov:2015naa}
S.~Rychkov and Z.~M. Tan, ``{The $\epsilon$-expansion from conformal field
  theory},'' \href{http://dx.doi.org/10.1088/1751-8113/48/29/29FT01}{{\em J.
  Phys.} {\bfseries A48} no.~29, (2015) 29FT01},
\href{http://arxiv.org/abs/1505.00963}{{\ttfamily arXiv:1505.00963 [hep-th]}}.

\bibitem{Calabrese:2002bm}
P.~Calabrese, A.~Pelissetto, and E.~Vicari, ``{Multicritical phenomena in
  O(n(1)) + O(n(2)) symmetric theories},''
  \href{http://dx.doi.org/10.1103/PhysRevB.67.054505}{{\em Phys. Rev.}
  {\bfseries B67} (2003) 054505},
\href{http://arxiv.org/abs/cond-mat/0209580}{{\ttfamily arXiv:cond-mat/0209580
  [cond-mat]}}.

\bibitem{DePrato:2003yd}
M.~De~Prato, A.~Pelissetto, and E.~Vicari, ``{Third harmonic exponent in
  three-dimensional N vector models},''
  \href{http://dx.doi.org/10.1103/PhysRevB.68.092403}{{\em Phys. Rev.}
  {\bfseries B68} (2003) 092403},
\href{http://arxiv.org/abs/cond-mat/0302145}{{\ttfamily arXiv:cond-mat/0302145
  [cond-mat]}}.

\bibitem{Caselle:1997gf}
M.~Caselle and M.~Hasenbusch, ``{The Stability of the O(N) invariant fixed
  point in three-dimensions},''
  \href{http://dx.doi.org/10.1088/0305-4470/31/20/004}{{\em J. Phys.}
  {\bfseries A31} (1998) 4603--4617},
\href{http://arxiv.org/abs/cond-mat/9711080}{{\ttfamily arXiv:cond-mat/9711080
  [cond-mat]}}.

\bibitem{Carmona:1999rm}
J.~M. Carmona, A.~Pelissetto, and E.~Vicari, ``{The N component Ginzburg-Landau
  Hamiltonian with cubic anisotropy: A Six loop study},''
  \href{http://dx.doi.org/10.1103/PhysRevB.61.15136}{{\em Phys. Rev.}
  {\bfseries B61} (2000) 15136--15151},
\href{http://arxiv.org/abs/cond-mat/9912115}{{\ttfamily arXiv:cond-mat/9912115
  [cond-mat]}}.

\bibitem{Shao:2019dbi}
H.~Shao, W.~Guo, and A.~W. Sandvik, ``{Monte Carlo Renormalization Flows in the
  Space of Relevant and Irrelevant Operators: Application to Three-Dimensional
  Clock Models},''
\href{http://arxiv.org/abs/1905.13640}{{\ttfamily arXiv:1905.13640
  [cond-mat.str-el]}}.

\bibitem{Nachtmann:1973mr}
O.~Nachtmann, ``{Positivity constraints for anomalous dimensions},''
\href{http://dx.doi.org/10.1016/0550-3213(73)90144-2}{{\em Nucl.Phys.}
  {\bfseries B63} (1973) 237--247}.

\bibitem{Komargodski:2012ek}
Z.~Komargodski and A.~Zhiboedov, ``{Convexity and Liberation at Large Spin},''
  \href{http://dx.doi.org/10.1007/JHEP11(2013)140}{{\em JHEP} {\bfseries 1311}
  (2013) 140},
\href{http://arxiv.org/abs/1212.4103}{{\ttfamily arXiv:1212.4103 [hep-th]}}.

\bibitem{Costa:2017twz}
M.~S. Costa, T.~Hansen, and J.~Penedones, ``{Bounds for OPE coefficients on the
  Regge trajectory},'' \href{http://dx.doi.org/10.1007/JHEP10(2017)197}{{\em
  JHEP} {\bfseries 10} (2017) 197},
\href{http://arxiv.org/abs/1707.07689}{{\ttfamily arXiv:1707.07689 [hep-th]}}.

\bibitem{Fitzpatrick:2012yx}
A.~L. Fitzpatrick, J.~Kaplan, D.~Poland, and D.~Simmons-Duffin, ``{The Analytic
  Bootstrap and AdS Superhorizon Locality},''
  \href{http://dx.doi.org/10.1007/JHEP12(2013)004}{{\em JHEP} {\bfseries 12}
  (2013) 004},
\href{http://arxiv.org/abs/1212.3616}{{\ttfamily arXiv:1212.3616 [hep-th]}}.

\bibitem{Meltzer:2018tnm}
D.~Meltzer, ``{Higher Spin ANEC and the Space of CFTs},''
  \href{http://dx.doi.org/10.1007/JHEP07(2019)001}{{\em JHEP} {\bfseries 07}
  (2019) 001},
\href{http://arxiv.org/abs/1811.01913}{{\ttfamily arXiv:1811.01913 [hep-th]}}.

\bibitem{SlavaUnpublished}
{Rychkov, S.} {\it unpublished work}.

\bibitem{Park2017GeneralHF}
J.~H. Park and S.~Boyd, ``General heuristics for nonconvex quadratically
  constrained quadratic programming,''
\newblock 2017.

\bibitem{iterativerankpenalty}
C.~Sun and R.~Dai, ``An iterative rank penalty method for nonconvex
  quadratically constrained quadratic programs,''
  \href{http://dx.doi.org/10.1137/17M1147214}{{\em SIAM Journal on Control and
  Optimization} {\bfseries 57} (01, 2019) 3749–3766}.

\bibitem{Barber96thequickhull}
C.~B. Barber, D.~P. Dobkin, and H.~Huhdanpaa, ``The quickhull algorithm for
  convex hulls,'' {\em ACM TRANSACTIONS ON MATHEMATICAL SOFTWARE} {\bfseries
  22} no.~4, (1996) 469--483.

\bibitem{Calabrese:2004ca}
P.~Calabrese and P.~Parruccini, ``{Harmonic crossover exponents in O(n) models
  with the pseudo-epsilon expansion approach},''
  \href{http://dx.doi.org/10.1103/PhysRevB.71.064416}{{\em Phys. Rev.}
  {\bfseries B71} (2005) 064416},
\href{http://arxiv.org/abs/cond-mat/0411027}{{\ttfamily arXiv:cond-mat/0411027
  [cond-mat]}}.

\bibitem{Katz:2014rla}
E.~Katz, S.~Sachdev, E.~S. S{\o}rensen, and W.~Witczak-Krempa, ``{Conformal
  field theories at nonzero temperature: Operator product expansions, Monte
  Carlo, and holography},''
  \href{http://dx.doi.org/10.1103/PhysRevB.90.245109}{{\em Phys.Rev.}
  {\bfseries B90} no.~24, (2014) 245109},
\href{http://arxiv.org/abs/1409.3841}{{\ttfamily arXiv:1409.3841
  [cond-mat.str-el]}}.

\bibitem{Iliesiu:2018fao}
L.~Iliesiu, M.~Kologlu, R.~Mahajan, E.~Perlmutter, and D.~Simmons-Duffin,
  ``{The Conformal Bootstrap at Finite Temperature},''
  \href{http://dx.doi.org/10.1007/JHEP10(2018)070}{{\em JHEP} {\bfseries 10}
  (2018) 070},
\href{http://arxiv.org/abs/1802.10266}{{\ttfamily arXiv:1802.10266 [hep-th]}}.

\bibitem{6866038}
J.~Towns, T.~Cockerill, M.~Dahan, I.~Foster, K.~Gaither, A.~Grimshaw,
  V.~Hazlewood, S.~Lathrop, D.~Lifka, G.~D. Peterson, R.~Roskies, J.~Scott, and
  N.~Wilkins-Diehr, ``Xsede: Accelerating scientific discovery,''
  \href{http://dx.doi.org/10.1109/MCSE.2014.80}{{\em Computing in Science and
  Engineering} {\bfseries 16} no.~05, (Sep, 2014) 62--74}.

\end{thebibliography}\endgroup
\bibliographystyle{utphys}
\end{document}